\documentclass[aps, pra, reprint, superscriptaddress, floatfix]{revtex4-1} 

\usepackage[free-standing-units=true]{siunitx}
\usepackage{xspace,amsmath} 
\usepackage{graphicx}
\usepackage{dsfont}
\usepackage{epstopdf}
\usepackage{bm, amsmath, amssymb, bbold}
\usepackage{float}
\usepackage{placeins}
\usepackage{xcolor}
\usepackage[T1]{fontenc}
\usepackage[separate-uncertainty=true]{siunitx}
\usepackage{boldline,multirow}
\usepackage{makecell}
\usepackage[hidelinks]{hyperref}

\newcommand{\ket}[1]{\left|#1\right>}
\newcommand{\bra}[1]{\left<#1\right|}

\newcommand{\citeMethods}{(see Methods)}

\newcommand{\omegasq}{\omega_{\mathrm{sq}}}

\newcommand{\omegaqubit}{\omega_{\mathrm{a}}}
\newcommand{\omegacav}{\omega_{\mathrm{b}}}
\newcommand{\omegacqr}{\omega_{\mathrm{zro}}}
\newcommand{\omegadiss}{\omega_{\mathrm{diss}}}

\newcommand{\omegage}{\omega_{\mathrm{01}}}
\newcommand{\omegaef}{\omega_{\mathrm{12}}}
\newcommand{\omegaRabi}{\omega_{\mathrm{R},ij}}
\newcommand{\Rabikq}[1]{\Omega_{{#1}}}
\newcommand{\omegarf}{\omega_\mathrm{r}}
\newcommand{\omegakq}[2]{\omega_{\mathrm{#1}}^{\mathrm{#2}}}
\newcommand{\Deltab}{\Delta_{\mathrm{b}}}
\newcommand{\omegaz}{\Omega_{\mathrm{z}}/2\pi}
\newcommand{\omegazval}{(5.05 \pm 0.09) \, \mathrm{MHz}}
\newcommand{\rdct}[1]{A_{#1}}
\newcommand{\rdctitalic}[1]{A_{#1}}
\newcommand{\omegaprobe}{\omega_{\mathrm{probe}}}

\newcommand{\Hkcq}{\hat{H}_{\mathrm{KCQ}}}
\newcommand{\Hcqr}{\hat{H}_{\mathrm{zro}}}
\newcommand{\Hz}{\hat{H}_{\mathrm{Z}}}
\newcommand{\Hk}{\hat{H}_{\mathrm{K}}}
\newcommand{\Hdisp}{\hat{H}_{\mathrm{disp.}}}
\newcommand{\Hstark}{\hat{H}_{\mathrm{Stark}}}
\newcommand{\epstwo}{\varepsilon_{2}}
\newcommand{\epsz}{\varepsilon_{\mathrm{Z}}}
\newcommand{\epszval}{\SI{494}{\kilo \hertz}}
\newcommand{\epszvalOne}{\SI{307}{\kilo \hertz}}
\newcommand{\gcqr}{g_{\mathrm{zro}}}
\newcommand{\gdiss}{g_{\mathrm{diss}}}

\newcommand{\chidisp}{\chi_{\mathrm{ab}}}
\newcommand{\Deltacrit}{\Delta_\mathrm{c}}
\newcommand{\alphaD}{\alpha_{\Delta}}
\newcommand{\dE}{\Delta E}
\newcommand{\Zdiss}{\langle Z\rangle}
\newcommand{\Zdissbg}{\langle Z\rangle_\mathrm{bg}}
\newcommand{\Hdiss}{\hat{H}_{\mathrm{KCQ,diss}}}

\newcommand{\phiz}{\phi_\mathrm{Z}}
\newcommand{\kappaqubit}{\kappa_{\mathrm{a}}}
\newcommand{\kappacav}{\kappa_{\mathrm{b}}}

\newcommand{\kappaoneeff}{\kappa_{\mathrm{a}}^{\mathrm{eff}}}
\newcommand{\kappaphieff}{\kappa_{\phi}^{\mathrm{eff}}}
\newcommand{\kappaoneeffval}{\SI{4.2}{\kilo\hertz}}
\newcommand{\kappaphieffval}{\SI{21.2}{\kilo \hertz}}

\newcommand{\kappaphi}{\kappa_\mathrm{\phi}}

\newcommand{\kappadiss}{\kappa_{\mathrm{diss}}}

\newcommand{\BigGamma}[1]{\Gamma_{\mathrm{#1}}}
\newcommand{\gammarabikq}{\gamma_{\mathrm{Rabi}}}
\newcommand{\gammarabival}{((2.91 \pm 0.5)\, \mathrm{\mu s})^{-1}}
\newcommand{\kappagdiss}{\kappa_{\mathrm{tot}}}

\newcommand{\kappatwophoton}{\kappa_{\mathrm{a}}^{(2)}}
\newcommand{\nthtwophoton}{n_{\mathrm{th,a}}^{(2)}}

\newcommand{\Tone}{T_{1}}
\newcommand{\TtwoR}{T_{2\mathrm{R}}}
\newcommand{\TtwoE}{T_{2\mathrm{E}}}
\newcommand{\Tz}{T_{\mathrm{Z}}}
\newcommand{\Tx}{T_{\mathrm{X}}}
\newcommand{\Ty}{T_{\mathrm{Y}}}

\newcommand{\Tonekq}[1]{T_{\mathrm{1}}^{{#1}}}
\newcommand{\Tphikq}[1]{T_{\mathrm{\phi}}^{{#1}}}
\newcommand{\kappaonekq}[1]{\kappa_{\mathrm{1}}^{#1}}
\newcommand{\kappaheatingkq}[1]{\kappa_{\uparrow}^{#1}}
\newcommand{\kappaphikq}[1]{\kappa_{\mathrm{\phi}}^{#1}}

\newcommand{\kappaphikqv}{\kappa_{\mathrm{\phi}}}

\newcommand{\nth}{n_{\mathrm{th,a}}}
\newcommand{\nthb}{n_{\mathrm{th,b}}}
\newcommand{\nthbval}{0.025}
\newcommand{\nthbvalfigthree}{0.004 }
\newcommand{\nthvalsim}{0.025}
\newcommand{\Tcav}{T_\mathrm{b}}
\newcommand{\Tcavval}{\SI{116}{\milli \kelvin}}
\newcommand{\Tcavvalfigthree}{\SI{78}{\milli \kelvin}}
\newcommand{\Tkcq}{T_\mathrm{a}}
\newcommand{\Tkcqval}{\SI{82}{\milli \kelvin}}
\newcommand{\nthplus}{n_{\mathrm{th,a}}^{+}}

\newcommand{\tauwait}{\tau_{\mathrm{wait}}}
\newcommand{\taurampsq}{\tau_{\mathrm{ramp,sq}}}
\newcommand{\taurampD}{\tau_{\mathrm{ramp,\Delta}}}
\newcommand{\pulseamp}{A}
\newcommand{\pige}{\pi_{\mathrm{01}}}
\newcommand{\pief}{\pi_{\mathrm{12}}}
\newcommand{\Adiss}{A_{\mathrm{diss}}}
\newcommand{\dt}{\Delta t}

\newcommand{\tauxgate}{\tau_{\pi/2}^{\mathrm{X}}}
\newcommand{\tauzgate}{\tau_{\mathrm{gate}}^{\mathrm{Z}}}

\newcommand{\tauRij}{\tau_{\mathrm{R},ij}}
\newcommand{\taucav}{\tau_{\mathrm{cav}}}
\newcommand{\taudelay}{\tau_{\mathrm{delay}}}
\newcommand{\Vsq}{V_{\mathrm{sq}}}

\newcommand{\pg}{p_{0}}
\newcommand{\pe}{p_{1}}
\newcommand{\pf}{p_{2}}

\newcommand{\pesim}{\pe^{\mathrm{sim}}}

\newcommand{\pfvalfit}{1.33}

\newcommand{\peqhval}{7.0\%}
\newcommand{\peval}{(9.2\pm0.3)\%}
\newcommand{\pevalratio}{(9.9\pm0.4)\%}
\newcommand{\epstr}{\varepsilon_{\mathrm{2,th}}}
\newcommand{\Fcqr}{\mathcal{F}^{\mathrm{zro}}}
\newcommand{\Qcqr}{\mathcal{Q}^{\mathrm{zro}}}
\newcommand{\bitflipprob}{\xi_{\mathrm{Z}}}

\newcommand{\gstate}{\psi_{\mathrm{0}}^{\pm}}
\newcommand{\estate}{\psi_{\mathrm{1}}^{\pm}}
\newcommand{\fstate}{\psi_{\mathrm{2}}^{\pm}}
\newcommand{\hstate}{\psi_{\mathrm{3}}^{\pm}}

\newcommand{\estateflip}{\psi_{\mathrm{1}}^{\mp}}
\newcommand{\fstateflip}{\psi_{\mathrm{2}}^{\mp}}
\newcommand{\hstateflip}{\psi_{\mathrm{3}}^{\mp}}
\newcommand{\transge}{\ket{\gstate} \leftrightarrow \ket{\estateflip}}
\newcommand{\transef}{\ket{\estate} \leftrightarrow \ket{\fstateflip}}
\newcommand{\transfh}{\ket{\fstate} \leftrightarrow \ket{\hstateflip}}
\newcommand{\transeg}{\ket{\estateflip} \leftrightarrow \ket{\gstate}}
\newcommand{\transfe}{\ket{\fstateflip} \leftrightarrow \ket{\estate}}
\newcommand{\transhf}{\ket{\hstateflip} \leftrightarrow \ket{\fstate}}
\newcommand{\rhokq}[2]{\hat{\rho}_{\mathrm{#1}}^{\mathrm{#2}}}
\newcommand{\rhoss}{\hat{\rho}_\mathrm{ss}}
\newcommand{\rhof}{\hat{\rho}_\mathrm{f}}

\newcommand{\Nsnail}{N_{\mathrm{SNAIL}}}
\newcommand{\Ec}{E_{\mathrm{c}}}

\newcommand{\Lj}{L_{\mathrm{j}}}
\newcommand{\Llin}{L_{\mathrm{lin.}}}
\newcommand{\jjasym}{\alpha}
\newcommand{\nbar}{\bar{n}}

\newcommand{\rdckq}[1]{A_{#1}}
\newcommand{\rdsgkq}[1]{\mathcal{M}_{#1}}

\newcommand{\relpeak}[1]{\mathcal{D}_{#1}}
\newcommand{\relnbar}[1]{\eta_{\mathrm{#1}}}

\newcommand{\fluxpoint}{\SI{0.202}{}\xspace}
\newcommand{\fqubit}{\SI{6.371}{\giga\hertz}\xspace}
\newcommand{\fres}{\SI{9.018}{\giga\hertz}\xspace}
\newcommand{\kerr}{\SI{1.74}{\mega\hertz}\xspace}
\newcommand{\Ecval}{\SI{135}{\mega\hertz}\xspace}
\newcommand{\bareTone}{\SI{55.7 \pm 0.7}{\micro\second}\xspace}
\newcommand{\bareTphi}{\SI{13.2 \pm 0.5}{\micro\second}\xspace}
\newcommand{\bareTecho}{\SI{40.9 \pm 1.3}{\micro\second}\xspace}
\newcommand{\kappacavval}{\SI{680}{\kilo\hertz}\xspace}
\newcommand{\dispshiftval}{\SI{180}{\kilo\hertz}\xspace}
\newcommand{\epstwofigtwo}{{2.4}}
\newcommand{\Deltafigtwo}{{8}}
\newcommand{\Deltafigthree}{{7}}

\newcommand{\cqrfidelity}{99.4\%}
\newcommand{\tauwaitval}{\SI{300}{\micro\second}\xspace}

\newcommand{\sigmaval}{\SI{7.14}{\milli\volt}}

\newcommand{\nthval}{0.007} 
\newcommand{\kappadissfigthree}{\SI{120}{\kilo\hertz}\xspace}

\newcommand{\kerrgatetimeest}{\SI{143}{\nano\second}}
\newcommand{\kerrgatetimemmt}{\SI{132}{\nano\second}}
\newcommand{\Ljval}{\SI{0.52}{\nano H}\xspace}
\newcommand{\Llinval}{$\approx \SI{180}{\pico H}\xspace$}
\newcommand{\gthreeval}{\SI{-6.5}{\mega\hertz}\xspace}
\newcommand{\gCQRval}{ (800 \pm 5)\, \text{kHz}}
\newcommand{\alphasnail}{$\approx \, 0.085$}
\newcommand{\kappaout}{\SI{524 \pm 6}{\kilo\hertz}\xspace}
\newcommand{\kappaother}{\SI{157 \pm 7}{\kilo\hertz}\xspace}

\newcommand{\gdissval}{\SI{166}{\kilo\hertz}}

\newcommand{\nthplusval}{0.018}
\newcommand{\penthval}{7.6\%}
\newcommand{\kappaphival}{\SI{16.4}{\hertz}}

\newcommand{\timeresOPX}{4 \, \mathrm{ns}}
\newcommand{\cqrnshots}{5\times 10^5}
\newcommand{\taurampval}{\SI{1}{\micro\second}}
\newcommand{\tauchirpval}{\SI{5.6}{\micro\second}}
\newcommand{\sigmarampval}{\SI{200}{\nano\second}}
\newcommand{\sigmachirpval}{\SI{1.12}{\micro\second}}

\newcommand{\taucavval}{\SI{1.2}{\micro \second}}

\newcommand{\taupulseval}{\SI{2}{\micro \second}}

\newcommand{\calC}{\mathcal{C}}
\newcommand{\calN}{\mathcal{N}}
\newcommand{\aop}{\hat{a}}
\newcommand{\aod}{\hat{a}^\dagger}
\newcommand{\aodtwo}{\hat{a}^{\dagger 2}}
\newcommand{\aoptwo}{\hat{a}^2}
\newcommand{\bo}{\hat{b}}
\newcommand{\bod}{\hat{b}^\dagger}

\newcommand{\Deltaab}{\Delta_{\mathrm{ab}}}
\newcommand{\gthre}{g_3}
\newcommand{\gfour}{g_4}
\newcommand{\gfoureff}{g_{4}^\mathrm{eff}}
\newcommand{\epssq}{\epsilon_{\mathrm{sq}}}
\newcommand{\epscqr}{\epsilon_{\mathrm{zro}}}
\newcommand{\xisq}[1]{\xi_{\mathrm{sq}}^{\mathrm{#1}}}
\newcommand{\xicqr}[1]{\xi_{\mathrm{zro}}^{\mathrm{#1}}}
\newcommand{\ypm}{\ket{\calC_{\alpha}^{\pm i}}}

\newcommand{\catp}{\ket{\calC_{\alpha}^+}}
\newcommand{\catm}{\ket{\calC_{\alpha}^-}}
\newcommand{\catpm}{\ket{\calC_{\alpha}^\pm}}

\newcommand{\psikq}[2]{\psi_{#1}^{#2}}
\newcommand{\psii}{\ket{\psi_i^\pm}}

\newcommand{\KQ}{KCQ}
\newcommand{\CQR}{ZRO}

\newcommand{\fixedket}[2]{\left| #1 \vphantom{#2} \right\rangle}
\newcommand{\fixedbra}[2]{\left\langle #1 \vphantom{#2} \right|}

\newcommand{\figwidth}{89mm} %3.375in or \columnwidth,
\newcommand{\figwidthWide}{183mm}

\newcommand{\figone}{1}
\newcommand{\figtwo}{2}
\newcommand{\figthree}{3}
\newcommand{\figfour}{4}
\newcommand{\figfive}{5}
\newcommand{\workingdelta}{$\Delta = 8K$}
\newcommand{\workingepstwo}{$\epstwo = 2.4K$}
\newcommand{\workingpoint}{\workingepstwo, \workingdelta}

\newcommand{\epstwodevtwo}{5.14}
\newcommand{\Deltadevtwo}{8}
\newcommand{\nthvaldevtwo}{0.017}
\newcommand{\nthbvaldevtwo}{0.0242}
\newcommand{\nthplusvaldevtwo}{0.175}
\newcommand{\kappaphivaldevtwo}{\SI{65}{\hertz}}
\newcommand{\Tkcqvaldevtwo}{\SI{165}{\milli\kelvin}}
\newcommand{\gdissmaxdevtwo}{\SI{120}{\kilo\hertz}}
\newcommand{\pezerodevtwo}{15.0\%}
\newcommand{\pesatdevtwo}{3.7\%}
\newcommand{\pelossdevtwo}{1.6\%}
\newcommand{\penthdevtwo}{3.3\%}

\newenvironment{ac}
{\begin{center}
		\textsc{author contributions}\\[1ex]
	\end{center}
}

\newenvironment{ack}
{\begin{center}
		\textsc{acknowledgments}\\[1ex]
	\end{center}
}

\newenvironment{meth}
{\begin{center}
		\textsc{methods}\\[1ex]
	\end{center}
}

\newenvironment{ai}
{\begin{center}
		\textsc{author information}\\[1ex]
	\end{center}
}

\newenvironment{datav}
{\begin{center}
		\textsc{data and code availability statement}\\[1ex]
	\end{center}
}

\begin{document}

\title{Enhancing Kerr-Cat Qubit Coherence with Engineered Dissipation}

\author{F.~Adinolfi}
\thanks{These authors contributed equally.}
\affiliation{PSI Center for Photon Science, 5232 Villigen PSI, Switzerland}
\author{D.~Z.~Haxell}
\thanks{These authors contributed equally.}
\affiliation{PSI Center for Photon Science, 5232 Villigen PSI, Switzerland}
\author{A.~Bruno}
\affiliation{PSI Center for Photon Science, 5232 Villigen PSI, Switzerland}
\affiliation{Swiss Nanoscience Institute, University of Basel, Klingelbergstrasse 82, 4056 Basel, Switzerland}
\author{L.~Michaud}
\affiliation{PSI Center for Photon Science, 5232 Villigen PSI, Switzerland}
\author{V.~Hasanuzzaman~Kamrul}
\affiliation{PSI Center for Photon Science, 5232 Villigen PSI, Switzerland}
\author{P.~Pandey}
\affiliation{PSI Center for Photon Science, 5232 Villigen PSI, Switzerland}
\author{A.~Grimm}
\affiliation{PSI Center for Photon Science, 5232 Villigen PSI, Switzerland}
\email{alexander.grimm@psi.ch}

\date{\today}
\begin{abstract}
Quantum computing crucially relies on maintaining quantum coherence for the duration of a calculation. Bosonic quantum error correction protects this coherence by encoding qubits into superpositions of noise-resilient oscillator states. In the case of the Kerr-cat qubit (\KQ), these states derive their stability from being the quasi-degenerate ground states of an engineered Hamiltonian in a driven nonlinear oscillator. \KQ s are experimentally compatible with on-chip architectures and high-fidelity operations, making them promising candidates for a scalable bosonic quantum processor. However, their bit-flip time must increase further to fully leverage these advantages. Here, we present direct evidence that the bit-flip time in a \KQ~is limited by leakage out of the qubit manifold and experimentally mitigate this process. We coherently control the leakage population and measure it to be $>9\%$, twelve times higher than in the undriven system. We then cool this population back into the \KQ~manifold with engineered single-photon dissipation, identify conditions under which this suppresses bit-flips, and demonstrate increased bit-flip times up to 3.6 milliseconds. 
By employing both Hamiltonian confinement and engineered dissipation, our experiment realizes dissipative stabilization of a \KQ.
Our results elucidate the interplay between these stabilization processes and indicate a path towards fully realizing the potential of these qubits for quantum error correction.
\end{abstract}
\maketitle

Building increasingly complex quantum systems while maintaining their coherence properties is a key challenge in quantum information processing and a crucial requirement for developing a quantum computer. Digital quantum error correction (QEC) pursues this goal by robustly encoding quantum information into concatenated physical two-level systems~\cite{napp_optimal_2013,fowler_surface_2012}. This approach is, however, expected to require very large numbers of physical qubits~\cite{fowler_surface_2012,gidney_how_2021,reiher_elucidating_2017}. An emerging, more hardware-efficient, alternative is bosonic QEC, which instead robustly encodes quantum information in a multi-level oscillator. This allows for the suppression of errors in a single physical system prior to concatenation~\cite{cochrane_macroscopically_1999,mirrahimi_dynamically_2014,michael_new_2016,gottesman_encoding_2001,puri_engineering_2017,goto_universal_2016,grimsmo_quantum_2020,menicucci_fault-tolerant_2014}.

One of the most widely used bosonic qubits is the two-component Schrödinger-cat qubit, whose basis states are superpositions of approximate opposite-phase coherent states~\cite{cochrane_macroscopically_1999,mirrahimi_dynamically_2014,puri_engineering_2017,goto_universal_2016}. Coherent states are robust against typical oscillator noise processes, such as excitation loss, and the $Z$-axis of a qubit Bloch sphere spanned by them is therefore protected. This results in noise bias, meaning that bit-flip errors are much less likely than phase-flip errors. Higher-order concatenation QEC codes can efficiently exploit this by focusing mainly on phase-flips~\cite{tuckett_ultrahigh_2018,bonilla_ataides_xzzx_2021,guillaud_error_2021,guillaud_repetition_2019,darmawan_practical_2021,ruiz_ldpc-cat_2025,putterman_hardware-efficient_2025,tuckett_fault-tolerant_2020,chamberland_building_2022}.

To implement a Schrödinger-cat qubit, its basis states need to be stabilized. This can be done through autonomous engineered dynamics, following two paradigms. i) Dissipative stabilization, in which the qubit states are the attractors of an engineered loss process~\cite{mirrahimi_dynamically_2014,leghtas_confining_2015,touzard_coherent_2018,lescanne_exponential_2020,guillaud_error_2021,guillaud_repetition_2019,ruiz_ldpc-cat_2025,Rousseau25,putterman_hardware-efficient_2025, Gertler2021,Albert2019,Marquet2024,reglade_quantum_2024}. Two-component dissipative cat qubits have been realized in harmonic oscillators subject to engineered two-photon loss~\cite{mirrahimi_dynamically_2014,leghtas_confining_2015}. ii) Hamiltonian confinement, where the qubit states are quasi-degenerate ground states that are separated from all other eigenstates by an energy gap. In Kerr-cat qubits (\KQ s), such Hamiltonian stabilization is realized in a two-photon-driven Kerr-nonlinear oscillator~\cite{cochrane_macroscopically_1999,puri_bias-preserving_2020,puri_engineering_2017,puri_stabilized_2019,goto_bifurcation-based_2016,goto_universal_2016,Grimm2020,frattini_observation_2024,venkatraman_driven_2024,albornoz_oscillatory_2024,iyama_observation_2024,qing_benchmarking_2024,Ding2024,Putterman2022,darmawan_practical_2021,xu_engineering_2022,hajr_high-coherence_2024}.

\KQ s are attractive for scaling in higher-order QEC codes because of their particularly simple implementation without the need for specialized auxiliary modes~\cite{Grimm2020}. Furthermore, high-fidelity initialization, measurement~\cite{frattini_observation_2024} and gate operations~\cite{qing_benchmarking_2024} have been experimentally demonstrated, facilitated by the large Hamiltonian energy gap that minimizes nonadiabatic errors during operations~\cite{xu_engineering_2022,darmawan_practical_2021}. 
However, the experimentally observed bit-flip protection in \KQ s~\cite{Grimm2020,frattini_observation_2024,hajr_high-coherence_2024,qing_benchmarking_2024} is weaker than expected from models considering only direct transitions between computational states~\cite{puri_engineering_2017,cochrane_macroscopically_1999,goto_universal_2016}. 
The leading hypothesis is that this is due to leakage to states outside of the qubit manifold~\cite{frattini_observation_2024,Putterman2022,Ruiz2023,Ding2024, BlaisRescueKCQ}.

Overcoming this limitation is the main challenge faced by \KQ s, and it was proposed to combine the \KQ\ with dissipative stabilization to profit from long bit-flip times as well as adiabatic high-fidelity operations~\cite{Grimm2020,puri_bias-preserving_2020,Gautier2022,Gravina2023,Putterman2022,Ruiz2023,Masuda2025}. This approach is supported by indirect measurements, showing that dissipation reduces dephasing of a microwave cavity coupled to a \KQ~\cite{Ding2024}.
However, direct measurements of spurious leakage in a \KQ\ are lacking, limiting the scope to understand and mitigate the noise processes responsible for leakage.
Most importantly, it remains an open question if reducing such a leakage population will increase \KQ~coherence.

Here, we answer this question in the affirmative by experimentally characterizing and interpreting the conditions under which engineered single-photon dissipation improves the bit-flip time $\Tz$ of a \KQ. 
We identify \KQ\ parameter regimes where engineered dissipation counter-intuitively degrades rather than improves $\Tz$, and relate the threshold for effective engineered dissipation to the temperature of the dissipative bath.
We correlate improvements in $\Tz$ with a reduction in the leakage population obtained by direct measurements, made possible by coherently controlling transitions to higher-lying states in the driven oscillator spectrum. We furthermore demonstrate that the engineered single-photon dissipation does not decrease the coherence times $\Tx$ and $\Ty$ of the axes spanned by the Schrödinger-cat states due to its frequency selectivity~\cite{Putterman2022}.

We implement the Kerr-nonlinear oscillator through a capacitively-shunted SNAIL array~\cite{frattini_3-wave_2017} with resonance frequency $\omegaqubit/2\pi = \fqubit$, single-photon lifetime ${\Tone = \bareTone}$, and Ramsey coherence time ${\TtwoR = \bareTphi}$. The thermal photon number in the undriven oscillator is $\nth=\nthval$. The oscillator is coupled to a microwave cavity with frequency $\omegacav/2\pi = \fres$ and total linewidth ${\kappacav/2\pi = \kappacavval}$ which is used for readout and as a waste mode for the engineered dissipation (Figs.~\ref{fig1}a,b) (see Supplementary Information Sections~I.A and~II).

To realize a \KQ, we apply a pump tone at frequency $\omegasq \approx 2\omegaqubit $ to the oscillator, which generates a two-photon drive through parametric downconversion. This results in the effective Hamiltonian
\begin{equation}
    \Hkcq/\hbar = \Delta \hat{a}^{\dag} \hat{a} - K \hat{a}^{\dag2} \hat{a}^{2} + \epstwo \left(\hat{a}^{\dag2} + \hat{a}^{2}\right),
    \label{eq:hamiltonian}
\end{equation}
where $\Delta=\omegaqubit-\omegasq/2$ is the mode frequency in the rotating frame of the two-photon drive, $K/2\pi=\kerr$ is the Kerr nonlinearity, and $\epstwo$ is the amplitude of the two-photon drive (see Supplementary Information Section~III.A). Equation~\ref{eq:hamiltonian} defines the characteristic double-well quasi-potential in oscillator phase space, which is populated by quantized eigenstates $\ket{\psi_i^\sigma}$, grouped in pairs with index $i$ (see Fig.~\ref{fig1}c). Here, $\sigma = + (-)$ indicates that only even (odd) Fock states are occupied. Throughout the text, we refer to the subspace spanned by each pair of states as a manifold $\ket{\psi_i^\pm}$.
We encode the \KQ~in the ground-state manifold by defining its $Z$-axis states as $\ket{\pm Z}\equiv(\ket{\psi^{+}_0}\pm\ket{\psi^{-}_0})/\sqrt{2}$, corresponding to opposite-phase approximate coherent states (see Wigner function representations in Fig.~\ref{fig1}d).
All other manifolds $\ket{\psi_{i>0}^{\pm}}$ are hereon referred to as leakage manifolds.

The bit-flip suppression of the \KQ~derives from the distance of the $\ket{\pm Z}$ states in oscillator phase space and the quasi-potential barrier separating them, both of which increase with $\epstwo$ and $\Delta$~\cite{cochrane_macroscopically_1999,puri_engineering_2017,goto_bifurcation-based_2016,frattini_observation_2024,hajr_high-coherence_2024}. Bit-flips are further suppressed at detunings $\Delta=2nK$, for integer $n$, where the qubit states become exactly degenerate due to destructive interference of inter-well tunneling paths~\cite{venkatraman_driven_2024,marthaler_quantum_2007}. However, state pairs within a leakage manifold in general have a finite inter-well tunnel coupling $\dE_i$ and face a lower quasi-potential barrier, facilitating transitions between wells~\cite{marthaler_switching_2006,chavez-carlos_driving_2023,Su2024, albornoz_oscillatory_2024,frattini_observation_2024}. For these reasons, it is expected that leakage out of the \KQ~manifold opens up additional bit-flip mechanisms.
 
\begin{figure}
    \includegraphics[angle = 0, width = \figwidth]{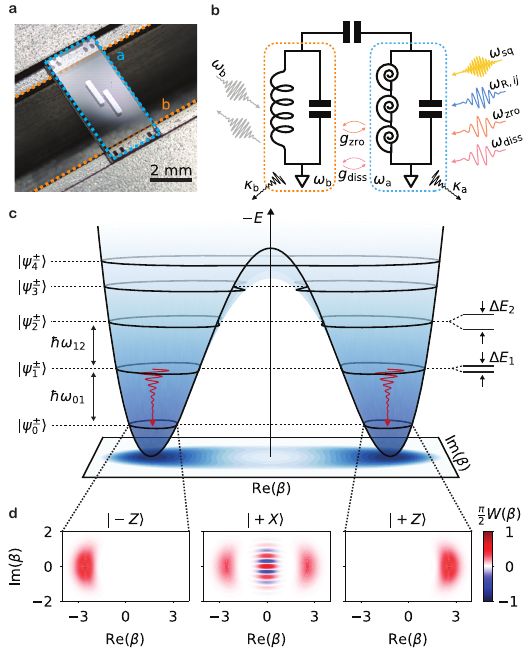}
    \caption{\label{fig1} \textbf{Experiment concept and implementation. a}, Photograph of the superconducting nonlinear oscillator chip ($a$, blue) placed inside one half of the microwave cavity ($b$, orange). 
    \textbf{b}, Circuit schematic of the nonlinear oscillator and microwave cavity system, with their respective frequencies $\omega_{\mathrm{a},\mathrm{b}}$ and loss rates $\kappa_{\mathrm{a},\mathrm{b}}$ indicated. The nonlinear oscillator consists of an array of three SNAILs~\cite{frattini_3-wave_2017} in parallel with a shunting capacitance. Microwave drives applied to the oscillator-cavity system, as well as parametric interactions, are respectively represented by colored pulses and two-way arrows (see text for description).
    \textbf{c}, Sketch of the quasi-potential energy $E$ of the Hamiltonian (Eq.~\ref{eq:hamiltonian}) as a function of the complex phase-space coordinate $\beta$ (blue). Black contour lines correspond to energies of the eigenstates $\ket{\psi_{i}^{\pm}}$ of Eq.~\ref{eq:hamiltonian}, with transition frequencies in the rotating frame, $\omega_{ij}$, indicated. An energy splitting $\dE_{1}$ ($\dE_{2}$) between the $\ket{\estate}$ ($\ket{\fstate}$) states is schematically indicated. Engineered dissipation (red arrows) brings population from $\ket{\estate}$ to $\ket{\gstate}$. 
    \textbf{d}, Wigner function representations of the states $\ket{\pm Z}$ and $\ket{+X}$ in the \KQ~manifold, for $\epstwo=\epstwofigtwo K$ and $\Delta=\Deltafigtwo K$.}
\end{figure}

\begin{figure*}
    \includegraphics[angle = 0, width = \figwidthWide]{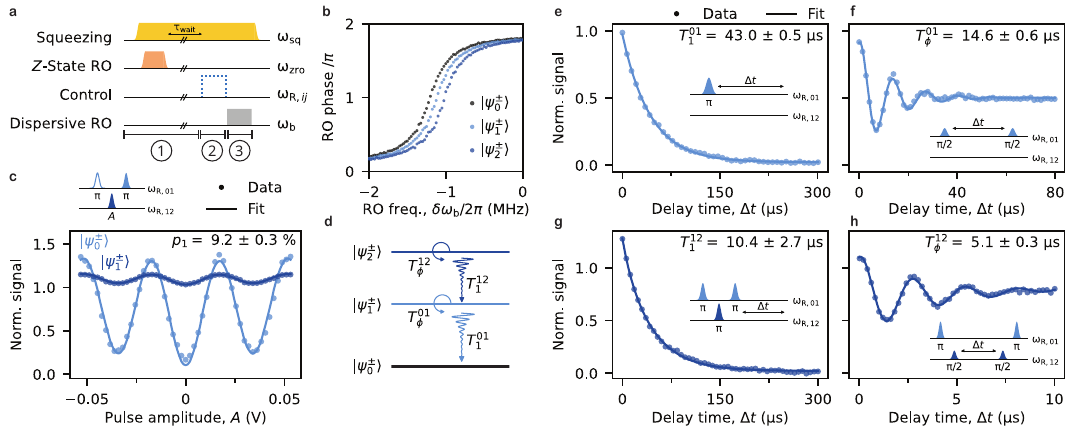}
    \caption{\label{fig2} \textbf{Coherent control and population measurement of leakage manifolds. a}, Pulse sequence to perform the following functions: (1) initialize a steady-state population across qubit and leakage manifolds, with wait time $\tauwait$ indicated; (2) coherently drive transitions between $\ket{\psi_{i}^{\pm}}$ and $\ket{\psi_{j}^{\pm}}$ manifolds ($i\neq j$), where the blue dotted box is a placeholder for pulses depicted in the panel insets; and (3) measure the $\ket{\psi_{i}^{\pm}}$-manifold-dependent readout (RO) cavity response. 
   \textbf{b}, Phase of reflected readout signal with the oscillator in $\ket{\gstate}$ (black), $\ket{\estate}$ (light blue) and $\ket{\fstate}$ (dark blue). The horizontal axis shows the detuning $\delta \omegacav = \omegaprobe - \omegacav$, where $\omegaprobe$ is the readout signal frequency and $\omegacav$ is the cavity resonance in the absence of the squeezing drive. 
   \textbf{c}, Rabi oscillations for the $\transef$ transition as a function of pulse amplitude $\pulseamp$ in units of the arbitrary waveform generator output voltage. Measurements performed with (without) an initial $\pige$-pulse are shown by light (dark) blue markers. The conditional $\pige$-pulse is indicated by a white fill in the pulse sequence. Solid lines are fits to extract the leakage population $\pe$ (see Supplementary Information Section~IV.G).
   \textbf{d}, Schematic representation of the relaxation times $T_{1}^{ij}$, and pure dephasing times $T_{\phi}^{ij}$ between $\ket{\psi_{i}^{\pm}}$ and $\ket{\psi_{j}^{\pm}}$ manifolds. 
   \textbf{e--h} Coherence measurements of $\ket{\estate}$ (\textbf{e,f}) and $\ket{\fstate}$ (\textbf{g,h}) manifolds. Experimental data (dots) are plotted alongside an analytical fit (solid line). Measurements were performed for $\epstwo=\epstwofigtwo K$ and $\Delta=\Deltafigtwo K$. The y-axis in panels \textbf{c},\textbf{e},\textbf{f},\textbf{g} and \textbf{h} is normalized with respect to the $\transge$ Rabi contrast (see Supplementary Information Section IV.C).}
    \end{figure*}

To better understand this process, we characterize the steady-state population $\pe$ in the $\ket{\estate}$ manifold. We use a self-calibrated Rabi-contrast protocol~\cite{Geerlings2013,Jin2015}, which consists of driving Rabi oscillations between the $\ket{\estate}$ and $\ket{\fstate}$ manifolds with and without an initial $\pige$-pulse (see Figs.~\ref{fig2}a,c) and comparing their oscillation amplitudes to extract $\pe$.
To realize this protocol, we need three ingredients: (1) initialization into the steady state, (2) coherent control over manifold populations, and (3) a readout that is sensitive to changes in population between different manifolds. Each is described in turn in the following. 

For this experiment, we initialize the system at ${\epstwo = \epstwofigtwo K}$ and ${\Delta = \Deltafigtwo K}$. We choose this working point for its strong bit-flip suppression and large energy gap even at moderate pump power, as has been shown for high-coherence \KQ\ operation~\cite{frattini_observation_2024,venkatraman_driven_2024,hajr_high-coherence_2024,qing_benchmarking_2024}. At such large values of $\Delta$, the system can become trapped in a local minimum near the origin of phase space during initialization~\cite{Gravina2023}. To prevent this, we adiabatically ramp both $\epstwo$ and $\Delta$ over durations of $\taurampval$ and $\tauchirpval$ respectively, for all measurements presented in this work (see Supplementary Information Section~III.C).
This is followed by a projective measurement in the \KQ~$Z$-basis~\cite{Grimm2020}~\citeMethods, to maintain consistency with subsequent experiments. We then wait for a duration ${\tauwait = \tauwaitval \gg \Tone}$ to allow the $\psii$ populations to reach the steady state (see Supplementary Information Section~IV).

To coherently control manifold populations, we then apply calibrated Gaussian pulses at frequencies $\omegaRabi=\omegasq/2+\omega_{ij}$, with $\omega_{ij}$ the transition frequency between manifolds $\ket{\psi_{i}^{\pm}}$ and $\ket{\psi_{j}^{\pm}}$ (Fig.~\ref{fig1}c). These pulses drive Rabi oscillations between states in the different manifolds, $\ket{\psi_{i}^{\pm}}\leftrightarrow\ket{\psi_{j}^{\mp}}$~\citeMethods.

Finally, we use the photon-number-dependent dispersive shift between the oscillator and the cavity $\chidisp \hat{a}^{\dag}\hat{a}\hat{b}^{\dag}\hat{b}$, with ${\chidisp/2\pi\approx\dispshiftval}$, to detect changes in populations $p_i$. Each $\psii$ has a different average photon number, thereby shifting the cavity resonance frequency by a different amount (see Supplementary Information Section~IV.A). Figure~\ref{fig2}b shows the phase of the reflected cavity signal after $\ket{\gstate}$, $\ket{\estate}$, or $\ket{\fstate}$ is prepared via calibrated $\pi$-pulses.

The complete measurement protocol yields an oscillating response in the readout signal as a function of pulse-amplitude $A$, both with and without the initial $\pige$-pulse (Fig.~\ref{fig2}c). 
We extract $\pe=\pevalratio$ from the ratio of oscillation amplitudes, taking into account a finite population in the $\ket{\fstate}$ manifold $\pf = (\pfvalfit \pm 0.5)\%$ which we independently estimate with incoherent spectroscopy (see Supplementary Information Section~IV.B). 
Note that the signal in Fig.~\ref{fig2}c does not return to its initial value (at $A=0$) after one oscillation, which we attribute to inter-manifold decoherence. While the ratio of oscillation amplitudes is insensitive to noise processes that affect each oscillation equally, errors in the initial $\pige$-pulse can impact the extracted $\pe$ (see Supplementary Information Section~IV.C).

To fully model the data shown in Fig.~\ref{fig2}c, we characterize the decoherence between manifolds spanning the subspace $\{\ket{\gstate},\ket{\estate},\ket{\fstate}\}$ using the effective model shown in Fig.~\ref{fig2}d. Relaxation (Figs.~\ref{fig2}e,g) and Ramsey interference (Figs.~\ref{fig2}f,h) measurements for each manifold are well-described by fits of analytical solutions of rate equations defined in this subspace~(see Supplementary Information Sections~IV.E,F). From this, we extract the exponential relaxation ($T_{1}^{ij}$) and pure dephasing ($T_{\phi}^{ij}$) time constants shown in the insets of Figs.~\ref{fig2}e-h. Note that these times are significantly shorter than $\tauwaitval$, justifying our choice of $\tauwait$.

Using the extracted coherence times as fixed parameters, we fit the $\transge$ and $\transef$ Rabi oscillations to a Lindblad master equation simulation with the overall measurement contrast and $\pe$ as the only free parameters~(see Supplementary Information Section~IV.G). 
The result reproduces the decrease in Rabi contrast in Fig.~\ref{fig2}c and gives $\pe=\peval$, in good agreement with the value obtained from the ratio of oscillation amplitudes. 

The measured value of $\pe$ is significantly larger than $\nth$, confirming that the driven system is indeed subject to additional excitation processes compared with the undriven oscillator~\cite{marthaler_switching_2006,Su2024,BlaisRescueKCQ, frattini_observation_2024,albornoz_oscillatory_2024,Ding2024}.
To understand this elevated leakage population, we simulate the steady state of the oscillator under different noise processes~\citeMethods. We first include only single-photon loss with a rate $\kappaqubit=1/\Tone$. While for $\Delta = 0$ the \KQ~manifold is spanned by coherent states, which are eigenstates of the annihilation operator $\hat{a}$~\cite{cochrane_macroscopically_1999,puri_engineering_2017}, this is no longer the case for $\Delta \neq 0$. As a result, single-photon loss leads to quantum heating~\cite{Ruiz2023,marthaler_quantum_2007,marthaler_switching_2006,albornoz_oscillatory_2024}, generating a non-zero leakage population.
Leakage induced by quantum heating gives $\pesim=\peqhval$, accounting for a significant fraction of the measured $\pe$. 
When including $\nth$, this model based on independently-measured parameters yields $\pesim=\penthval$, still underestimating the measured $\pe$.
This indicates additional noise sources contributing to leakage, which might include an elevated oscillator temperature due to strong drives~\cite{frattini_observation_2024,venkatraman_driven_2024,hajr_high-coherence_2024,albornoz_oscillatory_2024}, dephasing noise~\cite{frattini_observation_2024,hajr_high-coherence_2024} or spurious multi-photon transitions~\cite{Venkatraman2024_nl_diss,BlaisRescueKCQ,Su2024} (see Supplementary Information Section~V.A).
We introduce an effective model to account for such effects, and find that $\pe$ can be reproduced with additional oscillator thermal photon number $\nthplus=\nthplusval$ (corresponding to an effective temperature $\Tkcq \approx \Tkcqval$), or equivalently a dephasing rate of $\kappaphi/2\pi=\kappaphival$ (see Supplementary Information Section~V.C). While the underlying noise processes are likely a combination of these effects, we consider $\nthplus$ here for simplicity.

\begin{figure}
    \includegraphics[angle = 0, width = \figwidth]{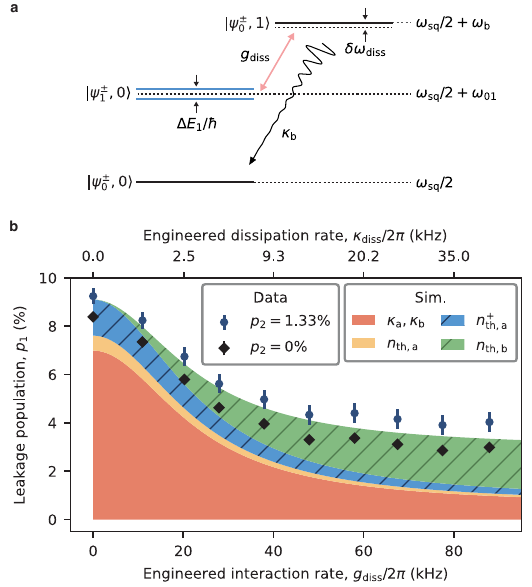}
    \caption{\label{fig3} \textbf{Effect of engineered dissipation on leakage population.}
     \textbf{a}, Schematic of the engineered dissipation process acting on the oscillator. The state of the oscillator-cavity system is indicated with $\ket{\psi_i^\pm,n}$, where $\psii$ labels an oscillator manifold and $\ket{n}$ the cavity photon number. The states $\ket{\estate}$ have an energy splitting $\dE_1$. A coherent photon exchange interaction with rate $\gdiss$ (pink) resonantly couples $\ket{\psi_{1}^{\mp},0}$ and $\ket{\gstate,1}$ when $\delta\omegadiss = 0$. The cavity relaxation, at rate $\kappacav$, transfers population from $\ket{\gstate,1}$ to $\ket{\gstate,0}$.
    \textbf{b}, Leakage population $\pe$ as a function of $\gdiss$ and $\kappadiss$. Experimental data (markers) for different values of $\pf$ are compared with simulation results (shaded regions), which include quantum heating (red), thermal heating of the undriven oscillator (yellow), additional effective thermal heating of the driven oscillator (blue hatched) and of the cavity (green hatched). Drive-induced leakage populations (hatched regions) are derived from effective models, each with one parameter fitted to the experimental data (see text).
    Error bars correspond to an uncertainty of one standard deviation obtained from the fitting procedure~(see Supplementary Information Section~IV.G).
    Measurements were performed for $\epstwo=\epstwofigtwo K$ and $\Delta=\Deltafigtwo K$.}
\end{figure}

\begin{figure*}
    \includegraphics[angle = 0, width = \figwidthWide]{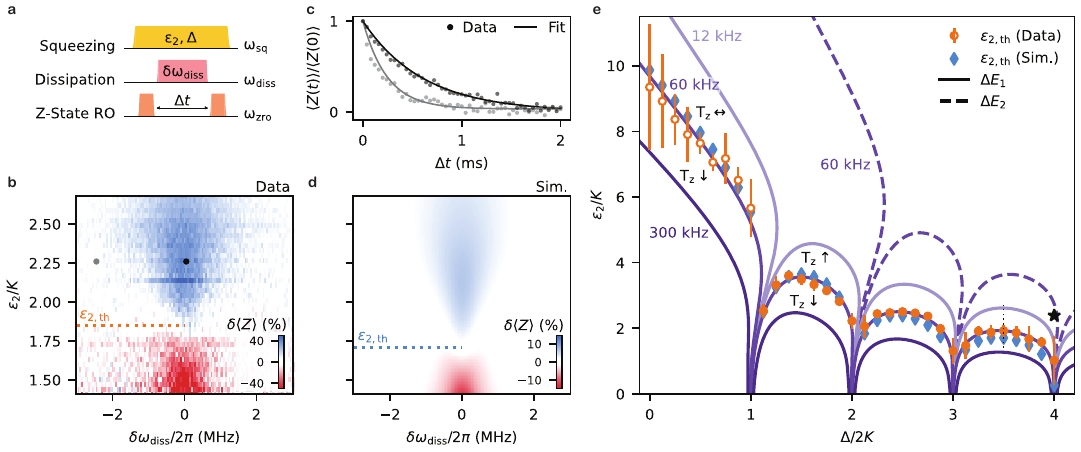}
    \caption{\label{fig4}  \textbf{Impact of engineered dissipation on bit-flip time, $\Tz$. a}, Pulse sequence to measure the change in $\Tz$ due to engineered dissipation. Variables inside pulses indicate parameters swept to obtain the data shown in panels \textbf{b}, \textbf{c} and \textbf{e}.
    \textbf{b}, Relative change in $Z$-state readout contrast, $\delta\Zdiss$, as a function of $\delta\omegadiss$ and $\epstwo$, for $\Delta=\Deltafigthree K$. The orange dotted line identifies $\epstr$. Black and gray dots indicate the parameters used in \textbf{c}. 
    \textbf{c}, Average $Z$-state readout contrast as a function of delay time $\dt$, for $\epstwo=2.26K$ and $\Delta=\Deltafigthree K$. Experimental data for on- (off-) resonant engineered dissipation are shown as black (gray) dots. Solid lines show exponential fits to the data.
    \textbf{d}, Simulation of the experiment in \textbf{b}, as explained in the text. The blue dotted line identifies the simulated $\epstr$ value.
    \textbf{e}, Threshold values $\epstr$ for different $\Delta$, extracted from measurements as in \textbf{b} (orange circles), or simulations as in \textbf{d} (blue diamonds).
    Filled markers denote a transition from negative to positive $\delta\Zdiss$ (as in \textbf{b},\textbf{d}), indicating a transition from decreased ($\downarrow$) to increased ($\uparrow$) $\Tz$. Open markers denote measurements for which $\delta\Zdiss$ saturates to zero above the transition ($\Tz\leftrightarrow$). Error bars on measured $\epstr$ correspond to one standard deviation and account for both the uncertainty in the $\epstwo$ calibration and statistical noise in the measurement data. Solid (dashed) lines show isolines of the energy splitting $\dE_{1}$ for $\ket{\estate}$ ($\dE_{2}$ for $\ket{\fstate}$) with values given in frequency units ($\dE_{i}/h$). The values of $\epstwo$ and $\Delta$ used for \textbf{b} are indicated by the vertical black dotted line, and measurements in Figs.~\ref{fig2},~\ref{fig3} and~\ref{fig5} are performed for parameters indicated by the star.}
\end{figure*}

After confirming an elevated leakage population, we engineer a frequency-selective single-photon dissipation channel acting on $\ket{\estate}$ and demonstrate that it reduces $\pe$. 
To engineer the dissipation, we use a coherent photon exchange interaction $\gdiss ( \hat{a} \hat{b}^{\dagger} + \hat{a}^{\dagger} \hat{b})$ between the oscillator and cavity. This interaction arises from a combination of the native three-wave mixing capability of our device~\cite{frattini_3-wave_2017,Grimm2020} with a drive at frequency ${\omegadiss = \omegacav - (\omegasq/2 + \omegage) + \delta\omegadiss}$. 
For $\delta\omegadiss=0$, the drive fulfills the frequency condition for selective population transfer between $\ket{\psi_{1}^{\mp},0}$ and $\ket{\gstate,1}$, where the second state refers to the readout cavity. Subsequent cavity relaxation, at rate $\kappacav$, transfers population to $\ket{\gstate,0}$. This process is illustrated by the schematic in Fig.~\ref{fig3}a.
If the engineered interaction rate is significantly smaller than the cavity loss rate, $\gdiss \ll \kappacav$, this oscillator-cavity interaction results in an effective single-photon loss channel acting on $\ket{\estate}$ with loss rate $\kappadiss=4\gdiss^2/\kappacav$~\cite{Putterman2022,Ruiz2023,Ding2024} (see Methods).

We measure $\pe$ in the presence of engineered dissipation by applying the pulse sequence shown in Fig.~\ref{fig2}a with the dissipation drive activated during the wait time $\tauwait$. 
The results are shown in Fig.~\ref{fig3}b for two assumed $\pf$ values: $\pf=(\pfvalfit \pm 0.5)\%$ (blue dots), as measured at $\gdiss=0$; and $\pf=0\%$ (black diamonds), to account for an expected reduction in $\pf$ at larger $\gdiss$. 
In both cases, $\pe$ decreases with increasing engineered interaction rate and saturates once $\gdiss/2\pi$ exceeds $\SI{50}{\kilo\hertz}$ ($\kappadiss/2\pi \approx \SI{15}{\kilo\hertz}$). For $\pf=0\%$, the saturation value is $\pe\approx3\%$. 

We model the dissipation-rate dependence of $\pe$ using the method described for Fig.~\ref{fig2}, with the addition of engineered dissipation~\citeMethods. The results are presented as shaded regions in Fig.~\ref{fig3}b. We find that quantum heating accounts for a significant fraction of the measured $\pe$ for all dissipation rates, with additional contributions coming from the undriven oscillator thermal photon number ($\nth$) and an effective additional thermal photon number ($\nthplus$). However, this does not fully reproduce the experimentally measured $\pe$ up to large $\gdiss$, because the engineered dissipation should fully evacuate the oscillator for sufficiently large dissipation rate. We obtain quantitative agreement at large $\gdiss$ by including an effective thermal photon number in the readout cavity of $\nthb=\nthbval$ (corresponding to a temperature $\Tcav \approx \Tcavval$). In this case, a reduction of $\pe$ in the oscillator is limited by a finite cavity population, resulting in a saturation of $\pesim$ at large $\gdiss$.
Although the $\nthb$ used here is fitted to the experimental data, direct measurements of $\nthb$ under similar experimental conditions return a comparable mode temperature that self-consistently describes $\pe$ at large $\gdiss$ (see Supplementary Information Section~V.D).

We now identify the conditions under which engineered dissipation can increase the bit-flip time $\Tz$.
To quantify the effect of dissipation on $\Tz$ as a function of $\epstwo$, $\Delta$, and $\delta\omegadiss$, we perform the measurement outlined by the pulse sequence in Fig.~\ref{fig4}a. We ramp on the squeezing drive for a given $\epstwo$ and $\Delta$, then initialize the \KQ~in $\ket{+Z}$ using a projective $Z$-state readout~\cite{Grimm2020}~\citeMethods. 
We then apply a $\SI{50}{\micro \second}$-long dissipation pulse with an engineered interaction rate $\gdiss/2\pi  = \gdissval$ (corresponding to $\kappadiss/2\pi=\kappadissfigthree$). Finally, we carry out a second $Z$-state measurement, which yields an expectation value $\Zdiss$.

As an indicative example, we show in Fig.~\ref{fig4}b the result of this measurement for $\Delta = \Deltafigthree K$ as a function of $\epstwo$ and $\delta\omegadiss$. To isolate the effect of engineered dissipation at each $\epstwo$, we plot $\delta\Zdiss$, the change in $\Zdiss$ relative to a fitted background value~\citeMethods.  
For large values of $\delta\omegadiss$, the dissipative process is off-resonant and has no effect ($\delta\Zdiss=0$). However, for $\delta\omegadiss=0$, $\delta\Zdiss$ goes from negative to positive for increasing $\epstwo$, indicating that dissipation reduces $\Tz$ at small $\epstwo$ but enhances $\Tz$ at larger $\epstwo$. Of particular interest is the threshold value $\epstr$, corresponding to the smallest $\epstwo$ where dissipation no longer reduces $\Tz$ (orange dotted line in Fig.~\ref{fig4}b).
We confirm the increase in $\Tz$ by explicit measurements at $\epstwo=2.26K>\epstr$ and $\delta\omegadiss =0$ (Fig.~\ref{fig4}c), where we see an increase to $\SI{610}{\micro\second}$, compared to $\SI{240}{\micro\second}$ when the interaction is off resonant (see Supplementary Information Fig.~S16 for the full $\delta\omegadiss$-dependence).

To better understand this threshold behavior, we extract $\epstr$ for different $\Delta$ and plot the result in Fig.~\ref{fig4}e (orange circles). We identify two distinct responses: filled markers correspond to datasets where $\delta \Zdiss$ is positive for $\epstwo > \epstr$, as in Fig.~\ref{fig4}b; open markers indicate cases where it saturates to zero beyond the threshold. The trend of $\epstr$ with $\Delta$ has a distinctive pattern, with minima at $\Delta=2nK$ for integer $n$. Such a trend is also evident in the energy splitting $\dE_{1}$ within the $\ket{\estate}$ manifold, which is illustrated by the solid purple lines obtained by numerical diagonalization of Eq.~\ref{eq:hamiltonian} with the addition of a pump-induced AC-Stark shift~\citeMethods. Similar to states in the $\ket{\gstate}$ manifold~\cite{marthaler_quantum_2007,venkatraman_driven_2024}, this energy splitting is suppressed with increasing quasi-potential-well depth, as well as for specific values of $\Delta$ due to destructive interference between inter-well tunneling paths. The measured $\epstr$ follows the $\dE_{1}/h=\SI{60}{\kilo\hertz}$ isoline.
The data is uniquely described by the energy splitting in the $\ket{\estate}$ manifold due to the minimum at $\Delta=2K$, which is not present for higher leakage manifolds. This is illustrated by the dashed line in Fig.~\ref{fig4}e, which shows the $\dE_{2}/h=\SI{60}{\kilo\hertz}$ isoline.

We model the threshold behavior in Figs.~\ref{fig4}b,e by simulating the engineered interaction between the oscillator and cavity modes. 
The simulation mimics the experiment of Fig.~\ref{fig4}b: we initialize in $\ket{+Z}$, then we turn on the frequency-dependent oscillator-cavity interaction with a duration and strength defined by the experiment, and finally we evaluate the $Z$-state projection in the \KQ-basis.
In addition to the thermal photon number in the oscillator introduced above ($\nth+\nthplus=\nthvalsim$), we include a finite thermal cavity photon number $\nthb$~\citeMethods.
The simulation corresponding to Fig.~\ref{fig4}b is shown in Fig.~\ref{fig4}d. It reproduces the characteristic behavior of the experiment, namely a decrease (increase) in $\Tz$ for $\epstwo<\epstr$ ($\epstwo>\epstr$).
We find that the simulated $\epstr$ depends on $\nthb$. The closest match to the experimental data across the full range of $\Delta$ is obtained for $\nthb=\nthbvalfigthree$ (see blue diamonds, Fig.~\ref{fig4}e). 
This corresponds to a lower effective temperature than in Fig.~\ref{fig3}b ($\Tcav\approx\Tcavvalfigthree$), which we attribute to a lower duty cycle of the measurement in Fig.~\ref{fig4}.

We interpret the results of Fig.~\ref{fig4} as a competition between the inter-well tunneling rate $\dE_{1}$ in the $\ket{\estate}$ manifold and the dissipative dynamics of the $\transeg$ transition in the presence of a cavity thermal photon number $\nthb>0$. All measurements presented in Fig.~\ref{fig4} are taken for $\kappadiss\gg\kappaqubit$. The cooling rate between the $\ket{\estate}$ and $\ket{\gstate}$ manifolds is therefore dominated by the engineered dissipation rate, $\kappadiss$. However, the engineered interaction also introduces a spurious excitation rate $\nthb\kappadiss$, because the finite cavity population can be transferred to the oscillator~\citeMethods. Since $\nthb\ll1$, the cooling rate dominates over the excitation rate such that the average $\pe$ decreases. However, this does not necessarily result in a longer bit-flip time, since each excitation event causes transient leakage population. For $\kappadiss\ll \dE_1/\hbar$, this leakage population can tunnel before it is cooled back to the $\ket{\gstate}$ manifold, increasing the rate of transitions between the wells and hence decreasing $\Tz$. In contrast, when $\kappadiss\gg \dE_1/\hbar$, the engineered dissipation suppresses leakage population faster than the inter-well tunneling can transfer it to the other well, and hence $\Tz$ increases.
The threshold $\epstr$ occurs for a value of $\dE_1$ at which these two processes exactly compensate each other, such that the inter-well tunneling rate is unaffected by the engineered dissipation. This balance depends sensitively on the spurious excitation rate, and therefore on $\nthb$. This is evident in Fig.~\ref{fig4}e, where even a small effective $\nthb$ can lead to a significant threshold $\epstr$. See Supplementary Information Section~V.F for more details about the dependence of $\epstr$ on $\nth$ and $\nthb$.

\begin{figure}
    \includegraphics[angle = 0, width = \figwidth]{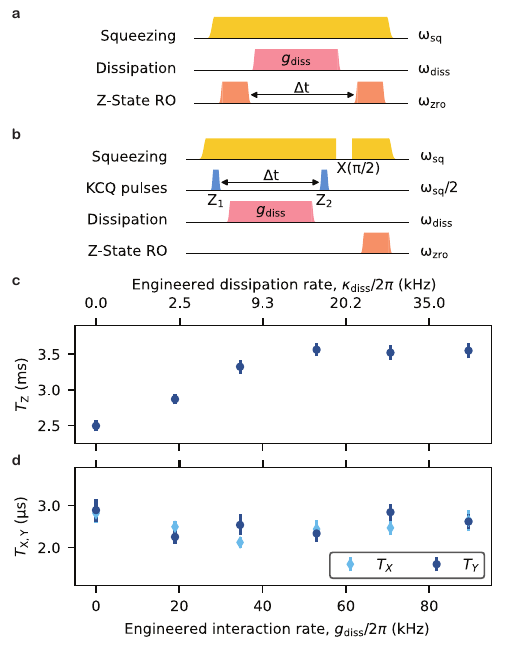}
    \caption{\label{fig5} \textbf{Dependence of \KQ~coherence times on dissipation rate.} 
    \textbf{a}, Pulse sequence to measure bit-flip time, $\Tz$, with engineered dissipation. The swept delay time $\dt$ and engineered interaction rate $\gdiss$ are indicated. 
     \textbf{b}, Pulse sequence to measure $X$- and $Y$-state coherence times, $T_{\mathrm{X,Y}}$. We initialize the \KQ~in $\ket{+X}$ or $\ket{+Y}$ by ramping on the squeezing drive followed by a conditional gate $Z_{1}$. After the delay time, the gates $Z_{2}$ and $X(\pi/2)$ rotate the decayed state to the $Z$-axis for readout~\citeMethods. 
    \textbf{c}, Measured $\Tz$ as a function of $\gdiss$, and the corresponding rate $\kappadiss$. 
    \textbf{d}, Measured $\Tx$ ($\Ty$) as a function of $\gdiss$ and $\kappadiss$ indicated by light blue diamonds (dark blue dots). 
    Measurements were performed for $\epstwo=\epstwofigtwo K$ and $\Delta=\Deltafigtwo K$.
    }
\end{figure}

We now combine the insights into the effect of engineered dissipation to demonstrate an enhancement in \KQ~coherence. 
To this end, we return to $\epstwo=\epstwofigtwo K$ and $\Delta=\Deltafigtwo K$ (as in Figs.~\ref{fig2} and~\ref{fig3}) where we expect high \KQ\ coherence that can be further improved by engineered dissipation ($\epstwo>\epstr$, indicated by the star in Fig.~\ref{fig4}e).
 
We begin by measuring the bit-flip time $\Tz$ as a function of $\gdiss$. We initialize the \KQ~in $\ket{+Z}$ via a projective $Z$-state measurement with a fidelity of $\cqrfidelity$ (see Supplementary Information Section~III.D), and then apply a dissipation pulse of variable duration $\dt$, followed by a second $Z$-state measurement (see Fig.~\ref{fig5}a). 
The result is shown in Fig.~\ref{fig5}c. As $\gdiss$ increases, we observe a clear enhancement of $\Tz$ to a value of $\SI{3.6}{\milli\second}$. Our measurements display a distinct correlation between the reduction in $\pe$ and the increase in $\Tz$, demonstrating the effectiveness of our method.

We now characterize the coherence times along the $X$- and $Y$-axes of the \KQ~Bloch sphere, $\Tx$ and $\Ty$, as a function of $\gdiss$ using the sequence in Fig.~\ref{fig5}b~\citeMethods. 
As we increase $\gdiss$, we might expect a reduction in $T_{\mathrm{X,Y}}$, since phase-flips are induced by single-photon loss~\cite{cochrane_macroscopically_1999, puri_engineering_2017,mirrahimi_dynamically_2014}.
However, the linewidth $\kappacav$ of the engineered dissipation is much smaller than the frequency gap $\omegage$, making the process selective to the leakage manifold $\ket{\estate}$ and leaving the \KQ~manifold unchanged~\citeMethods.
As a result, $\Tx$ and $\Ty$ remain approximately constant at $\langle T_{\mathrm{X,Y}} \rangle \approx \SI{2.5}{\micro\second}$ for increasing $\gdiss$ (see Fig.~\ref{fig5}d), consistent with a slightly increased single-photon loss rate $\kappaoneeff/2\pi = \kappaoneeffval$ in the driven system~\cite{lescanne_exponential_2020,frattini_observation_2024,Rousseau25}. Note that, if the maximum dissipation rate of $\kappadiss^{\mathrm{max}}/2\pi=\SI{43}{\kilo\hertz}$ was not frequency selective, these coherence times would have been reduced to a value of $\SI{0.26}{\micro\second}$.

Our results demonstrate an enhancement in \KQ~bit-flip time correlated with a reduction in leakage out of the qubit manifold, achieved through engineered single-photon dissipation.
We report the highest \KQ~bit-flip time to date, while preserving the phase-flip time due to the frequency selectivity of the dissipation. Our engineered dissipation increases both overall coherence and noise bias, the latter reaching a maximum value estimated as $\Tz/T_{\mathrm{X,Y}}\approx 1400$. Moreover, our qubit remains compatible with high readout fidelity ($\approx 99.4\%$) and low gate errors (see Supplementary Information Section~III.F for estimates of the latter). 
We provide a physical understanding of the parameter values for which this approach is helpful rather than harmful, making it a simple yet effective strategy for improving the error robustness in \KQ s. 
Further, in contrast to engineered two-photon dissipation, the process presented here is particularly well-suited to suppressing single-photon leakage. This leakage flips photon-number parity, which is then corrected by our single-photon dissipation such that we do not introduce additional phase-flip errors.
Single-photon leakage can arise in \KQ s due to quantum heating~\cite{frattini_observation_2024,venkatraman_driven_2024}, pump-induced thermal heating~\cite{frattini_observation_2024,venkatraman_driven_2024,hajr_high-coherence_2024,albornoz_oscillatory_2024}, dephasing~\cite{frattini_observation_2024,hajr_high-coherence_2024} and non-adiabatic gate errors~\cite{Putterman2022}, covering a broad range of noise sources that limit \KQ\ coherence.
Our system is therefore a promising realization of a hybrid Hamiltonian-dissipative cat-qubit stabilization scheme.

The bit-flip time saturated for larger dissipation rates in our experiment. 
Although our numerical simulations do not fully reproduce the observed saturation, they indicate that the cavity thermal photon number $\nthb$ and leakage to higher manifolds both limit the maximum $\Tz$ (see Supplementary Information Section~V.I).
In the former case, a finite $\nthb$ limits the minimum $\pe$ achievable with engineered dissipation, leaving a residual leakage population that can facilitate bit flips.
In the latter case, dissipation targeting the $\transeg$ transition does not correct for leakage into higher manifolds.
The remaining limitation to the bit-flip time not captured by our model could involve multi-photon transitions to higher leakage manifolds~\cite{Su2024,BlaisRescueKCQ,Ruiz2023, Dai_Hazra_2025}.
These limitations could be overcome with natural extensions to the implementation presented here. A dedicated waste mode can be designed to minimize $\nthb$, and spurious multi-photon transitions can be suppressed via careful design of the microwave environment~\cite{Dai_Hazra_2025,BlaisRescueKCQ}. Remaining leakage to higher manifolds can then be addressed using a comb of dissipation frequencies, or by coupling the oscillator to a dissipative environment via a band-pass filter that only targets leakage states~\cite{Putterman2022}. 
The latter approach might also allow for higher dissipation rates, which in our experiment were limited by the cavity linewidth $\kappacav$.
These strategies could build on the results presented here to further increase bit-flip times towards the second-timescales observed in strongly-damped systems~\cite{Beaulieu2025}, without compromising qubit coherence.

Beyond improving \KQ~performance, our results demonstrate coherent control of transitions to higher-lying quantum states in a parametrically-driven Duffing oscillator~\cite{DykmanBook}, to our knowledge for the first time.
We use this to directly measure a leakage population in $\ket{\estate}$ consistent with significant quantum heating~\cite{marthaler_switching_2006}, an effect expected to be present in many driven systems described by a static-effective quasienergy spectrum. 
Moreover, the double-well potential that describes our driven oscillator is commonly invoked to model a wide range of physical phenomena, including chemical reactions~\cite{cabral_roadmap_2024} and two-level defects in solid-state devices~\cite{lisenfeld_decoherence_2016}.
We anticipate that the coherent excitation and engineered dissipation presented here will expand the toolbox available for analog quantum simulations of such systems.
Finally, the impact of the bath temperature on the system dynamics identified in our work will be relevant for dissipative stabilization of any system with quasi-degenerate energy levels~\cite{brooks_protected_2013,earnest_realization_2018,kalashnikov_bifluxon_2020}.
From these examples, we expect that, in addition to quantum computing, our results will be applied in the study of fundamental effects in quantum nonlinear oscillators and to simulate complex quantum dynamics.

\bibliographystyle{naturemag}
\bibliography{bibliography}

\begin{ac}
AG, FA developed the idea for the experiment. 
FA designed the experiments with input from DZH and AG. 
AB, VHK, FA, fabricated the sample. 
FA, DZH built the experimental setup with help from LM, PP. 
FA carried out initial experiments with help from PP. 
DZH, FA performed the measurements and analyzed the data. 
FA performed the simulations and developed the theoretical analysis with inputs from DZH and AG. 
AG supervised the project. 
FA, DZH, AG wrote the manuscript with input from all authors.
\end{ac}

\begin{ack}
We acknowledge useful discussions with Ian Yang and Yiwen Chu, as well as help from Tristan Kuttner in the fabrication of the quantum limited amplifier used in this project.
We thank the cleanroom operations teams of the Paul Scherrer Institute and of the Binnig and Rohrer Nanotechnology Center for their help and
support. This work was supported by the Swiss National Science Foundation Grant No. 200021\_1972551, the Swiss Nanoscience Institute Fellowship Grant No. P2101 and the Swiss State Secretariat for Education, Research and Innovation (SERI).
\end{ack}

\begin{meth}
\textbf{Characterizing the energy spectrum of the oscillator}\\
Here, we describe the method used to extract the transition frequencies ($\omega_{ij}$) between manifolds $\psii$ and $\ket{\psi_{j}^{\pm}}$, to calibrate $\epstwo$ as a function of the squeezing-drive-pump voltage amplitude $V_{\mathrm{sq}}$ generated by our digital-analog-converter (DAC) instrument, and to determine the energy splitting $\dE_{1}$ (Fig.~\ref{fig4}e). 

We first perform an incoherent spectroscopy measurement as a function of $V_{\mathrm{sq}}$ for fixed detuning $\Delta$~\cite{frattini_observation_2024}. We then fit the experimentally-obtained transition frequencies using the eigenenergies of the effective Hamiltonian
\begin{equation}
\begin{split}
    \Hkcq/\hbar+\Hstark/\hbar = (\Delta-4K|\xisq{eff}|^{2}) \hat{a}^{\dag} \hat{a} \\
    - K \hat{a}^{\dag2} \hat{a}^{2} + \epstwo \left(\hat{a}^{\dag2} + \hat{a}^{2}\right),
    \label{eq:ham_stark_methods}
\end{split}
\end{equation}
which includes the AC-Stark shift induced by the squeezing-drive pump, $\Hstark$ (see Supplementary Information Section~III.A). Here, $\xisq{eff}=\epstwo/3g_{3}$ is the effective linear squeezing-drive displacement amplitude written in terms of the squeezing-drive strength $\epstwo$ and the third-order nonlinearity of the SNAIL $g_{3}$~\cite{frattini_3-wave_2017,Grimm2020}. The only free fit parameter is a conversion factor between $V_{\mathrm{sq}}$ and $\epstwo$. This defines the calibrated $\epstwo$ values used throughout this work and the corresponding transition frequencies $\omega_{ij}=(\omega_{ij}^{+}+\omega_{ij}^{-})/2$, where the superscript $+$ ($-$) denotes an initial state with even (odd) parity. Transition frequencies used in Figs.~\ref{fig2},~\ref{fig3}, and~\ref{fig5} are further refined using Ramsey interference measurements. The energy isolines for $\dE_{1}$ and $\dE_{2}$ shown in Fig.~\ref{fig4}e are computed using Eq.~\eqref{eq:ham_stark_methods}.
\\
\\
\textbf{Experimental details of $Z$-axis readout}\\
The projective $Z$-state readout (\CQR) is used to both initialize and measure the \KQ~in the $Z$-basis (Figs.~\ref{fig4} and~\ref{fig5}). The \CQR~is realized using the three-wave mixing capability of the oscillator, in combination with a drive at frequency $\omegacqr=\omegacav-\omegasq/2$. 
This generates a parametric oscillator-cavity interaction with rate $\gcqr$ given by
\begin{equation*}
    \Hcqr/\hbar = \gcqr ( \hat{a} \hat{b}^{\dagger} + \hat{a}^{\dagger} \hat{b})
\end{equation*}
in a frame rotating at the cavity frequency $\omegacav$. When projected into the \KQ~basis, $\Hcqr$ induces opposite-phase coherent displacements of the readout cavity dependent on the $Z$-state of the \KQ. This process acts as a quantum non-demolition measurement~\cite{Grimm2020}. 

For the measurements presented in Figs.~\ref{fig2},~\ref{fig4} and~\ref{fig5}, the $Z$-state readout pulse is a flat-top Gaussian with a constant length of $\SI{2}{\micro\second}$ and a rise time of $\SI{20}{\nano\second}$ ($\sigma=\SI{8}{\nano\second}$). Further details on the $Z$-state readout are given in Supplementary Information Section~III.D.
\\
\\
\textbf{Regime of utility of $\pe$ measurement protocol}\\
The protocol to measure $\pe$ described here works well for values of $\epstwo$ and $\Delta$ where the leakage states are localized inside the quasi-potential wells and are therefore quasi-degenerate. This coincides with the regime where long bit-flip times in the \KQ~are expected, making our method a useful tool for studying leakage processes in this system. 
In contrast, the projective $Z$-state measurement is not a viable readout method for these parameters, as it measures quadrature amplitudes (``which-well'' information) and thus cannot easily distinguish populations in different manifolds $\psii$. 
Moreover, unlike dispersive population measurements, our method yields a quantitative estimate of $\pe$ in the regime $\chidisp\lesssim\kappacav$ where \KQ~experiments typically operate to limit cavity-mediated Purcell loss~\cite{frattini_observation_2024,Ding2024}.
\\
\\
\textbf{Experimental details of $\pe$ measurement (Figs.~\ref{fig2} and~\ref{fig3})}\\
Here, we describe details of the $\pe$ measurement of Figs.~\ref{fig2}c and~\ref{fig3}b, as well as the manifold-coherence measurements of Figs.~\ref{fig2}e-h.
We coherently transfer population between manifolds $\psii$ and $\ket{\psi_{j}^{\pm}}$ by applying a microwave drive at a frequency $\omegaRabi$ with a Gaussian envelope with total length $\tauRij=\SI{2}{\micro\second}$ ($\sigma=\SI{332}{\nano\second}$). We set $\omegage/2\pi=\SI{-25.83}{\mega\hertz}$ and $\omegaef/2\pi=\SI{-21.65}{\mega\hertz}$ for $\transge$ and $\transef$, respectively and calibrate the pulse amplitudes with Rabi oscillation measurements. Note that the pulses additionally switch the superscript $+/-$ of the states because the coherent drive does not conserve photon-number parity. This does however not influence the manifold population measurement.  We use an additional $\pige$-pulse for $\ket{\fstate}$ measurements (Figs.~\ref{fig2}c,g,h,~\ref{fig3}b) to increase the readout contrast~\cite{Geerlings2013}. Data in Figs.~\ref{fig2}c,e,f,g and h are normalized to the $\transge$ Rabi contrast, which are explicitly shown in Supplementary Information Section~IV.C.
\\
\\
\textbf{Extraction of threshold value $\epstr$ (Fig.~\ref{fig4})}\\
Here, we describe the procedure used to extract $\epstr$ in Figs.~\ref{fig4}b,e.
For each value of $\epstwo$, we fit $\Zdiss$ with a Lorentzian profile as a function of $\delta\omegadiss$ to extract the background $\Zdissbg$ and peak amplitudes. 
We then define $\delta\Zdiss = (\Zdiss-\Zdissbg)/\Zdissbg$ which we plot in Fig.~\ref{fig4}b.
Note that we observe a sharp change in $\delta\Zdiss$ around $\epstwo = 2.1 K$ which we attribute to a sharp decrease in the background $\Tz$, potentially due to an accidental transition to a higher-lying state~\cite{albornoz_oscillatory_2024,BlaisRescueKCQ}. 
We determine $\epstr$ using the fitted peak values as a function of $\epstwo$. In the case where $\delta\Zdiss$ is positive above the threshold, we extract $\epstr$ by identifying where the fitted peak value changes sign. This corresponds to filled markers in Fig.~\ref{fig4}e. If no positive region of $\delta \Zdiss$ is observed, $\epstr$ corresponds to the smallest $\epstwo$ where the fitted peak value goes below the standard deviation of the measurement trace. This corresponds to the open markers in Fig.~\ref{fig4}e.  For more details on the extraction of $\epstr$ and example measurements corresponding to Fig.~\ref{fig4}b at other values of $\Delta$, see Supplementary Information Section~V.E.
\\
\\
\textbf{Experimental details of engineered dissipation measurements (Figs.~\ref{fig4} and~\ref{fig5})}\\
Here, we describe details of engineered dissipation measurements. The dissipation pulse used in the experiment is an approximate square pulse with a rise time of approximately $\SI{4}{\nano\second}$ set by equipment limitations. All experiments including engineered dissipation feature a delay time of $\taucav=\SI{1.2}{\micro\second}$ between the dissipation pulse and any subsequent pulses, to account for the decay of residual photons in the cavity.

We calibrate the engineered interaction rate $\gdiss$ by measuring the decay dynamics of the undriven nonlinear oscillator. To do this, we perform a $\Tone$~measurement while applying an engineered-dissipation pulse with variable amplitude $\Adiss$ and duration $\Delta t$. By fitting the resulting decay curves, we extract the corresponding value of $\gdiss$ for each $\Adiss$. For more details, see Supplementary Information Section~V.B.

The pulse sequence to measure $\Tx$ ($\Ty$) in the presence of engineered dissipation is shown in Fig.~\ref{fig5}b.
We first initialize the system in the \KQ~state $\ket{+X}$~\cite{Grimm2020, Gravina2023} (see main text and Supplementary Information Sections~III.B,C). We then apply the gate $Z_{1}=Z(0)$ ($Z(\pi/2)$) to prepare the \KQ~state $\ket{+X}$ ($\ket{+Y}$). This is followed by a dissipation pulse of variable duration $\dt$ and the gates $Z_2 = Z(\pi/2)$ ($Z(0)$) and $X(\pi/2)$ to map the decayed $\ket{+X}$ ($\ket{+Y}$) state onto the $Z$-basis for readout.
The $Z$ gates are implemented using flat-top Gaussian pulses at frequency $\omegasq/2$ with a ramp duration of $\SI{160}{\nano\second}$ ($\sigma = \SI{32}{\nano\second}$), followed by a constant-amplitude segment of approximately $\SI{100}{\nano\second}$ for $Z(\pi/2)$ and approximately $\SI{50}{\nano\second}$ for $Z(0)$.  The $X(\pi/2)$ gate has a duration of approximately $\SI{132}{\nano\second}$. See Supplementary Information Section~III.F for details on the gate calibration and estimates of the gate fidelity.
\\
\\
\textbf{Numerical simulation of engineered dissipation (Figs.~\ref{fig3} and~\ref{fig4})}\\
Here, we describe the model used to simulate the effect of engineered dissipation on the driven oscillator. These simulations are used in the results presented in Figs.~\ref{fig3}b and~\ref{fig4}d,e.

As described in the main text, we realize the engineered dissipation using a coherent oscillator-cavity exchange interaction. We describe this interaction using the effective Hamiltonian
\begin{equation}
\begin{split}
    \Hdiss/\hbar = &\Hkcq/\hbar + \Hstark/\hbar \\
    &+\Delta_{\mathrm{b}}\hat{b}^{\dagger}\hat{b} + \gdiss ( \hat{a} \hat{b}^{\dagger} + \hat{a}^{\dagger} \hat{b}),
    \label{eq:diss_ham_methods}
\end{split}
\end{equation}
where $\Hkcq$ is the Kerr-cat Hamiltonian (Eq.~\ref{eq:hamiltonian}) and $\Hstark$ is the AC-Stark shift Hamiltonian induced by the squeezing-drive pump.
We choose here a rotating frame for the cavity mode such that the interaction term has no time dependence.
For the simulations of Figs.~\ref{fig3} and~\ref{fig4}, we set $\Delta_{\mathrm{b}}=\omegage+\delta\omegadiss$ (see Supplementary Information Sections~V.B,C,F,I).

We use this effective Hamiltonian to model the engineered dissipation with the Lindblad master equation
\begin{equation}
\begin{split}
    \frac{\mathrm{d}\hat{\rho}}{\mathrm{d}t} &= -\frac{i}{\hbar} [\Hdiss, \hat{\rho}] \\ 
    &+\kappaqubit(1+\nth+\nthplus)D[\aop]\hat{\rho} \\
    &+ \kappaqubit (\nth+\nthplus) D[\aod]\hat{\rho} \\
    &+\kappacav(1+\nthb) D[\hat{b}] \hat{\rho} \\
    &+ \kappacav \nthb D[\bod] \hat{\rho},
    \label{eq:diss_master_methods}
\end{split}
\end{equation}
where $\rho$ is the density matrix of the joint oscillator-cavity system, $D[\hat{O}]\hat{\rho}=\hat{O}\hat{\rho}\hat{\rho}^{\dagger}-\frac{1}{2}\hat{O}^{\dagger}\hat{O}\hat{\rho}-\frac{1}{2}\hat{\rho}\hat{O}^{\dagger}\hat{O}$, $\kappaqubit$ and $\kappacav$ are the respective single-photon loss rates for the oscillator and cavity (which we set to the experimentally-measured values unless stated otherwise) and $\nth+\nthplus$, $\nthb$ are thermal photon numbers in the oscillator and in the cavity, respectively. 

We first explain the simulation used to estimate the engineered dissipation rate $\kappadiss$ on the $\transge$ transition ($\delta \omegadiss = 0$) for each measured value of $\gdiss$. We initialize the oscillator-cavity system in $\ket{\estate}\otimes\ket{0}$, where the second state refers to the cavity. We then evolve the oscillator-cavity system under Eq.~\eqref{eq:diss_master_methods}, with $\kappaqubit=0$ and $\nth=\nthb=0$. After a variable delay time $\Delta t$, we evaluate the population in the $\ket{\estate}$ manifold of the oscillator, $\pesim$. This is given by
\begin{equation}
    \pesim(\Delta t) = \sum_{\pm}\bra{\estate}\hat{\rho}_{\mathrm{a}}(\Delta t)\ket{\estate},
    \label{eq:pesim_methods}
\end{equation}
where $\hat{\rho}_{\mathrm{a}}$ is the reduced density matrix of the oscillator mode after tracing out the cavity mode. We extract $\kappadiss$ from the $1/e$ decay time of $\pesim (\Delta t)$. For more details on the simulation of $\kappadiss$, see Supplementary Information Section~V.B.

Next, we describe the simulation used to obtain the leakage population $\pe$ as a function of the engineered interaction rate $\gdiss$, presented in Fig.~\ref{fig3}b. Here, we set $\delta\omegadiss = 0$ to target the $\transeg$ transition. We first perform a steady-state simulation of Eq.~\eqref{eq:diss_master_methods} at fixed engineered interaction rate $\gdiss$. The result corresponds to the state of the system after the dissipation pulse has been applied for a time $\tauwait$ in the experiment. Using this state as the initial condition, we then evolve the system under Eq.~\eqref{eq:diss_master_methods} with $\gdiss=0$, for an evolution time $\taudelay$. This accounts for a delay time in the experiment after the dissipation pulse, which is composed of $\taucav$ (as described above) and the coherent-control pulses applied during the measurement sequence, $\tauRij$. We set $\taudelay=\taucav+\tau_{\mathrm{R},01}+\tau_{\mathrm{R},12}/2=\SI{4.2}{\micro\second}$. We then evaluate $\pesim$ using Eq.~\eqref{eq:pesim_methods}. For this simulation, we set $\nthplus=\nthplusval$ and $\nthb=\nthbval$. For more details, see Supplementary Information Section~V.C.

We now explain how we obtain the simulation results presented in Fig.~\ref{fig4}d, which shows the impact of engineered dissipation on $\delta \Zdiss$. We first set an initial oscillator-cavity state of $\ket{+Z}\otimes\ket{0}$, where the second state refers to the cavity. Then, we evolve the system for a time $\tau_{\mathrm{sim.}}=\SI{50}{\micro\second}$ under Eq.~\eqref{eq:diss_master_methods}, with $\nth=\nthvalsim$ and $\nthb=\nthbvalfigthree$. The final state of the system is then $\hat{\rho}_{\mathrm{final}}(\delta\omegadiss,\epstwo)$, for each set of parameters $\delta\omegadiss$ and $\epstwo$. We compute the projection on the oscillator $Z$-axis using the operator
\begin{equation*}
    \hat{O}_{\mathrm{Z}} = [(\ket{+Z}\bra{+Z}-\ket{-Z}\bra{-Z})\otimes I_{\mathrm{b}}],
\end{equation*}
to get the simulated projection value $\Zdiss = \mathrm{tr}[\hat{O}_{\mathrm{Z}}\hat{\rho}_{\mathrm{final}}]$. We repeat the simulation for different $\delta\omegadiss$ and $\epstwo$. Note that the change in simulated contrast plotted in Fig.~\ref{fig4}d, $\delta\Zdiss$, is calculated relative to the value at ${\delta\omegadiss/2\pi=\SI{-3}{\mega\hertz}}$.

For the results presented in Fig.~\ref{fig4}e, we perform a simulation similar to that used for Fig.~\ref{fig4}d. We initialize the system in $\ket{+Z}\otimes\ket{0}$, evolve it for a variable time $\Delta t$, and compute $\Zdiss$ as a function of $\Delta t$. For each pair of $\epstwo$ and $\Delta$, we extract $\Tz$ from an exponential fit of $\Zdiss(\Delta t)$, both in the presence of resonant engineered dissipation ($\delta\omegadiss = 0$) and without dissipation ($\gdiss = 0$). The threshold value $\epstr$ for a given $\Delta$ is defined as the smallest $\epstwo$ for which $\Tz$ with engineered dissipation becomes larger than $\Tz$ without it.

Our model reproduces the behavior of $\Tz$ for $\epstwo<\epstr$ for the entire measured range. For $\epstwo>\epstr$ and $\Delta>2K$ (solid markers in Fig.~\ref{fig4}e) this remains the case, but the saturation of $\Tz$ for $\epstwo>\epstr$ and $\Delta<2K$ is not captured. A possible cause of this saturation is spurious multi-photon excitations to higher-lying oscillator-cavity states unaffected by our dissipative process and not included in our model~\cite{albornoz_oscillatory_2024,BlaisRescueKCQ,Dai_Hazra_2025} (see Supplementary Information Fig.~S17).

We comment here on the $\nth$ and $\nthb$ values used for the simulation results presented in Figs.~\ref{fig4}d,e. Firstly, we fix $\nth=\nthvalsim$, as this is the same value used to reproduce the measured $\pe$ at $\gdiss=0$, shown in Figs.~\ref{fig2}c and~\ref{fig3}b. We then set $\nthb=\nthbvalfigthree$. This value is chosen such that the simulated $\epstr$ falls on the $\dE_{1}/h \approx \SI{60}{\kilo\hertz}$ isoline at $\Delta = 0$, the point where the dependence of $\epstr$ on $\nthb$ is strongest. This choice provides good agreement with the experimental data over the full range of $\Delta$ in Fig.~\ref{fig4}e. Compared to the effective temperature used to reproduce the data in Fig.~\ref{fig3}, this results in a lower effective temperature $\Tcav\approx\Tcavvalfigthree$. We attribute this to a lower duty cycle of the measurement in Fig.~\ref{fig4}, or to a $\Delta$-dependence of the process described by the effective parameter $\nthb$. For more details on the simulation results presented in Fig.~\ref{fig4}e, see Supplementary Information Section~V.F.

Note that for all simulations in this work we use the eigenbasis of the driven oscillator to represent the oscillator state. We truncate the Hilbert space dimension to include at least the first two states outside the double-well quasi-potential. In case the number of states inside the double-well is smaller than eight, we still include the eight lowest energy eigenstates.
When including the cavity mode, we truncate its Hilbert space to three (two) Fock states for $\nthb >0$ $(\nthb = 0)$. This is based on the assumption that the average photon number in the cavity remains well below one as is the case for the adiabatic-elimination limit.
\\
\\
\textbf{Frequency-selectivity of the engineered dissipation process}\\
Here, we discuss the frequency selectivity of the engineered dissipation and explain why it has a negligible impact on the phase-flip time of the \KQ, despite being a single-photon process. 
For simplicity, we consider the case $\dE_i = 0$. The analysis can be extended to $\dE_i \neq 0$, with no change in the conclusions presented here.

We describe the effect of the oscillator-cavity interaction on transitions within the driven oscillator spectrum. To this end, we adiabatically eliminate the cavity mode in Eq.~\eqref{eq:diss_master_methods} for $\Delta_{\mathrm{b}}=\omegage+\delta\omegadiss$, following the method of Ref.~\cite{Breuer2007}. We thereby obtain transition-specific dissipators acting on an arbitrary single-photon transition $\ket{\psi_i^\pm}\leftrightarrow\ket{\psi_j^\mp}$
\begin{align}
    &\frac{\kappacav \, \gdiss^{2}}{\kappacav^{2}/4+(\omegage + \delta\omegadiss - \omega_{ij})^2}(1+\nthb)\, D[\Pi_j \aop \Pi_i],\\
    &\frac{\kappacav \, \gdiss^{2}}{\kappacav^{2}/4+(\omegage + \delta\omegadiss - \omega_{ij})^2}\, \nthb \, D[\Pi_i \aod \Pi_j],
\end{align}
where $\Pi_{i} = \sum_\pm \ket{\psi_{i}^\pm}\bra{\psi_{i}^\pm}$. We see from this equation that, for $\delta\omegadiss=0$, the effective dissipation rate is a Lorentzian centered around $\omegage$ with linewidth $\kappacav$. 
For the $\transge$ transition ($\omega_{ij}=\omegage$), the dissipators above correspond to cooling and, for finite $\nthb$, excitation processes. However, for all transitions $\psii \leftrightarrow\ket{\psi_{j}^{\mp}}$ where $|\omega_{ij}-\omegage|\gg\kappacav$, the process is suppressed. In particular, this is the case for the $\ket{\psi_0^\pm}\leftrightarrow\ket{\psi_0^\mp}$ transition, which corresponds to phase flips within the \KQ~manifold.
\\
\\
\textbf{Supplementary information content}\\
The accompanying Supplementary Information contains details on: the experimental setup (Section I, Supplementary Figs.~1,2); the full system parameters (Section II, Supplementary Fig.~3, Supplementary Table~1); the oscillator-cavity system Hamiltonian and Kerr-cat qubit calibration (Section III, Supplementary Figs.~4---7); modeling and calibrations of the $\pe$ measurements (Section IV, Supplementary Figs.~8---10, Supplementary Table~2); and calibration, analysis and modeling of engineered dissipation measurements (Section V, Supplementary Figs.~11---19 and Supplementary Tables~3---6).
\end{meth}

\begin{ai}
Correspondence should be addressed to Alexander Grimm (alexander.grimm@psi.ch).
\end{ai}

\begin{datav}
Numerical simulations were performed using a Python-based open source software (QuTiP). The data and code that support the findings of this study are available from the corresponding authors upon reasonable request.
\end{datav}

\setcounter{figure}{0}
\setcounter{table}{0}
\setcounter{equation}{0}

\renewcommand{\thefigure}{S\arabic{figure}}
\renewcommand{\thetable}{S\arabic{table}}
\renewcommand{\theequation}{S\arabic{equation}}

\clearpage

\onecolumngrid
\begin{center}
	\textsc{\Large{Supplementary Information}}
\end{center}

\section{Experimental Setup}
\subsection{Sample: description, fabrication, and design}

Our device consists of a superconducting nonlinear oscillator, which hosts the Kerr-cat qubit (\KQ), coupled to a three-dimensional (3D) microwave readout cavity (see Fig. \figone a of the main text).

\begin{figure*}[b]
    \includegraphics[angle = 0, width = \figwidthWide]{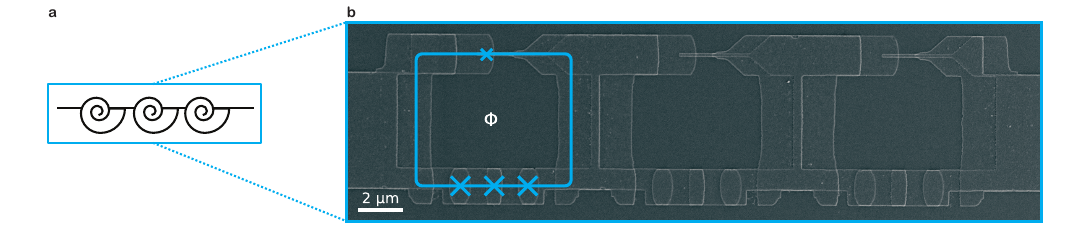}
    \caption{\label{fig_SI_Device} \textbf{Superconducting Nonlinear Asymmetric Inductive eLement (SNAIL) array. a}, Circuit schematic of the SNAIL array forming the nonlinear inductor element.  \textbf{b}, Scanning electron micrograph of the nonlinear inductor element, consisting of three SNAILs in series. Each SNAIL consists of four Josephson junctions in a superconducting loop~\cite{frattini_3-wave_2017}, as schematically indicated, and it is threaded by an external flux $\Phi$. }
    \end{figure*}

The nonlinear oscillator is patterned on a sapphire substrate and consists of a Superconducting Nonlinear Asymmetric Inductive eLement (SNAIL)~\cite{frattini_3-wave_2017} array (Fig.~\ref{fig_SI_Device}a) connecting two superconducting capacitive pads. The capacitor pads are made of tantalum, while the SNAIL array is made of aluminium and aluminium-oxide Josephson junctions.  
The device fabrication process follows the recipe described in Refs.~\cite{Place2021,Crowley2023}. We start from a sapphire wafer coated with a sputtered tantalum film purchased from Star Cryoelectronics. After cleaning the wafer in piranha solution, we define the capacitor geometry using optical lithography. The process involves spin-coating photoresist, patterning it with a laser writer, developing the resist, and performing an oxygen descum to remove resist residues in the developed areas. We then etch the tantalum layer in a hydrofluoric-acid-based solution. Finally, we remove the resist with solvent cleaning, followed by piranha-solution cleaning and a buffered oxide etch.  
We fabricate the SNAIL array using electron-beam lithography following the Dolan-bridge technique~\cite{Dolan1977}. A bilayer resist stack is spin-coated and covered by a thin gold charge-dissipation layer. After exposure and development in a deionized-water/isopropanol mixture, we perform argon ion milling to clean the developed areas and remove the native tantalum oxide. We then define the Josephson junctions via double-angle shadow evaporation of aluminium with an intermediate oxidation step. The resist stack is finally lifted off in N-methyl-2-pyrrolidone. Figure~\ref{fig_SI_Device}b shows a scanning electron micrograph of a representative SNAIL structure.

We use the fundamental mode of a rectangular 3D cavity for readout of the driven oscillator. To minimize ohmic losses while still allowing magnetic flux biasing of the SNAIL array, the cavity is composed of two halves: one made of aluminium and the other of copper coated with indium (with a thickness of $\approx \SI{2}{\micro\meter}$)~\cite{Brecht2015}. We intentionally leave a narrow slit on the copper surface uncoated by applying a mask during the indium deposition. This exposed copper region allows magnetic flux to enter the cavity. We place an external superconducting coil outside the cavity and we use it to apply the flux bias $\Phi$ to the SNAIL array. To model flux biasing, we perform simulations in Ansys Maxwell. The cavity is defined as a perfect conductor except for a narrow slit modeled as copper, which is designed to allow the magnetic flux to enter the cavity. We simulate the magnetic field profile of the external coil and optimize the slit geometry to minimize internal losses while ensuring we can thread at least half a flux quantum ($\Phi_0/2$) through the SNAIL array.

We couple the readout cavity to an aluminium waveguide that acts as a Purcell filter. This increases the coupling of the readout cavity to the readout transmission line without degrading the $\Tone$ of the nonlinear oscillator. The waveguide is designed with a cutoff frequency above that of the nonlinear oscillator but below that of the readout cavity. It is coupled to the cavity via a small aperture and is terminated on the opposite side with a matched coupler to the readout line. We use the waveguide to deliver both the cavity readout tone and the microwave pump that activates the squeezing drive for the \KQ. Additionally, we insert a weakly-coupled coaxial pin into the cavity. We use this port to apply drives for the \KQ~coherent control, for the engineered dissipation, and for the $Z$-state readout (\CQR).

To design the nonlinear oscillator, we first use an in-house Python script to analytically compute its frequency $\omegaqubit$ and the third- and fourth-order nonlinearities, $g_3$ and $g_4$ respectively, following the methods in Refs.~\cite{frattini_3-wave_2017,Grimm2020,frattini_observation_2024}. This procedure defines the oscillator charging energy $\Ec$, the number of SNAILs $\Nsnail$, the inductance of the large junctions $\Lj$ and the asymmetry factor $\jjasym$ between large and small junctions~\cite{frattini_3-wave_2017}. We then simulate the nonlinear oscillator and the readout cavity using Ansys HFSS to optimize the geometry of the oscillator capacitor pads. We adjust the pad size and spacing to achieve the target $\Ec$ and dispersive shift $\chidisp$ between the oscillator and the cavity. These quantities are extracted from the simulation using the PyEPR library~\cite{Minev2021}.

\subsection{Wiring diagram}

\begin{figure*}
    \includegraphics[angle = 0, width = \figwidthWide]{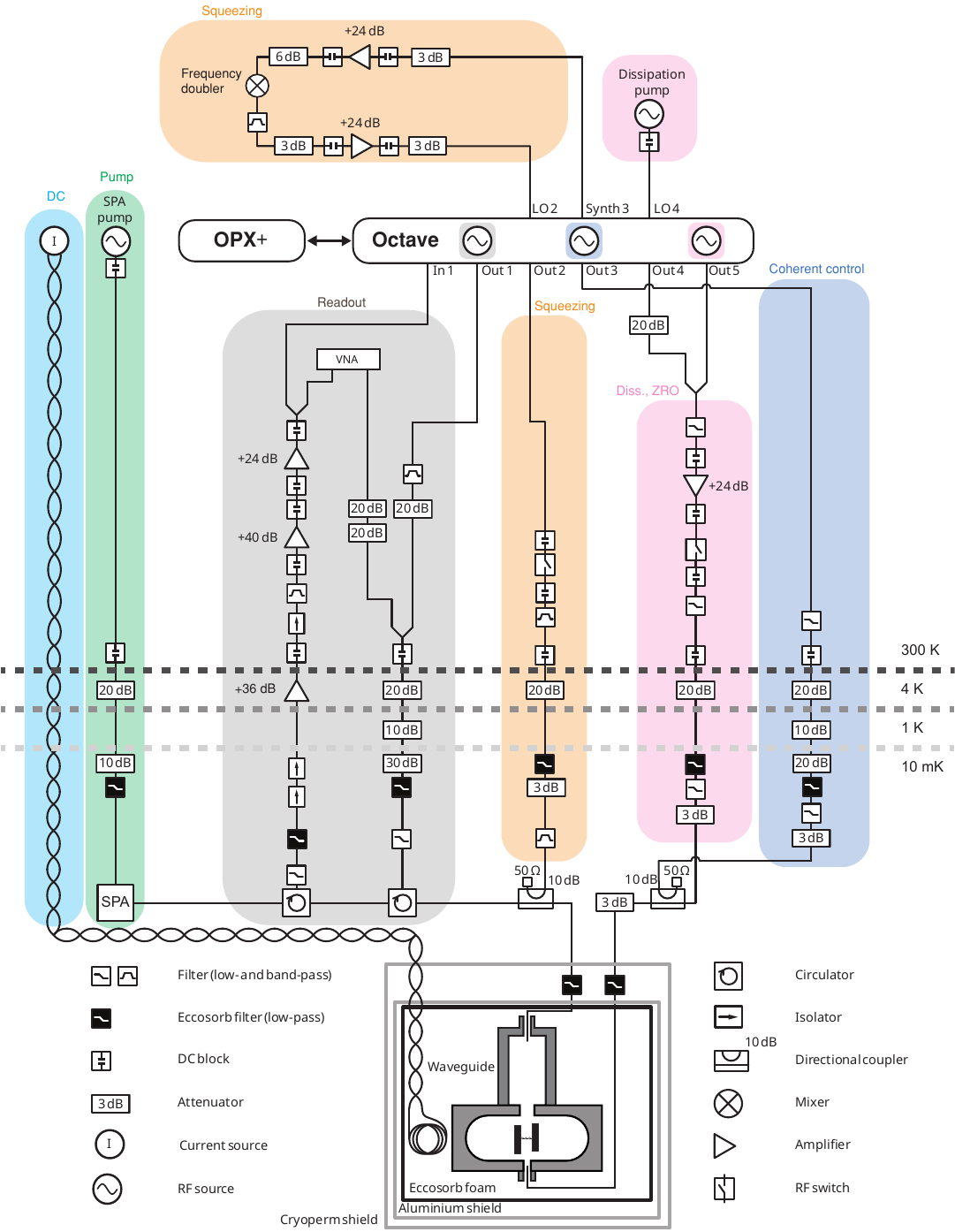}
    \caption{\label{fig_SI_Wiring} \textbf{Wiring Diagram.} Schematic of the experimental setup, including the nonlinear oscillator embedded in the readout cavity, the waveguide coupler, the coil to flux bias the SNAIL array, and the Cryoperm and aluminium shields. The various stages of the dilution refrigerator are indicated with gray dashed lines, with their corresponding temperatures on the right. Signal lines are color-coded, and a legend identifies the main components. }
    \end{figure*}
    
In this subsection, we present the wiring diagram of our experimental setup. Figure~\ref{fig_SI_Wiring} illustrates the diagram, where the different signal lines are grouped by color. From left to right: the DC line for flux biasing the SNAIL array (light blue), the pump line for the SNAIL-parametric amplifier (SPA)~\cite{Frattini2018} (green), the readout input-output line (gray), the squeezing-drive pump line (yellow), the \CQR~and engineered dissipation line (pink), and finally the line for coherent control of the nonlinear oscillator and \KQ~(blue).
We generate the control pulses and acquire the output signal using a Quantum Machines OPX+ and an Octave. All microwave switches present in the setup are triggered by the OPX+.

The DC line (light blue) connects a low-noise DC current source (Yokogawa GS200) at room temperature to the superconducting coil on the sample. This coil is used to apply a flux bias to the SNAIL array. The line is made of twisted-pair superconducting cables thermalized at various stages of the dilution refrigerator (DR).

The SPA pump line (green) delivers the pump tone to the SPA. A Keysight N5183B-520 RF source, triggered by the OPX+, generates the pump signal. The pump tone is then routed to the DR, where it is attenuated and filtered before reaching the SPA. 

On the readout line (gray), the input signal can either come from the Octave or from a vector network analyzer (VNA), which is used only in the initial tune-up of the experiment. The signal is then routed into the DR, reaching the mixing-chamber stage (at a temperature of approximately $\SI{10}{\milli\kelvin}$) where it is combined with the squeezing-drive pump line using a directional coupler. The reflected signal from the sample is first amplified by the SPA at the mixing-chamber stage, and then at the $\SI{4}{\kelvin}$ stage by a High-Electron-Mobility Transistor (HEMT) amplifier. Finally, the signal is further amplified at room temperature and then split towards the VNA and the Octave for demodulation. During \CQR~operations, we use the same readout line to acquire the output signal, but no input tone is applied on this line.

The squeezing-drive pump line (yellow) consists of two parts: one dedicated to generating the local oscillator (LO) of the squeezing-drive pump and the other for sending the modulated signal into the DR. For the first part, the coherent-control-line LO is amplified and routed to a frequency doubler. The upconverted signal is then filtered and amplified before being sent back to the Octave, where it is used as the LO for the squeezing-drive pump. The squeezing-drive pump signal is output from the Octave, sent to a microwave switch, a band-pass filter, and then to the DR. At the mixing-chamber stage, the signal passes through additional filtering stages, including a sharp band-pass filter centered around the squeezing-drive pump frequency. Finally, the squeezing-drive pump is combined with the readout signal through a directional coupler and arrives at the sample.

The \CQR~and frequency-selective dissipation lines are shown in pink. Both signals are generated by the Octave, on different output ports. Note that the dissipation tone uses an additional external LO (R\&S SGS100A SGMA). The two signals from the Octave are first combined, amplified, and then sent through a microwave switch and a low-pass filter before going to the DR. At the mixing-chamber stage, the signals go through further filtering stages, including a low-pass filter with a sharp cutoff. Finally, the signals are combined with the coherent-control line via a directional coupler before reaching the sample.

The coherent-control line is shown in blue. The signal is generated by the Octave, passes through a band-pass filter, and is then sent into the DR. At the mixing-chamber stage, the signal is further attenuated and filtered to suppress thermal noise before being combined with the \CQR~and dissipation signals through a directional coupler. The combined signal is then routed to the sample.

We mount the sample at the mixing-chamber stage of the DR, using a copper bracket that is thermally anchored to the mixing-chamber plate. To suppress infrared radiation and magnetic fields, we wrap the device in Eccosorb foam and enclose it in aluminium and Cryoperm shields.

\newpage
\section{System Parameters}
\label{sec:params}

In this section, we discuss the system parameters shown in Table~\ref{tab:params}, together with the methods used to estimate or measure them. 

\begin{figure}[b]
    \centering
    \includegraphics[width=\linewidth]{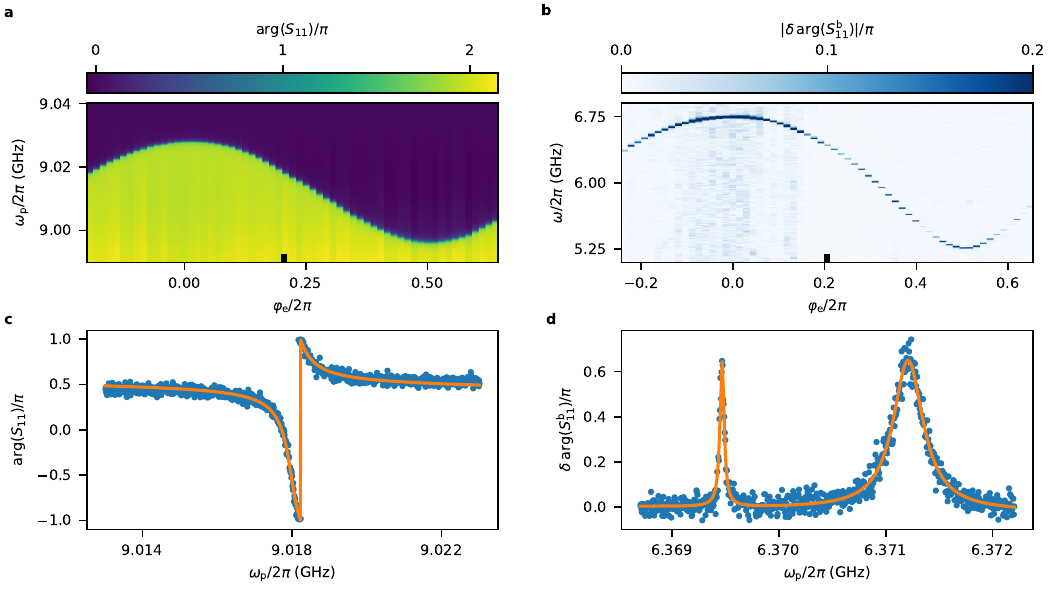}
    \caption{\textbf{Cavity and oscillator spectroscopy as a function of the reduced applied magnetic flux, $\varphi_{\mathrm{e}}=\Phi/\Phi_{0}$.}
    \textbf{a,} Phase of the cavity reflection coefficient $\mathrm{arg}(S_{11})$ as a function of $\varphi_{\mathrm{e}}$ and the probe frequency $\omega_{\mathrm{p}}$.
    \textbf{b,} Two-tone spectroscopy of the nonlinear oscillator as a function of $\varphi_{\mathrm{e}}$. The colorscale shows $|\delta\mathrm{arg}(S_{11}^{\mathrm{b}})|$, the absolute change in the phase of $S_{11}$ measured at the flux-dependent cavity frequency $\omegacav(\varphi_{\mathrm{e}})$ as the probe frequency $\omega_{\mathrm{p}}$ is swept. The rectangles on the $\varphi_{\mathrm{e}}$ axis in panels \textbf{a} and \textbf{b} mark the flux operating point $\varphi_{\mathrm{e}}/2\pi=0.202$.
    \textbf{c,} Cavity phase response at the operating point.
    \textbf{d,} Two-tone oscillator spectroscopy at the operating point. The broader peak corresponds to the $\ket{0}\leftrightarrow\ket{1}$ transition and the narrow peak to the $\ket{0}\leftrightarrow\ket{2}$ transition. In panels \textbf{c} and \textbf{d}, markers show the experimental data and solid lines are fits to a single (\textbf{c}) or two separate (\textbf{d}) Lorentzian profiles.}
    \label{fig:si_cav_qubit_spec}
\end{figure}

To characterize the nonlinear oscillator, we first sweep the external flux bias $\Phi$ applied to the SNAIL array and measure the nonlinear oscillator frequency using two-tone spectroscopy (see Fig.~\ref{fig:si_cav_qubit_spec}b). Given the designed values of the charging energy $\Ec$ and the number of SNAILs $\Nsnail$, we fit the measured nonlinear-oscillator flux-dependent frequency response and extract an approximate value for the inductance of the large Josephson junctions ($\Lj$) and for the asymmetry factor $\jjasym$ between the large and small junctions in the SNAIL. We set the flux bias to $\Phi/\Phi_0 = \fluxpoint$, which is the operating point for the experiment.

We first characterize the readout cavity at this bias point. We perform a VNA measurement of the cavity reflection coefficient as a function of probe frequency $\omega_{\mathrm{p}}$ (see Fig.~\ref{fig:si_cav_qubit_spec}c). By fitting the response, we extract the cavity resonance frequency $\omegacav$, the external coupling rate to the readout line $\kappa_{\mathrm{b,out}}$, and the total additional loss rate $\kappa_{\mathrm{b,l}}$. The last term includes both the internal losses of the cavity, $\kappa_{\mathrm{b,int}}$, and the losses associated with the additional coupling pin used to drive the oscillator, $\kappa_{\mathrm{b,drive}}$. We estimate the contribution from the drive pin to be approximately $\kappa_{\mathrm{b,drive}}/2\pi \approx \SI{30}{\kilo\hertz}$.

We then characterize the nonlinear oscillator at this bias point. The oscillator frequency $\omegaqubit$ and Kerr nonlinearity $K$ are extracted via two-tone spectroscopy by identifying the $\ket{0} \leftrightarrow \ket{1}$ and $\ket{0} \leftrightarrow \ket{2}$ transitions of the undriven ($\epstwo = 0$) oscillator (see Fig.~\ref{fig:si_cav_qubit_spec}d). The three-wave mixing strength $g_3$ of the oscillator is numerically estimated using an analytical model for the nonlinear oscillator~\cite{frattini2021phd} and fitting the spectroscopy measurements of the driven oscillator~\cite{frattini_observation_2024}. To extract the thermal occupation in the $\ket{1}$ state of the undriven oscillator, $\nth$, we perform a self-calibrated Rabi-contrast measurement on the $\ket{1}\leftrightarrow\ket{2}$ transition with and without first applying a $\pi_{01}$-pulse to the $\ket{0}\leftrightarrow\ket{1}$ transition~\cite{Geerlings2013,Jin2015}. We characterize the oscillator coherence by measuring the single-photon relaxation time $\Tone$, the Ramsey coherence time $\TtwoR$, and the Hahn-echo coherence time $\TtwoE$ using standard transmon-measurement protocols. These values remained stable over the four-day measurement period required to collect the data presented in this work. However, we observe significant fluctuations on longer timescales, which we attribute to coupling between the oscillator mode and microscopic two-level fluctuators in the substrate and in the Josephson junction tunneling barriers~\cite{Lisenfeld2015}.

We measure the dispersive shift $\chidisp$ between the oscillator and the cavity by applying a $\pi_{01}$-pulse to the $\ket{0} \leftrightarrow \ket{1}$ transition of the oscillator and subsequently measuring the cavity resonance frequency $\omegacav$. The shift in the cavity frequency approximately corresponds to the dispersive shift $\chidisp$~\cite{Schuster2007}.

The \CQR~interaction rate $\gcqr$ between the nonlinear oscillator and the readout cavity is measured by performing a $\Tone$ experiment on the oscillator while activating the single-photon interaction during the wait time~\cite{Grimm2020}. The calibration procedure is identical to that used for determining the interaction rate $\gdiss$. See Sections III.D and V.B for further details.

\begin{table}[h]
	\centering
    \renewcommand{\arraystretch}{1.2}
	\begin{tabular}{l|c|l}
		\textbf{Parameter} 
		& \textbf{Value}
		& \textbf{Method of estimate or measurement}\\ \hlineB{2}   
        Nonlinear oscillator charging energy $\Ec/2\pi$   & \Ecval & Design and Simulation \\ \hline
        Nonlinear oscillator number of SNAILs $\Nsnail$   & 3 & Design \\ \hline
        Nonlinear oscillator linear inductance $\Llin$  & \Llinval & Design and fitting of flux curve  \\ \hline
        Asymmetry factor in the SNAIL $\jjasym$  & \alphasnail & Design and fitting of flux curve \\ \hline
        Large Josephson junction inductance in SNAIL $\Lj$ & \Ljval & Design and fitting of flux curve  \\ \hline
        SNAIL flux bias point $\Phi/\Phi_0$ & \fluxpoint & Flux curve \\ \hline
        
        Nonlinear oscillator frequency $\omegaqubit/2\pi$   & \fqubit & Two-tone spectroscopy \\ \hline
        Kerr-nonlinearity $K/2\pi$ & \kerr & Two-tone spectroscopy of $\ket{0} \leftrightarrow \ket{2}$ transition  \\ \hline
        Third-order nonlinearity $g_3/2\pi$ & $\approx\gthreeval$ & \makecell[l]{Design, fitting of flux curve, \\ spectroscopy of driven oscillator} \\ \hline
        Undriven nonlinear oscillator thermal photon number $\nth$ & \nthval & Self-calibrated Rabi-contrast measurement ($\epstwo=0$) \\ \hline
        Nonlinear oscillator single-photon decay time $\Tone$   & \bareTone & Standard relaxation measurement \\ \hline
        Nonlinear oscillator Ramsey coherence time $\TtwoR$   & \bareTphi & Standard Ramsey coherence measurement \\ \hline
        Nonlinear oscillator Ramsey coherence time (echo) $\TtwoE$   & \bareTecho & Standard Hahn-echo coherence measurement \\ \hline
        Readout cavity frequency $\omegacav/2\pi$  & \fres & Direct RF reflection measurement \\ \hline 
        Readout cavity linewidth (output coupling) $\kappa_{\mathrm{b,out}}/2\pi$   & \kappaout & Direct RF reflection measurement \\ \hline
        Readout cavity linewidth (other losses) $\kappa_{\mathrm{b,l}}/2\pi$   & \kappaother & Direct RF reflection measurement \\ \hline
        Cavity --- nonlinear oscillator dispersive shift $\chidisp/2\pi$ & $\approx \dispshiftval$ & Cavity spectroscopy with (without) $\pi_{01}$-pulse \\ \hline
        \CQR~interaction rate $\gcqr/2\pi$ & $\gCQRval$ & \makecell[l]{$\Tone$ experiment on the oscillator ($\epstwo = 0$) \\ with $\gcqr$ interaction}  \\ \hline
        
	\end{tabular}  
	\caption{\textbf{System parameters.} Summary of the main system parameters. }
	\label{tab:params}
\end{table}

\newpage
\section{System Hamiltonian and Kerr-Cat Qubit Calibration}

\subsection{Hamiltonian of the nonlinear oscillator coupled to the readout cavity}
In this subsection, we define the Hamiltonian describing the nonlinear oscillator, the readout cavity, their coupling, as well as the microwave drives that activate both the squeezing drive and \CQR. In addition, a resonant single-photon drive can be applied to perform a Z-rotation on the qubit.

The full Hamiltonian describing the system is
\begin{equation}
\hat{H}_{0} = \hat{H}_{\mathrm{a}} + \hat{H}_{\mathrm{b}} + \hat{H}_{\mathrm{i}} + \hat{H}_{\mathrm{d}}.
\end{equation}
The first term is the Hamiltonian of the nonlinear SNAIL oscillator, $\hat{H}_{\mathrm{a}}/\hbar = \omegaqubit^0 \aod_0 \aop_0 + \gthre (\aod_0 + \aop_0)^{3} + g_{4} (\aod_0 + \aop_0)^{4}$ with bare-mode annihilation operator $\aop_0$, bare-mode frequency $\omegaqubit^0$, and respective third- and fourth-order nonlinearities $\gthre$ and $\gfour$~\cite{frattini_3-wave_2017}. The second term is the Hamiltonian of the readout cavity, $\hat{H}_{\mathrm{b}}/\hbar = \omegacav^0 \bod_0 \bo_0$, with bare-mode frequency $\omegacav^0$ and annihilation operator $\bo_0$. The third term describes the interaction between the two modes, $\hat{H}_{\mathrm{i}}/\hbar = g(\aod_0+\aop_0)(\bod_0+\bo_0)$, parametrized by a coupling strength $g$. Finally, external drives are described by the Hamiltonian $\hat{H}_{\mathrm{d}}/\hbar = 2 \left[ \epssq \, \mathrm{Re}(e^{i \omegasq t}) + \epscqr \, \mathrm{Re}(e^{i \omegacqr t}) + \epsilon_\mathrm{Z} \, \mathrm{Re}(e^{i \omegasq t/2 + i\phiz}) \right] (\aod_0 + \aop_0)$. Slowly-varying amplitudes $\epssq$, $\epscqr$, $\epsilon_\mathrm{Z}$ respectively correspond to drives that activate the squeezing drive, \CQR~interactions and the $Z$-axis rotation in the \KQ, at frequencies \( \omegasq \), \( \omegacqr \) and $\omegasq/2$. Here $\phiz$ is the phase of the $Z$-axis rotation drive.

We now perform a series of transformations on $\hat{H}_{0}$ to obtain a static-effective Hamiltonian for the driven nonlinear oscillator based on standard techniques used to analyze parametric processes in circuit quantum electrodynamics. These calculations are described in more detail in Refs.~\cite{Grimm2020,frattini_observation_2024} and only briefly summarized here. We first obtain a Hamiltonian for the dressed modes by performing a first-order rotating-wave approximation (RWA) of the interaction Hamiltonian $\hat{H}_{\mathrm{i}}$ and diagonalizing the linear coupling. This gives the dressed-mode frequencies \( \omegaqubit = \omegaqubit^0 - 2g^{2}/\Deltaab\) and \( \omegacav = \omegacav^0 + 2g^{2}/\Deltaab \), where \( \Deltaab = \omegacav^0 - \omegaqubit^0 \) is the difference between bare-mode frequencies. Then, after applying a displacement transformation to eliminate the linear pump drives~\cite{Grimm2020, frattini2021phd}, we move into the rotating frames of the driven nonlinear oscillator (\( \omegasq/2 \)) and readout cavity (\( \omegacav \)). Finally, we expand the nonlinear terms and perform a first-order RWA. The resulting static-effective Hamiltonian can be written in terms of a \KQ~effective Hamiltonian (Eq.~1 of the main text, reproduced here for readability), drive-induced Stark shifts, an oscillator-cavity dispersive shift, a \CQR~interaction and the Z-axis rotation drive
\begin{equation}
    \hat{H} = \Hkcq + \Hstark + \Hdisp + \Hcqr + \Hz,
\end{equation}
with
\begin{equation}
     \Hk/\hbar = -K \, \aodtwo \aoptwo,
     \label{eq:Ham_k}
\end{equation}
\begin{equation}
    \Hkcq/\hbar = \Delta \, \aod \aop + \Hk/\hbar + \epstwo ( \aoptwo + \aodtwo ), 
    \label{eq:Ham_kcq}
\end{equation}
\begin{equation}
    \Hstark/\hbar = -4K\left( |\xisq{eff}|^2 + |\xicqr{eff}|^2 \right) \aod \aop, 
    \label{eq:Ham_stark}
\end{equation}
\begin{equation}
    \Hdisp/\hbar = \chidisp \, \aod \aop \, \bod \bo,
    \label{eq:Ham_disp}
\end{equation}
\begin{equation}
    \Hcqr/\hbar = \gcqr \left( \aop \bod + \aod \bo \right),
    \label{eq:Ham_cqr}
\end{equation}
\begin{equation}
    \Hz/\hbar = \epsz^* \hat{a} +\epsz \hat{a}^{\dag}.
    \label{eq:Hamzrot}
\end{equation}
The Kerr nonlinearity is \( K = -6 \gfoureff \), where \( \gfoureff = \gfour - 5 \gthre^2/ \omegaqubit \) is the effective four-wave mixing coefficient obtained via second-order perturbation theory~\cite{frattini2021phd}. The squeezing drive is parametrized by a frequency detuning \( \Delta = \omegaqubit - \omegasq/2 \) and amplitude \( \epstwo = 3 \gthre \, \xisq{eff} \), 
where $\xisq{eff} = \epssq\left( \frac{1 }{\omegasq-\omegaqubit} -\frac{ 1}{\omegasq+\omegaqubit} \right)$ 
is the effective linear squeezing-drive displacement amplitude. 
The squeezing drive and Kerr nonlinearity combine to give the \KQ~Hamiltonian, $\Hkcq$. In the presence of parametric drives, the nonlinear oscillator has a Stark shift, which is a function of the displacement amplitudes \( \xisq{eff} \) and $\xicqr{eff} =  \epscqr\left( \frac{1}{\omegacqr-\omegaqubit } -\frac{1}{\omegacqr+\omegaqubit } \right) $ for the squeezing-drive and \CQR~pumps, respectively. The dispersive shift between the nonlinear oscillator and the readout cavity is given by \( \chidisp = 24 \gfoureff (g/\Deltaab)^2 \).  The \CQR~drive generates a beam-splitter interaction \(\Hcqr\) with coupling strength 
\(\gcqr = 6 \gthre \, \xicqr{eff} \, (g/\Deltaab)\). Finally, the single-photon drive, with complex amplitude \(\epsz\ = \epsilon_\mathrm{Z} e^{-i\phiz}\), implements the Z-axis rotation of the \KQ.

\subsection{Kerr-cat qubit basis state definitions}

In this subsection, we define the \KQ~basis states in terms of the ground states of Eq.~\eqref{eq:Ham_kcq}, denoted by $\ket{\psi_0^{\pm}}$, and describe how to map the $\ket{0}$ ($\ket{1}$) state of the undriven oscillator ($\epstwo = 0$) onto $\ket{\psi_0^{+}}$ ($\ket{\psi_0^{-}}$). The superscript of $\ket{\gstate}$ refers to photon number parity, such that $\ket{\psi_{0}^{+}}$ ($\ket{\psi_{0}^{-}}$) is given by a superposition of Fock states which only have an even (odd) number of photons. 

The \KQ~basis states are given by
\begin{align*}
    \ket{\pm X} &= \ket{\gstate},\\
    \ket{\pm Y} &= \frac{1}{\sqrt{2}}\left(\ket{\psi^{+}_0}\mp i\ket{\psi^{-}_0}\right),\\
    \ket{\pm Z} &= \frac{1}{\sqrt{2}}\left(\ket{\psi^{+}_0}\pm\ket{\psi^{-}_0}\right).
\end{align*}

In the specific case of $\Delta=0$, the ground states are the Schr{\"o}dinger cat states $\ket{\gstate}=\catpm\equiv\calN_{\alpha}^{\pm}(\ket{+\alpha}\pm\ket{-\alpha})$, where $\alpha=\sqrt{\epstwo/K}$ is the amplitude of coherent states $\ket{\pm\alpha}$ and $\calN_{\alpha}^{\pm}=1/\sqrt{2(1\pm e^{-2|\alpha|^2})}$ is a normalization coefficient. There, the \KQ~basis states are $\ket{\pm X}=\catpm$, $\ket{\pm Y}=\ypm$ and $\ket{\pm Z}=(\catp\pm\catm)/\sqrt{2}\approx\ket{\pm\alpha}$, where the $Z$-states are exactly given by $\ket{\pm\alpha}$ in the limit of large $\alpha$~\cite{puri_engineering_2017}.

To initialize the system in the \KQ~manifold, we start from the $\ket{0}$ ($\ket{1}$) state of the undriven oscillator ($\epstwo = 0$). We then adiabatically ramp the squeezing-drive amplitude and detuning, thereby mapping the $\ket{0}$ ($\ket{1}$) state onto the $\ket{\psi_0^+}$ ($\ket{\psi_0^-}$) state of the driven oscillator~\cite{Grimm2020}. This mapping is possible because the states $\ket{0}$ ($\ket{1}$) and $\ket{\psi_0^+}$ ($\ket{\psi_0^-}$) share the same even (odd) parity, which is conserved by the squeezing drive.

\subsection{Initialization with detuning}
In this subsection, we describe the initialization protocol we follow to reliably prepare the oscillator in the \KQ~manifold when $\Delta >0$.

The detuned \KQ~is described by the Hamiltonian of a Kerr-nonlinear oscillator under a squeezing drive, $\Hkcq$ (Eq.~\eqref{eq:Ham_kcq}).
This system exhibits a phase transition as a function of the detuning $\Delta$. For small $\Delta$, the Hamiltonian supports two stable attractors corresponding to opposite-phase approximate coherent states $\ket{\pm \alphaD}$ which define the two \KQ~basis states $\ket{\pm Z}$. However, when $\Delta$ exceeds a critical value $\Deltacrit = 2\epstwo$, the vacuum state at the origin of phase space becomes a stable attractor together with $\ket{\pm\alphaD}$~\cite{venkatraman_driven_2024,eichler2023parametric,Wustmann2019}.
In the presence of single-photon loss, the vacuum becomes the steady state, while the states $\ket{\pm \alphaD}$ become metastable~\cite{Gravina2023}.
For the \KQ, operating at large $\Delta>0$ is desirable to enhance the bit-flip time ($\Tz$), as it increases the depth of the double-well potential without requiring excessively large squeezing amplitudes $\epstwo$, that could induce spurious excitations in the oscillator~\cite{Venkatraman2024_nl_diss, BlaisRescueKCQ}. However, when $\Delta > \Deltacrit$, initializing the system becomes nontrivial: a simple ramp-up of the squeezing-drive amplitude would leave the system trapped at the origin of the phase space.

\begin{figure*}[b]
    \includegraphics[angle = 0, width = \figwidthWide]{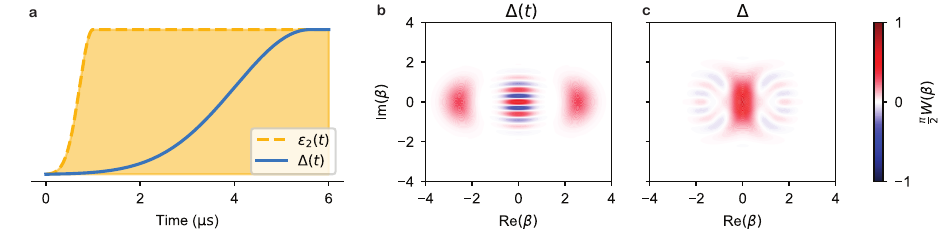}
    \caption{\label{fig_SI_Chirp} \textbf{Initialization with a ramped squeezing-drive detuning. a}, Profiles of the squeezing-drive amplitude $\epstwo(t)$ (yellow dashed line) and the detuning $\Delta(t)$ (blue solid line) as a function of time during the initialization sequence.
    \textbf{b,c}, Simulated Wigner functions of the final state when ramping up $\epstwo$ starting from the vacuum state, for the case with and without ramping of $\Delta$, respectively. We perform the simulations for \workingpoint.}
\end{figure*}

To address this, we implement an initialization protocol in which both the squeezing-drive amplitude $\epstwo(t)$ and the detuning $\Delta(t)$ are ramped from zero~\cite{Gravina2023}. This is depicted in Fig.~\ref{fig_SI_Chirp}a. Starting from the vacuum state, we ramp up $\epstwo(t)$ with a Gaussian profile of duration $\taurampsq = \taurampval$ $(\sigma = \sigmarampval)$. At the same time, we dynamically change the frequency of the squeezing drive by ramping the detuning with a Gaussian profile, from $\Delta(0) = 0$ to $\Delta(\taurampD) = \Delta$ over a duration $\taurampD = \tauchirpval$ $(\sigma = \sigmachirpval)$. This protocol allows us to adiabatically map the vacuum state of the undriven nonlinear oscillator into the \KQ~manifold.

To support this approach, we perform numerical simulations (neglecting losses) of Eq.~\eqref{eq:Ham_kcq} starting from the vacuum state and ramping up the squeezing-drive amplitude with and without ramping up the detuning. We perform the simulation for \workingpoint. Figures~\ref{fig_SI_Chirp}b,c show the Wigner functions of the final states obtained in both cases. With a ramped squeezing-drive frequency detuning, the system reaches the desired \KQ~$\ket{\gstate}$ manifold with a fidelity of 93\%, whereas without ramping, the fidelity drops dramatically to 0.001\%.

\subsection{$Z$-state readout rate, QND-ness and fidelity}
\label{sec:CQR}
In this subsection, we describe the calibration measurements performed to characterize the \CQR. The \CQR~is used to both initialize and readout the \KQ~in the $Z$-basis, and is an integral part of the measurements in Figs.~\figfour~and \figfive~of the main text.

This readout is a parametrically-activated process realized by driving the system at the frequency difference between the \KQ~and the readout cavity, $\omegacqr=\omegacav-\omegasq/2$. The drive activates a three-wave mixing process in the nonlinear oscillator, resulting in the effective Hamiltonian $\Hcqr$ (Eq.~\eqref{eq:Ham_cqr}). In the \KQ~basis, we can write Eq.~\eqref{eq:Ham_cqr} as

\begin{equation}
\hat{H}_\mathrm{zro,b}/\hbar =\gcqr \hat{\sigma}_z( \alpha_{\Delta} \hat{b}^{\dag} +  \alpha_{\Delta}^* \hat{b}),
\end{equation}
where $\hat{\sigma}_z$ is the Pauli-$Z$ operator in the \KQ~basis, and $\gcqr$ denotes the rate of the frequency-converting beam-splitter interaction. 
This Hamiltonian induces a qubit-state-dependent displacement of the readout cavity: photons in the nonlinear oscillator with quadrature amplitude $\pm\alpha_\Delta$ displace the cavity by $\pm2i\gcqr\alpha_\Delta/\kappacav$~\cite{Grimm2020}. Because the displacement is conditional on the $Z$-state of the \KQ, measuring the cavity implements a quantum non-demolition (QND) readout that projects the qubit onto the $Z$-basis~\cite{Grimm2020,frattini_observation_2024}.

\begin{figure*}
    \includegraphics[angle = 0, width = \figwidthWide]{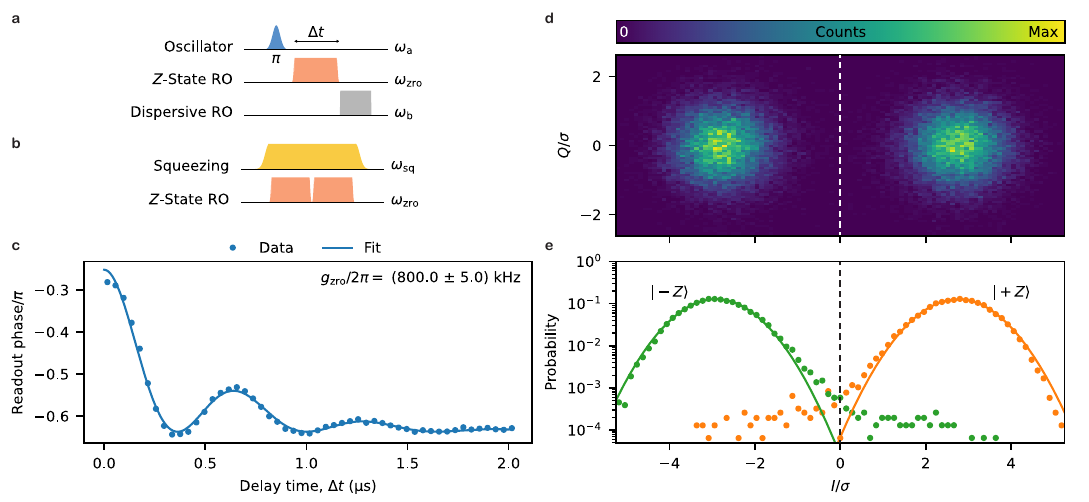}
    \caption{
    \label{fig_SI_qnd} 
    \textbf{Calibration of $Z$-state readout (\CQR).} 
    \textbf{a}, Pulse sequence used to extract the coupling rate $\gcqr$. 
    \textbf{b}, Pulse sequence for measuring the QND-ness and fidelity of the \CQR. 
    \textbf{c}, Measured readout phase normalized by $\pi$ (blue dots) as a function of delay time $\dt$. A fit (solid line) gives a coupling strength $\gcqr/2\pi = \SI{800.0\pm5.0}{\kilo\hertz}$. 
    \textbf{d}, Histogram of the $I$ and $Q$ quadratures of the cavity field, measured over $\cqrnshots$ shots for the first readout. Two distinct distributions along the $I$-axis correspond to measurement outcomes $\ket{+Z}$ and $\ket{-Z}$. The vertical dashed line at $I = 0$ indicates the measurement threshold. Both axes are normalized by the data standard deviation of one distribution $\sigma = \sigmaval$.
    \textbf{e}, Conditional probability distributions of the second \CQR~readout given the first measurement outcome is $\ket{-Z}$ (green) or $\ket{+Z}$ (orange). The markers represent the data and the solid line is a Gaussian fit.
    Measurements in d,e are performed for \workingpoint~and $\gcqr/2\pi = \gCQRval$.
    }
\end{figure*}

To calibrate $\gcqr$, we perform a time-domain measurement on the nonlinear oscillator in absence of the squeezing drive ($\epstwo = 0$). The pulse sequence, illustrated in Fig.~\ref{fig_SI_qnd}a, is similar to a standard transmon $\Tone$ experiment: we first apply a $\pi$-pulse to the nonlinear oscillator, then apply the $\gcqr$ interaction during a variable delay time $\dt$ before readout. This interaction induces a coherent single-photon exchange between the nonlinear oscillator and the readout cavity, leading to damped oscillations in the nonlinear-oscillator population, as shown in Fig.~\ref{fig_SI_qnd}c. We fit these oscillations using an analytical model (see Section V.B) to extract the coupling rate. From this fit, we determine $\gcqr/2\pi = \SI{800.0\pm5.0}{\kilo\hertz}$, with the uncertainty given by the standard error of the fit.

After calibrating $\gcqr$, we characterize the QND-ness and fidelity of the \CQR~at the working point discussed in Figs.~\figtwo, \figthree~and \figfive~of the main text (\workingpoint). To do so, we perform the pulse sequence shown in Fig.~\ref{fig_SI_qnd}b. We start by adiabatically ramping up the squeezing drive, then we apply two consecutive \CQR~pulses, and we finally adiabatically ramp down the squeezing drive. Each \CQR~pulse is a flat-top Gaussian with a constant length of $\SI{2}{\micro\second}$ and a rise time of $\SI{20}{\nano\second}$ ($\sigma=\SI{8}{\nano\second}$). Figure~\ref{fig_SI_qnd}d shows a histogram of the results from the first \CQR, plotted in the $IQ$ plane of the readout-cavity field. The data clusters into two well-separated regions along the $I$ quadrature, corresponding to the two possible readout outcomes $\ket{\pm Z}$ for the \KQ. We set a threshold at $I = 0$ to discriminate between $\ket{+Z}$ and $\ket{-Z}$ results.

In Fig.~\ref{fig_SI_qnd}e, we display the conditional probability distributions of the $I$-quadrature result for the second \KQ~measurement, given that the first measurement yielded either $\ket{+Z}$ (orange) or $\ket{-Z}$ (green). We fit these distributions with two Gaussian curves (solid lines) and obtain a standard deviation $\sigma = \sigmaval$ of the voltage signal on our analog-to-digital converter, which we use to normalize the axis in Figs.~\ref{fig_SI_qnd}d,e. We extract the readout fidelity and QND-ness directly from the experimental data following Refs.~\cite{Grimm2020,frattini_observation_2024}, and obtain
\begin{align}
\Fcqr &= 1 - p(\mathrm{+Z|{-Z}}) - p(\mathrm{-Z|{+Z}}) = 99.41\%, \\
\Qcqr &= \frac{p(\mathrm{+Z|+Z}) + p(\mathrm{-Z|-Z})}{2} = 99.71 \% .
\end{align}

\subsection{Characterizing the \KQ\ energy spectrum}

In this section, we show the spectroscopy measurements that we use to map the leakage spectrum of the \KQ\ and thereby calibrate the two-photon-drive amplitude $\epstwo$ as a function of the squeezing-drive pump amplitude $\Vsq$. This follows the description provided in the Methods.
For this measurement, we ramp on the two-photon drive at fixed detuning $\Delta$, initialize the system in $\ket{+Z}$ with a first \CQR\ measurement, and then apply a single-photon probe pulse of fixed duration $\tau$ at a variable frequency $\omegaprobe = \omegasq/2 + \delta\omegaprobe$. We choose $\tau$ to be smaller than or comparable to the bit-flip time $\Tz$ to maximize the spectroscopic contrast, while keeping the probe amplitude much smaller than $\epstwo$ so that the probe does not significantly change the eigenstates of the system. A second \CQR\ measurement then reads out the final state. When the probe is resonant with a transition from the ground manifold $\ket{\gstate}$ to a leakage manifold $\psii$, it populates that manifold, which reduces the confinement of the $\ket{+Z}$ state and induces inter-well transitions that lower the measured signal~\cite{frattini_observation_2024}.

\begin{figure}
    \centering
    \includegraphics[width=\linewidth]{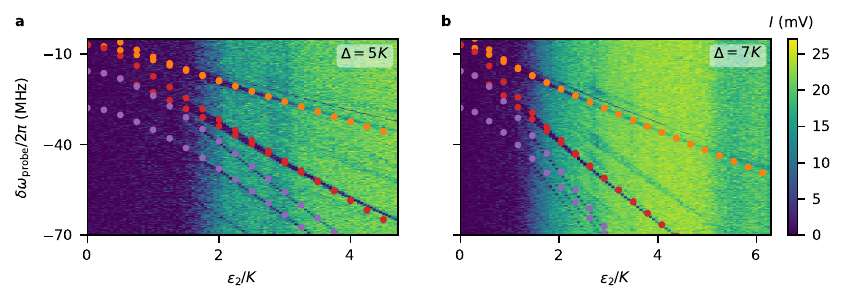}
    \caption{\textbf{Spectroscopy of the driven-oscillator energy levels.} Measured readout-signal amplitude $I_{\mathrm{ZRO}}$ as a function of the two-photon-drive amplitude $\epstwo$ and the probe-frequency detuning $\delta\omegaprobe$, for $\Delta=5K$ (\textbf{a}) and $\Delta=7K$ (\textbf{b}). In both panels the probe pulse has duration $\tau=\SI{50}{\micro\second}$ and amplitude $\epsilon_{\mathrm{probe}}/2\pi\approx\SI{300}{\kilo\hertz}$. Markers indicate the transition frequencies computed numerically from $\Hkcq+\Hstark$, with orange, red and purple corresponding to the $\transge$, $\ket{\gstate}\leftrightarrow\ket{\fstateflip}$ and $\ket{\gstate}\leftrightarrow\ket{\hstateflip}$ transitions, respectively. Additional features in the data correspond to transitions from $\ket{\estate}$ to higher leakage manifolds, which we omit here to avoid overcrowding the figure.}
    \label{fig:si_kcq_spectroscopy}
\end{figure}

Example measurements for $\Delta=5K$ and $\Delta=7K$ are shown in Fig.~\ref{fig:si_kcq_spectroscopy}. We convert the pump amplitude into the two-photon-drive amplitude through the linear relation $\epstwo=\eta\Vsq$, where the conversion factor $\eta$ is obtained by fitting the numerically computed transition frequencies of $\Hkcq+\Hstark$ to the measured spectra (markers in Fig.~\ref{fig:si_kcq_spectroscopy}). For each dataset at fixed $\Delta$, this single free parameter reproduces multiple transitions across a wide range of $\epstwo$. Note that the background signal grows with $\epstwo$, which is consistent with the simultaneous improvement of $\Tz$ and of the readout signal-to-noise ratio.

\subsection{Calibration of $X$-and $Z$-gates}

In this subsection, we describe how to calibrate the $X(\pi/2)$-gate (also referred to as the Kerr gate) and the $Z$-gate for \workingpoint, used for measurements presented in Fig.~\figfive~of the main text.

To calibrate the \( X(\pi/2) \)-gate, we need to determine both its duration and the associated phase rotation of the \KQ~\cite{Grimm2020, frattini_observation_2024}. We expect the $X(\pi/2)$-gate to generate an additional \( \pi/2 \) rotation in phase space, and to have a gate duration \( \tauxgate = \pi/2K \).  We estimate \( \tauxgate \approx \kerrgatetimeest\) using the Kerr nonlinearity extracted from incoherent spectroscopy of the undriven oscillator. For a more precise calibration of the gate time and phase rotation, we initialize the system in the \( \ket{+Y} \) state by ramping on the squeezing drive starting from the nonlinear oscillator state \( ( \ket{0} - i\ket{1} ) / \sqrt{2} \). We then set $\epstwo = 0$ for a variable time while also sweeping the phase updates for the squeezing drive and for the \CQR. Finally, we switch the squeezing drive back on and perform a \CQR. Since the state $\ket{+Y}$ is mapped to $\ket{+Z}$ after a successful $X(\pi/2)$ rotation, we identify the optimal parameters by maximizing the contrast of the final \CQR. This gives a calibrated gate time of \( \tauxgate \approx \kerrgatetimemmt\).

\begin{figure*}
    \includegraphics[angle = 0, width = \figwidthWide]{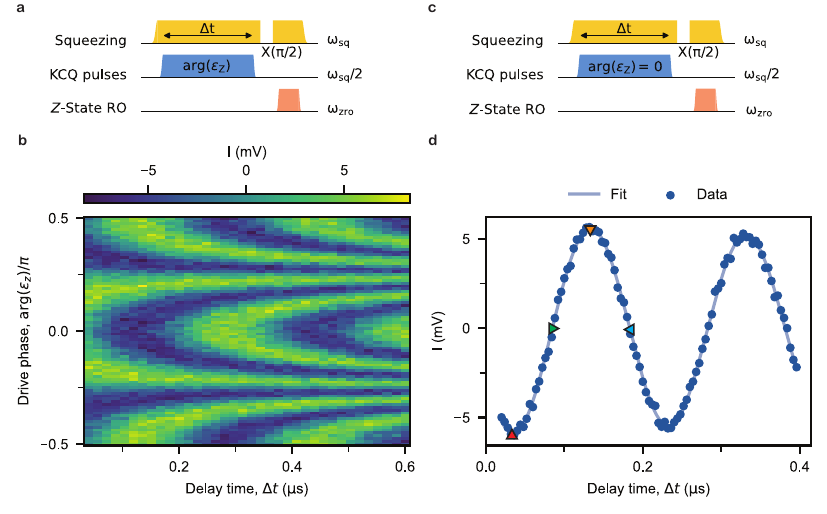}
    \caption{
    \textbf{Calibration of the \( Z \)-rotation gate.} 
    \textbf{a}, Pulse sequence for calibrating the phase of the \( Z \)-axis rotation drive \( \epsz \). 
    \textbf{b}, Measured demodulated voltage $I$ obtained from the \CQR, as a function of drive phase \( \arg(\epsz) \) and Rabi time \( \dt \). Here, the drive amplitude is set to \( \epsz/2\pi = \epszvalOne \).
    \textbf{c}, Pulse sequence for a Rabi experiment at \( \arg(\epsz)=0 \), similar to \textbf{a}.
    \textbf{d}, Measured Rabi oscillations (blue points) and corresponding fit (light blue line). Colored markers denote drive durations corresponding to specific \( Z \)-axis rotations: \( Z(0) \) (green right-pointing triangle), \( Z(\pi/2) \) (orange downward triangle), \( Z(\pi) \) (blue left-pointing triangle), and \( Z(3\pi/2) \) (red upward triangle).
    Measurements are performed for \workingpoint~with $\epsz/2\pi =\epszval$.
    }
    \label{fig:SI_KCQ_rabi}

\end{figure*}

We now calibrate the $Z$-axis rotation. The rotation is implemented by applying a drive at the \KQ~frequency $\omegasq/2$, which realizes the Hamiltonian $\Hz$ (Eq.~\ref{eq:Hamzrot}).
To calibrate the drive phase, we identify the value of $\arg(\epsz)$ that maximizes the Rabi rate. Following a procedure similar to Ref.~\cite{Grimm2020}, we ramp on the squeezing drive from vacuum and then apply a drive with variable phase $\arg(\epsz)$ and pulse duration $\dt$. We then apply an $X(\pi/2)$-gate and a \CQR. The pulse sequence and the measurement results are shown in Figs.~\ref{fig:SI_KCQ_rabi}a and b, respectively.
We select the phase that maximizes the Rabi rate, which is $\arg(\epsz) = 0$. We then perform a Rabi experiment at this optimal phase using a higher drive amplitude. The pulse sequence and the measured Rabi oscillations are shown in Figs.~\ref{fig:SI_KCQ_rabi}c and d, respectively. Colored triangular markers denote pulse durations corresponding to $R_{\mathrm{Z}}(0)$ (green, right-pointing), $R_{\mathrm{Z}}(\pi/2)$ (orange, down-pointing), $R_{\mathrm{Z}}(\pi)$ (blue, left-pointing) and $R_{\mathrm{Z}}(3\pi/2)$ (red, up-pointing) rotations.

To quantify the $Z$-gate error $\bitflipprob$, we follow the method described in Ref.~\cite{Rousseau25}.
We fit the Rabi oscillations with a decaying sinusoidal function, extracting a decay rate of $\gammarabikq = \gammarabival$ 
and an oscillation frequency of $\omegaz = \omegazval$. 
We then compute the $\bitflipprob$ associated with the maximum gate time $\tauzgate \approx \SI{100}{\nano\second}$, 
\begin{equation}
\bitflipprob = \frac{1 - e^{-\gammarabikq \tauzgate}}{2} = (1.67 \pm 0.26) \%.
\end{equation}

To estimate the fidelity of the $X(\pi/2)$-gate, we perform a simulation of the system evolution using the master equation
\begin{equation}
\frac{\mathrm{d}\hat{\rho}}{\mathrm{d}t} = -\frac{i}{\hbar}[\Hk, \hat{\rho}] + \kappaoneeff D[\hat{a}]\hat{\rho} + \kappaphieff D[\hat{a}^\dagger \hat{a}]\hat{\rho},
\end{equation}
where $\Hk$ is the Kerr Hamiltonian (Eq.~\eqref{eq:Ham_k}) and $\kappaoneeff$ and $\kappaphieff$ are effective loss and dephasing rates, respectively.
The parameters \( \kappaoneeff/2\pi = \kappaoneeffval \) and \( \kappaphieff/2\pi = \kappaphieffval \) are chosen, respectively, to reproduce the measured $X$-state coherence times at \workingpoint~and to reproduce the dephasing rate corresponding to the nonlinear oscillator \( T_{2} \). Note that \( \kappaoneeff \) differs from $\kappaqubit = 1/\Tone$, which we attribute to pump-induced heating of the oscillator due to the squeezing drive.
To evaluate the gate fidelity, we initialize the system in the state \( \rho(0) = \ket{+Z}\bra{+Z} \) and let it evolve under the Kerr Hamiltonian \( \Hk \) (Eq.~\eqref{eq:Ham_k}) for a duration \( \tauxgate \approx \pi/2K\). We then compare the final state \( \hat{\rho}_{\mathrm{losses}} \), which includes dissipative effects, to the ideal final state \( \hat{\rho}_{\mathrm{ideal}} \), obtained from unitary evolution. The fidelity is calculated as
\begin{equation}
\mathcal{F}_{X(\pi/2)} = \left[ \mathrm{tr}\left(\sqrt{ \sqrt{\hat{\rho}_{\mathrm{ideal}}} \, \hat{\rho}_{\mathrm{losses}} \, \sqrt{\hat{\rho}_{\mathrm{ideal}}} } \right)\right]^2.
\end{equation}
To account for the finite time resolution of the OPX+, which is $\timeresOPX$, we simulate the evolution for both $\tauxgate = \SI{140}{\nano\second}$  and $\SI{144}{\nano\second}$. This yields a range of fidelity values $\mathcal{F}_{X(\pi/2)} = 0.87-0.914$.

\section{Measuring the Leakage Population in the Driven Oscillator}
\label{sec:disp_shift_KCQ}
In this section, we present the key elements required to model and measure the leakage population in the $\ket{\estate}$ manifold of the driven oscillator, $\pe$. We first define the dispersive shift between the cavity and the oscillator eigenstates, and describe an incoherent spectroscopy method used to estimate the leakage population in the $\ket{\fstate}$ manifold, $\pf$. We then show how we model the decoherence processes for the $\ket{\gstate}, \ket{\estate}, \ket{\fstate}$ manifolds, focusing on energy-relaxation ($T_1^{\mathrm{01,12}}$) and pure-dephasing ($T_\phi^{\mathrm{01,12}}$) times. We outline the simulation and fitting procedure for the $\pe$ measurements. Finally we discuss the effect of the \CQR~on the $\pe$ measurement, and we assess the impact of neglecting excitation processes in our decoherence model when extracting $\pe$.

For this analysis, we express the Hamiltonian $\Hkcq$ (Eq.~\eqref{eq:Ham_kcq}) and the system density matrix $\hat{\rho}$ in the driven oscillator eigenbasis, such that
\begin{equation}
\begin{aligned}
    \Hkcq/\hbar &= \sum_{i}\sum_\pm E_i^\pm \, \fixedket{\psi_{i}^{\pm}}{\psi_{i}^{\pm}} \fixedbra{\psi_{i}^{\pm}}{\psi_{i}^{\pm}}, \\
    \hat{\rho} &= \sum_{i,j}\sum_{\alpha,\beta} \hat{\rho}_{ij}^{\alpha\beta} \, \fixedket{\psi_{i}^{\alpha}}{\psi_{i}^{\beta}} \fixedbra{\psi_{j}^{\beta}}{\psikq{j}{\beta}}, \\
    \hat{\rho}_{ij}^{\alpha\beta} &= \fixedbra{\psi_{i}^{\alpha}}{\psi_{j}^{\beta}} \, \hat{\rho} \, \fixedket{\psi_{j}^{\beta}}{\psi_{i}^{\alpha}},
\end{aligned}
\end{equation}
where $\{i,j\} = \{0,1,2,\dots\}$ label different manifolds of the oscillator, and $\{\alpha,\beta\} = \{+,-\}$ denotes the parity of the states.

For the analysis discussed here, we assume that $\hat{\rho}_{ii}^{+-}(0) = \hat{\rho}_{ii}^{-+}(0) = 0$. This condition applies to all experiments discussed in this section and in Figs.~\figtwo~and \figthree~of the main text, where we wait $\tauwait \gg \Tone$ to initialize the system in the steady state. During this time period, single-photon loss processes scramble the parity information within each manifold, generating an incoherent mixture of the states $\psii$ in each manifold $i$.

\subsection{Dispersive-shift readout}

\begin{figure*}[b]
    \includegraphics[angle = 0, width = \figwidthWide]{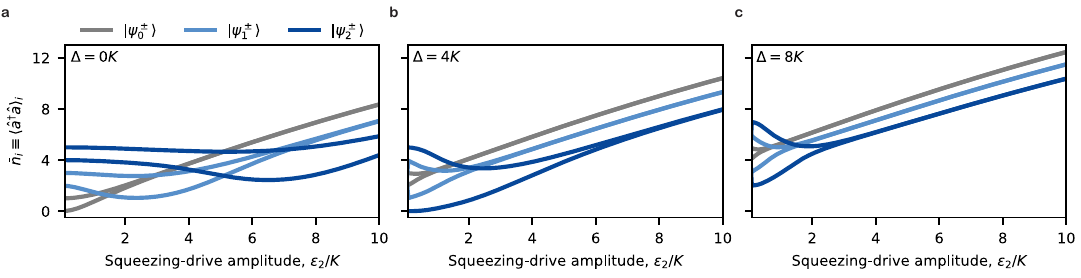}
    \caption{\label{fig_SI_avgnbar} \textbf{Average photon number of the driven oscillator states $\psii$.} Average photon number $\nbar_{i}\equiv\bra{\psi_{i}^\pm}\hat{a}^{\dagger}\hat{a}\ket{\psi_{i}^\pm}$ for $\ket{\gstate}$ (gray),  $\ket{\estate}$ (light blue) and  $\ket{\fstate}$ (dark blue). Values are obtained by numerical diagonalization of Eq.~\eqref{eq:Ham_kcqplusstark}, and are plotted as a function of squeezing-drive amplitude $\epstwo$. Results are given for squeezing-drive detuning values of $\Delta=0$ (\textbf{a}), $4K$ (\textbf{b}), and $8K$ (\textbf{c}).}
\end{figure*}

In this subsection, we describe in detail the photon-number-dependent dispersive-shift readout method used to measure both the leakage population $\pe$ and the coherence times of the $\ket{\gstate},\ket{\estate},\ket{\fstate}$ manifolds. This readout approach exploits the dependence of the cavity frequency on the average photon number of the driven oscillator eigenstates.

The average photon number in different eigenstates of the driven oscillator is illustrated in Fig.~\ref{fig_SI_avgnbar}. Results are obtained by numerical diagonalization of the Hamiltonian
\begin{equation} 
\label{eq:Ham_kcqplusstark}
\hat{H} = \Hkcq + \Hstark,
\end{equation}
which includes the \KQ~Hamiltonian $\Hkcq$ (Eq.~\eqref{eq:Ham_kcq}) and a term accounting for the Stark shift arising from the squeezing-drive pump, $\Hstark$ (Eq.~\eqref{eq:Ham_stark} with $\xicqr{eff}=0$). The resulting average photon numbers in the manifold $i$, $\nbar_{i}$, are plotted as a function of squeezing-drive amplitude $\epstwo$, for three values of squeezing-drive detuning: $\Delta=0$ (Fig.~\ref{fig_SI_avgnbar}a), $\Delta=4K$ (Fig.~\ref{fig_SI_avgnbar}b), and \workingdelta~(Fig.~\ref{fig_SI_avgnbar}c). The latter corresponds to the parameter regime of Figs.~\figtwo~,~\figthree~and~\figfive~of the main text. 

It can be seen from Fig.~\ref{fig_SI_avgnbar} that different manifolds generally have different average photon numbers $\nbar_{i}$. For increasing $\epstwo$, the average photon number of different states within a given manifold converges to a common value. This is even more evident for larger values of $\Delta$. We can understand this behavior from the energy spectrum of the driven oscillator. For increasing $\epstwo$, excited states become increasingly localized in the double-well quasienergy potential~\cite{frattini_observation_2024}, resulting in an exponentially-decreasing energy splitting within a manifold. This is correlated with the convergence in the average photon number within the same manifold. The squeezing-drive detuning also deepens the double-well potential~\cite{venkatraman_driven_2024}, leading to a similar effect. When states in the $\ket{\estate}$ manifold are strongly localized (see for example $\epstwo>4K$ for $\Delta = 4K$, Fig.~\ref{fig_SI_avgnbar}b), their average photon number is smaller compared with states in the $\ket{\gstate}$ manifold. This can be intuitively understood as a smaller phase-space displacement of the localized wavefunction compared to the \KQ-computation-manifold states, since the former is closer to the edge of the confined region of the double-well potential (see Fig.~\figone~of the main text).

\subsection{Incoherent spectroscopy of the driven oscillator}
\label{sec:incoherent_spec}

In this subsection, we independently estimate the populations $\pg$, $\pe$, and $\pf$ of the driven oscillator using incoherent spectroscopy. An accurate estimate of $\pf$ is useful to refine the result of the self-calibrated Rabi protocol used to extract $\pe$ (Figs.~\figtwo~and~\figthree~of the main text). This protocol is based on selectively exciting transitions between manifolds in the driven oscillator, resulting in a change in cavity response that is proportional to the population in these manifolds. 

\begin{figure*}[b]
    \includegraphics[angle = 0, width = \figwidthWide]{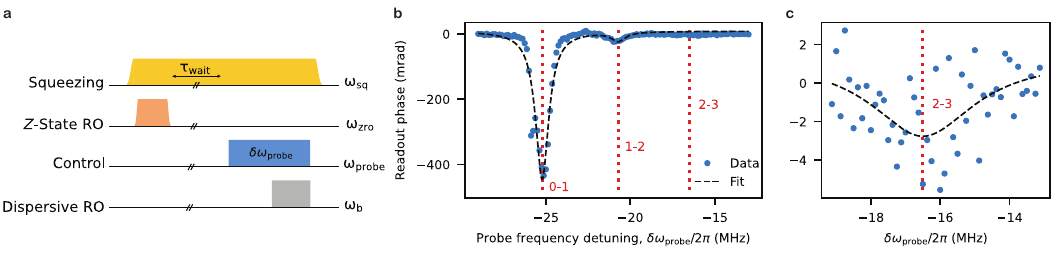}
    \caption{\label{fig_SI_spec_cavro} \textbf{Incoherent spectroscopy measurement. a}, Pulse sequence to perform the spectroscopy measurement.
    \textbf{b,c}, Readout-cavity phase response as a function of the probe-tone frequency detuning $\delta\omegaprobe$. Single-photon transitions between the $\ket{\gstate}, \ket{\estate}, \ket{\fstate}, \ket{\hstate}$ manifolds are indicated by red vertical dotted lines. Blue dots represent the measured data, and black dashed lines show independent Lorentzian fits for each panel. The phase response is offset to remove the signal background. Measurements are performed for \workingpoint.}
    \end{figure*}

The pulse sequence for the measurement is shown in Fig.~\ref{fig_SI_spec_cavro}a. We initialize the system  using the same protocol as outlined in Fig.~\figtwo~of the main text: we adiabatically ramp up the squeezing-drive amplitude and frequency detuning, apply a \CQR, and wait for a time $\tauwait=\SI{300}{\micro\second}$ to allow the system to reach the steady state. We then apply a weak probe tone with varying frequency $\omegaprobe=\omegasq/2+\delta\omegaprobe$ to the oscillator for $\SI{300}{\micro\second}$, followed by a readout pulse on the cavity to perform dispersive readout. The probe tone remains on during the readout to prevent relaxation during measurement.

The result of the measurement is shown in Figs.~\ref{fig_SI_spec_cavro}b and c, as a function of probe-tone frequency detuning $\delta\omegaprobe$. We offset the readout signal to remove the background. The dips in spectroscopy correspond to the transitions in the driven oscillator spectrum, with the expected transition frequencies indicated by red dotted lines. When the probe tone is resonant with a transition ($\delta\omegaprobe=\omega_{ij}$), it drives coherent oscillations between corresponding manifolds $\ket{\psi_{i}^\pm}$ and $\ket{\psi_{j}^\mp}$. For drive times which are long compared to $\Tonekq{ij}$ and $ \Tphikq{ij}$, these oscillations decay into an incoherent mixture with equal populations in both manifolds, $(p_i + p_j)/2$. This leads to a change in the cavity frequency due to the dispersive-shift interaction  $\chidisp$, and therefore to a change in the response to the readout pulse.

The method to estimate $\pg$,  $\pe$,  $ \pf$ is as follows. 
We associate the population $p_i$ in the $\psii$ manifold with the corresponding cavity readout signal $\rdct{i}$. To estimate $\rdct{i}$, we define the cavity reflection response when all the population is in the $\psii$ manifold ($p_i = 1$), $S_{11}^i(\omega)$, where $\omega$ denotes the readout frequency. The reflection response is computed using the independently measured $\kappa_{\mathrm{b,out}}$ and $\kappa_{\mathrm{b,l}}$, and includes the dispersive shift of the cavity induced by population in the $\psii$ manifold (see Section IV.A for details). We then define the readout signal as
\begin{equation}
    \rdct{i} = \arg\left[S_{11}^i(\omega_\mathrm{M})\right],
\end{equation}
where $\omega_\mathrm{M}$ corresponds to the cavity resonance frequency when $\pg = 1$. We assume that the total population is confined to the $\ket{\gstate}$,  $\ket{\estate}$, and  $\ket{\fstate}$~manifolds, such that
\begin{equation}
    \pg + \pe + \pf = 1.
    \label{eq:popsum}
\end{equation}
Moreover, we suppose that the steady-state populations are unaffected when the probe drive is off-resonant with any transition between manifolds, $\delta\omegaprobe\neq\omega_{ij}$. Under these conditions, the cavity readout signal $M$ for an off-resonant probe tone is given by
\begin{equation}
M(\delta\omegaprobe\neq\omega_{ij}) = \sum_i p_i\rdct{i} = \pg\rdct{0} + \pe\rdct{1} + \pf\rdct{2}.
\end{equation}
When the probe tone is resonant with the $\transge$ transition, we have
\begin{equation}
M(\delta\omegaprobe=\omegage) = \frac{(\pg+\pe)}{2}(\rdct{0}+\rdct{1}) + \pf\rdct{2},
\end{equation}
and similarly for the $\transef$ transition
\begin{equation}
M(\delta\omegaprobe=\omegaef) = \pg\rdct{0} + \frac{(\pe+\pf)}{2}(\rdct{1}+\rdct{2})
\end{equation}
and $\transfh$ transition
\begin{equation}
M(\delta\omegaprobe=\omega_{23}) = \pg\rdct{0} + \pe\rdct{1} + \frac{\pf}{2}(\rdct{2}+\rdct{3}).
\end{equation}
We then look at the readout contrast shift relative to the background value
\begin{equation}
\relpeak{ij} \equiv M(\delta\omegaprobe=\omega_{ij})-M(\delta\omegaprobe\neq\omega_{ij}),
\end{equation}
and normalize the relative contrast changes $\relpeak{12}$, $\relpeak{23}$ by $\relpeak{01}$ to obtain
\begin{align}
\frac{\relpeak{12}}{\relpeak{01}} &= \frac{\pe-\pf}{\pg-\pe}\relnbar{12} ,\label{eq:relpeak_ef} \\
\frac{\relpeak{23}}{\relpeak{01}}  &= \frac{\pf}{\pg-\pe}\relnbar{23} \label{eq:relpeak_fh},
\end{align}
with $\eta_{ij}=(\rdctitalic{j}-\rdctitalic{i})/(\rdct{1}-\rdct{0})$. 
By fitting the data in Figs.~\ref{fig_SI_spec_cavro}b~and~c with a Lorentzian function, we extract 
\begin{align}
    \relpeak{01} = (-453.7 \pm 14)\, \mathrm{mrad}, \quad \relpeak{12} = (-28.0 \pm 1.37)\,\mathrm{mrad}, \quad  \relpeak{23} = (-4.1 \pm 1.6)\,\mathrm{mrad} .
\end{align} 
We then solve Eqs.~\eqref{eq:popsum}, \eqref{eq:relpeak_ef}, and~\eqref{eq:relpeak_fh} to determine the steady-state populations $\pg$, $\pe$, and $\pf$. To account for uncertainties in the fitted data, we propagate the errors from $\relpeak{ij}$ using a Monte Carlo method~\cite{Zhang2020,Possolo2017}. This yields
\begin{equation}
    \pg = (90.98 \pm 2.84)\,\%, \quad \pe = (7.69 \pm 2.79)\,\%, \quad \pf = (1.33 \pm 0.52)\,\%.
\end{equation}
This extracted value of $\pe$ is consistent, within uncertainty, with the value obtained from Rabi-contrast measurements (see Fig.~\figtwo~of the main text). 
Note that the value of $\pf$ has significant uncertainty because the measurement noise is comparable in magnitude to $\relpeak{23}$ (see Fig.~\ref{fig_SI_spec_cavro}c).

\subsection{Rabi oscillations on $\transge$ transition}

Here, we present Rabi oscillations on the $\transge$ transition in the \KQ\ corresponding to the data presented in Fig.~\figtwo\ of the main text (\workingpoint), and discuss how decoherence during a $\pige$-pulse affects the extraction of $\pe$ in a Rabi-contrast protocol (see Fig.~\figtwo\ of the main text). 

\begin{figure}
    \centering
    \includegraphics[width=0.5\linewidth]{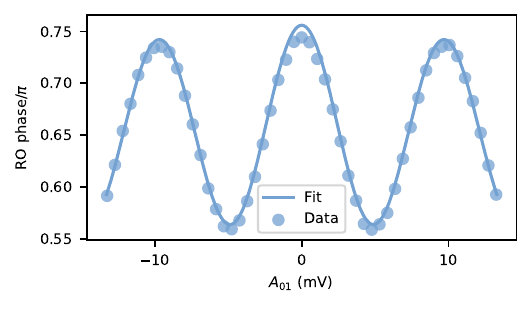}
    \caption{\textbf{Rabi oscillations for $\transge$ transition.} Phase of reflected readout signal as a function of pulse amplitude on the $\transge$ transition $A_{01}$, in units of the arbitrary waveform generator output voltage. Circles are measured data, and the solid line is a fit using a Lindblad master equation simulation (see text and Sec.~\ref{sec:p1_fitting}). Measurements were performed for $\epstwo=\epstwofigtwo K$ and $\Delta=\Deltafigtwo K$, and were interleaved with those presented in Fig.~\figtwo\ of the main text. The fit was performed with $\pe$ and the overall readout contrast values as fit parameters, together with the data in Figs.~\figtwo c and~\figthree b of the main text.}
    \label{fig:si_rabizeroone}
\end{figure}

Figure~\ref{fig:si_rabizeroone} shows the phase of the reflected readout signal as a function of pulse amplitude on the $\transge$ transition, $A_{01}$. The signal oscillates between the values corresponding to $\ket{\gstate}$ and $\ket{\estate}$, but does not return to the value at $A_{01}=0$ after one oscillation. As discussed in the main text, this is due to decoherence acting during the $\transge$ pulse. At $A_{01}=0$, decoherence does not affect the signal because the system stays in its initial state. However, for finite pulse amplitude ($A_{01}\neq0$), decoherence during the finite pulse length changes the state trajectory such that the final state is not fully pure, corresponding to a reduction in signal. This effect is present in both the $\transge$ and $\transef$ Rabi oscillations (see Fig.~\figtwo c of the main text), but is more pronounced in the $\transef$ Rabi oscillations because the decoherence is larger on these states. To account for decoherence effects, we fit the data with a Lindblad master equation simulation that accounts for the measured decoherence between the \KQ\ manifolds (see Sec.~\ref{sec:p1_fitting}). This fit takes $\pe$ and the overall readout contrast values as global fit parameters to the data in Figs.~\figtwo c and~\figthree b. 

Now, we examine how decoherence during the $\pige$-pulse affects the extraction of $\pe$. From the above discussion, a $\pige$ pulse does not fully swap population between $\ket{\gstate}$ and $\ket{\estate}$ due to decoherence during the pulse. The calculation of $\pe$ using the ratio of oscillation amplitudes assumes that the $\pige$ pulse fully swaps the population in $\ket{\gstate}$ and $\ket{\estate}$. If some population is not swapped by the $\pige$ pulse, this rescales the population participating in the reference Rabi oscillations (with $\pige$) and therefore introduces a systematic bias in the extraction of $\pe$. For this reason, we extract $\pe$ using a Lindblad master equation simulation that accounts for decoherence effects (Sec.~\ref{sec:p1_fitting}), rather than directly from the ratio of oscillation amplitudes.

\subsection{Rotating frame transformation in the Kerr-cat eigenbasis}

To simplify the analysis in the following subsections, we introduce a unitary rotating frame transformation defined by the eigenstates of the \KQ~Hamiltonian (Eq.~\eqref{eq:Ham_kcq}). We define
\begin{equation}
    U = \exp\left( -i  \omegarf t \sum_{n} \sum_\pm n \, \fixedket{\psi_n^\pm}{\psi_n^\pm} \fixedbra{\psi_n^\pm}{\psi_n^\pm} \right)
    \label{eq:rotframe}
\end{equation}
where the summations $\sum_{n} \sum_\pm$ run over the manifold indices and parity sectors, respectively.
The Hamiltonian in this new frame becomes
\begin{equation}
    \Hkcq^1 = U\Hkcq U^\dagger + i\hbar \frac{\mathrm{d}U}{\mathrm{d}t} U^\dagger.
\end{equation}
Since $\Hkcq$ and $U$ are both diagonal in the \KQ~eigenbasis, they commute, and hence we can write
\begin{equation}
    U \Hkcq U^\dagger = \Hkcq.
\end{equation}
The time derivative of the unitary operator $U$ gives
\begin{equation}
    i \hbar \frac{\mathrm{d}U}{\mathrm{d}t} U^\dagger = \hbar \omegarf \sum_{\pm} \sum_n n \, \fixedket{\psi_n^\pm}{\psi_n^\pm} \fixedbra{\psi_n^\pm}{\psi_n^\pm}.
\end{equation}
In this rotating frame, the Hamiltonian takes the form
\begin{equation}
    \Hkcq^1 = \sum_{\pm} \sum_n \left( E_n^\pm + n\hbar \omegarf \right) \fixedket{\psi_n^\pm}{\psi_n^\pm} \fixedbra{\psi_n^\pm}{\psi_n^\pm}
\end{equation}
where the effect of the unitary transformation is to shift the energy of each eigenstate $\ket{\psi_n^\pm}$ by an amount $ n\hbar \omegarf $.

\subsection{$\Tonekq{01}$ and $\Tphikq{01}$ for the $\transge$ transition}

In this subsection, we model the decay times $\Tonekq{01}$ and $\Tphikq{01}$ associated with the $\transge$ transition. We restrict the analysis to the subspace spanned by the $\ket{\gstate}$ and $\ket{\estate}$ states.

We make the approximation
$$
\omegakq{0,1}{+}- \omegakq{0,1}{-} \approx 0, \quad \text{with} \quad E_n^\pm = \hbar \omega_{n}^{\pm}.
$$
Note that the pulses we use to excite transitions between manifolds have a linewidth of $\SI{480}{\kilo\hertz}$ and the engineered dissipation has linewidth of $\kappacav/2\pi = \kappacavval$~\citeMethods. Hence, neither the excitation pulses nor the dissipation process can distinguish energy splittings smaller than $\approx \SI{500}{\kilo \hertz}$. This condition is satisfied for the \workingpoint (Figs. \figtwo, \figthree, \figfive), most of parameter space of Fig. \figfour e, and in general for desirable \KQ~operating points. 
Thus, the energy levels within each manifold, $\ket{\gstate}$ and $\ket{\estate}$, can be treated as effectively degenerate. Note that this assumption is not strictly required and similar analysis can be performed for  $\omegakq{0,1}{+}- \omegakq{0,1}{-} \neq 0$.

We now define the Lindblad master equation for this analysis. We begin by applying the unitary transformation $U$ to the \KQ~Hamiltonian $\Hkcq$ (Eq.~\eqref{eq:Ham_kcq}), defined in Eq.~\eqref{eq:rotframe}. We set $\omegarf = -(\omegakq{1}{\pm}-\omegakq{0}{\pm}) +\delta \omega $, with $\delta \omega$ the frequency detuning that we will use in the $\Tphikq{01}$ measurement.
We then introduce two dissipators: a single-photon decay dissipator, 
\begin{equation}
    \kappaonekq{01} D\left(\sum_{\pm} \fixedket{\psikq{0}{\pm}}{} \fixedbra{\psikq{1}{\mp}}{} \right),
    \label{eq:dissipator_one}
\end{equation}
and a dephasing dissipator,
\begin{equation}
    2\kappaphikq{01} D\left(\sum_{\pm} \fixedket{\psikq{1}{\pm}}{} \fixedbra{\psikq{1}{\pm}}{} \right),
    \label{eq:dissipator_phi}
\end{equation}
with $\kappaonekq{01} = 1/\Tonekq{01}$ and $\kappaphikq{01} = 1/\Tphikq{01}$.
These terms respectively describe incoherent decay between the $\ket{\estate}$ and $\ket{\gstate}$ manifolds, and dephasing of the $\ket{\estate}$ manifold.
The system evolution is then described by the Lindblad master equation
\begin{equation}
\frac{\mathrm{d}\hat{\rho}}{\mathrm{d}t} = 
\begin{pmatrix}
\kappaonekq{01}  \rhokq{11}{(2)} & \left( i \delta \omega - \BigGamma{01} \right) \rhokq{01}{(1)} \\
\left( -i \delta \omega - \BigGamma{01} \right) \rhokq{10}{(1)} & -\kappaonekq{01}  \rhokq{11}{(1)}
\end{pmatrix},
\label{eq:lind_twolevel}
\end{equation}
with

\begin{equation}
\hat{\rho}_{ij}^{(1)} = 
\begin{pmatrix}
\hat{\rho}_{ij}^{++} & \hat{\rho}_{ij}^{+-} \\
\hat{\rho}_{ij}^{-+} & \hat{\rho}_{ij}^{--}
\end{pmatrix},
\qquad
\hat{\rho}_{ij}^{(2)} = 
\begin{pmatrix}
\hat{\rho}_{ij}^{--} & \hat{\rho}_{ij}^{-+} \\
\hat{\rho}_{ij}^{+-} & \hat{\rho}_{ij}^{++}
\end{pmatrix}
\end{equation}
and $\BigGamma{01}  = \kappaonekq{01}/2 + \kappaphikq{01}$.

We now model the $\Tonekq{01}$ decay experiment. We consider the initial density matrix
\begin{equation} 
    \hat{\rho}(0) = \frac{1}{2} \left( \sum_{\pm} \fixedket{\psikq{1}{\pm}}{} \fixedbra{\psikq{1}{\pm}}{} \right).
\end{equation}
This initial state corresponds to a fully mixed state within the $\ket{\estate}$ manifold.
Starting from this initial condition, we solve the Lindblad master equation (Eq.~\eqref{eq:lind_twolevel}). The evolution of the density matrix elements is given by
\begin{equation}
\begin{aligned}
 \rhokq{11}{++}(t) &= \frac{1}{2} e^{-\kappaonekq{01} t}, \quad &  \rhokq{00}{--}(t) &= \frac{1}{2} -  \rhokq{11}{++}(t), \\
 \rhokq{11}{--}(t) &= \frac{1}{2} e^{-\kappaonekq{01} t}, \quad &  \rhokq{00}{++}(t) &= \frac{1}{2} -  \rhokq{11}{--}(t), \\
 \hat{\rho}_{xy}^{zf}(t) &= 0,
\end{aligned}
\end{equation}
where \( x, y \in \{ \mathrm{0}, \mathrm{1}, \mathrm{2}, \dots \} \) and \( z, f \in \{ +, - \} \) denote all other combinations of manifold and parity indices not explicitly listed above.
Here, the populations in the $\ket{\estate}$ manifold decay exponentially at a rate $\kappa_1$, while populations in $\ket{\gstate}$ increase correspondingly. All off-diagonal coherence terms remain zero throughout the evolution, reflecting the absence of initial coherence.
The readout signal \( \rdsgkq{\Tonekq{01}} (t) \) evolves as
\begin{equation}
\rdsgkq{\Tonekq{01}} (t) = \rdckq{1} e^{-\kappaonekq{01} t} + \rdckq{0} \left( 1 - e^{-\kappaonekq{01} t} \right),
\end{equation}
where \( \rdckq{1}  \) and \( \rdckq{0}  \) denote the readout response for population in the $\ket{\estate}$ and $\ket{\gstate}$ manifolds, respectively. As discussed in the main text and in Section IV.A, we measure the system using the dispersive shift between the cavity and the oscillator, meaning \( \rdckq{i}  \) is proportional to the average photon number in the corresponding manifold $i$. Explicitly,
\begin{equation}
\rdckq{1}  \propto \fixedbra{\psikq{1}{+}}{} \hat{a}^\dagger \hat{a} \fixedket{\psikq{1}{+}}{} \approx \fixedbra{\psikq{1}{-}}{} \hat{a}^\dagger \hat{a} \fixedket{\psikq{1}{-}}{},
\end{equation}
\begin{equation}
\rdckq{0}  \propto \fixedbra{\psikq{0}{+}}{} \hat{a}^\dagger \hat{a} \fixedket{\psikq{0}{+}}{} \approx \fixedbra{\psikq{0}{-}}{} \hat{a}^\dagger \hat{a} \fixedket{\psikq{0}{-}}{}.
\end{equation}

We now model a Ramsey experiment used to extract the dephasing time \( \Tphikq{01} \). The system is initialized in a statistical mixture of two orthogonal superposition states
\begin{equation}
\hat{\rho}(0) = \frac{1}{2} \left( \fixedket{\phi^\mathrm{A}_{\mathrm{01}}}{} \fixedbra{\phi^\mathrm{A}_{\mathrm{01}}}{} + \fixedket{\phi^\mathrm{B}_{\mathrm{01}}}{} \fixedbra{\phi^\mathrm{B}_{01}}{} \right),
\end{equation}
where
\begin{equation}
\fixedket{\phi^\mathrm{A}_{\mathrm{01}}}{} = \frac{1}{\sqrt{2}} \left( \fixedket{\psikq{0}{+}}{} + \fixedket{\psikq{1}{-}}{} \right),
\end{equation}
\begin{equation}
\fixedket{\phi^\mathrm{B}_{\mathrm{01}}}{} = \frac{1}{\sqrt{2}} \left( \fixedket{\psikq{0}{-}}{} + \fixedket{\psikq{1}{+}}{} \right).
\end{equation}
These states represent a coherent superposition between the $\ket{\gstate}$ and $\ket{\estate}$ manifolds with opposite parity. 
The solution of the Lindblad master equation (Eq. \eqref{eq:lind_twolevel}) for this initial condition is
\begin{equation}
\begin{aligned}
 \rhokq{11}{++}(t) &= \frac{1}{4} e^{-\kappaonekq{01} t}, \quad &  \rhokq{00}{++}(t) &= \frac{1}{2} -  \rhokq{11}{--}(t), \\
 \rhokq{11}{--}(t) &= \frac{1}{4} e^{-\kappaonekq{01} t}, \quad &  \rhokq{00}{--}(t) &= \frac{1}{2} -  \rhokq{11}{++}(t), \\
 \rhokq{01}{+-}(t) &= \rhokq{01}{-+}(t) = \frac{1}{4} e^{\left(i \delta \omega - \BigGamma{01}\right)t},   \quad &
 \rhokq{10}{+-}(t) &= \rhokq{10}{-+}(t) = \frac{1}{4} e^{\left(-i \delta \omega - \BigGamma{01}\right)t}, \\
 \hat{\rho}_{xy}^{zf}(t) &= 0,
\end{aligned}
\end{equation}
where \( x, y \in \{ 0,1,2, \dots \} \) and \( z, f \in \{ +, - \} \) denote all other combinations of manifold and parity indices not explicitly listed above.
Here, the diagonal terms describe the population decay from $\ket{\estate}$ to $\ket{\gstate}$ at rate \(\kappaonekq{01}\), while the off-diagonal terms encode the Ramsey fringes, oscillating at frequency \(\delta\omega\) and decaying with the total decoherence rate \(\BigGamma{01}=\kappaonekq{01}/2 + \kappaphikq{01}\). 
At the end of the Ramsey sequence, a final $\pige/2$-pulse is applied on the \(\transge\) transition before performing readout in the photon-number basis. 
We model the $\pige/2$-rotation as
\[
R^\mathrm{01}\left( \frac{\pi}{2} \right) = \frac{1}{\sqrt{2}}
\begin{pmatrix}
1 & 0 & 0 & -1 \\
0 & 1 & -1 & 0 \\
0 & 1 & 1 & 0 \\
1 & 0 & 0 & 1
\end{pmatrix}.
\]
We then compute the readout signal as
\begin{equation}
\rdsgkq{\Tphikq{01}} (t) = \frac{1}{2}(\rdckq{1}  + \rdckq{0} ) + \frac{1}{2}(\rdckq{1}  - \rdckq{0} ) \cos(\delta \omega\, t) \, e^{-\BigGamma{01}t}.
\end{equation}
The oscillatory component reflects the coherent evolution between manifolds, while the exponential decay captures both energy relaxation and dephasing.

\subsection{$\Tonekq{12}$ and $\Tphikq{12}$ for the $\transef$ transition}
In this subsection, we model the decay times $\Tonekq{12}$ and $\Tphikq{12}$ associated with the $\transef$ transition. We restrict the analysis to the  $\ket{\gstate},\ket{\estate}$ and $\ket{\fstate}$ manifolds.
Following the approach used in the previous subsection, we assume that the frequency splitting between the $\pm$ states is negligible, that is
\begin{equation}
\omegakq{0,1,2}{+}- \omegakq{0,1,2}{-} \approx 0.
\end{equation}

We now define the Lindblad master equation for the analysis in this subspace. We begin by applying the unitary transformation $U$ defined in Eq.~\eqref{eq:rotframe} to the \KQ~Hamiltonian $\Hkcq$ (Eq.~\eqref{eq:Ham_kcq}), . We set $\omegarf = -(\omegakq{2}{\pm} - \omegakq{1}{\pm}) + \delta \omega$ and, in addition to the dissipators defined in Eqs.~\eqref{eq:dissipator_one} and~\eqref{eq:dissipator_phi}, we introduce dissipators to model the decoherence processes associated with the $\transef$ transition
\begin{equation}
\kappaonekq{12} D\left(\sum_{\pm} \fixedket{\psikq{1}{\pm}}{\psikq{2}{\mp}} \fixedbra{\psikq{2}{\mp}}{} \right),
\quad \quad
2\kappaphikq{12} D\left(\sum_{\pm} \fixedket{\psikq{2}{\pm}}{} \fixedbra{\psikq{2}{\pm}}{} \right),
\end{equation}
where \(\kappaonekq{12} = 1/\Tonekq{12}\) and \(\kappaphikq{12} = 1/\Tphikq{12}\) are the energy relaxation rate from $\ket{\fstate}$ to $\ket{\estate}$ and the pure dephasing rate of the $\ket{\fstate}$ manifold, respectively. 
We then compute the full Lindblad master equation, incorporating dissipative terms for the $\ket{\gstate}$, $\ket{\estate}$ and $\ket{\fstate}$ manifolds, which gives

\begin{equation}
\frac{\mathrm{d}\hat{\rho}}{\mathrm{d}t} =
\begin{pmatrix}
\kappaonekq{01} \rhokq{11}{(2)} & (i\delta_1 - \BigGamma{01})\rhokq{01}{(1)} & (i\delta_2 - \BigGamma{12}) \rhokq{02}{(1)} \\
(-i\delta_1 - \BigGamma{01}) \rhokq{10}{(1)} & -\kappaonekq{01} \rhokq{11}{(1)} + \kappaonekq{12} \rhokq{22}{(2)} & (i\delta\omega - \BigGamma{01} - \BigGamma{12}) \rhokq{12}{(1)} \\
(-i\delta_2 - \BigGamma{12}) \rhokq{20}{(1)} &  (-i\delta\omega - \BigGamma{01} - \BigGamma{12}) \rhokq{21}{(1)} & -\kappaonekq{12} \rhokq{22}{(1)}
\end{pmatrix}
\label{eq:lind_threelevel}
\end{equation}
with $\BigGamma{12}  = \kappaonekq{12}/2 + \kappaphikq{12}, \quad
\delta_1 = \delta \omega + 2\omegakq{1}{\pm}  - \omegakq{2}{\pm}  - \omegakq{0}{\pm}, \quad
\delta_2 = 2 \delta \omega + 2\omegakq{1}{\pm}  - \omegakq{2}{\pm} - \omegakq{0}{\pm}  .$

We now model the $\Tonekq{12}$ experiment. The system is prepared in a fully mixed state within the $\ket{\fstate}$ manifold
\begin{equation}
\hat{\rho}(0) = \frac{1}{2} \left( \sum_{\pm} \fixedket{\psikq{2}{\pm}}{} \fixedbra{\psikq{2}{\pm}}{} \right).
\end{equation}
Due to the structure of both the Hamiltonian and the dissipators, the dynamics of the system separates into two independent subspaces
\begin{align}
S_1 &= \left\{ \fixedket{\psikq{0}{+}}{\psikq{2}{+}},\ \fixedket{\psikq{1}{-}}{\psikq{2}{+}},\ \fixedket{\psikq{2}{+}}{} \right\}, \\
S_2 &= \left\{ \fixedket{\psikq{0}{-}}{\psikq{2}{+}},\ \fixedket{\psikq{1}{+}}{\psikq{2}{+}},\ \fixedket{\psikq{2}{-}}{} \right\}.
\end{align}
This separation allows us to solve the dynamics independently within each subspace, significantly simplifying the analysis.
We therefore solve for the evolution in $S_1$ and $S_2$ separately. For the subspace $S_1$, the time evolution of the relevant density matrix elements is given by
\begin{equation}
\begin{aligned}
 \rhokq{22}{++}(t) &= \frac{1}{2} e^{-\kappaonekq{12} t}, \\
 \rhokq{11}{--}(t) &= \frac{\kappaonekq{12}/2}{\kappaonekq{01} - \kappaonekq{12}} \left( e^{-\kappaonekq{12} t} - e^{-\kappaonekq{01} t} \right), \\
 \rhokq{00}{++}(t) &= \frac{1}{2} -  \rhokq{11}{--}(t) -  \rhokq{22}{++}(t). \\
\end{aligned}
\end{equation}
The dynamics for $S_2$ follow identically, with the relations 
\[
 \rhokq{22}{++}(t) =  \rhokq{22}{--}(t), \quad  \rhokq{11}{++}(t) =  \rhokq{11}{--}(t), \quad  \rhokq{00}{++}(t) =  \rhokq{00}{--}(t).
\]
All other density matrix elements are given by $\hat{\rho}_{xy}^{zf}(t) = 0$, where \( x, y \in \{ \mathrm{0}, \mathrm{1}, \mathrm{2}, \dots \} \) and \( z, f \in \{ +, - \} \) denote all other combinations of manifold and parity indices not explicitly listed above.
After the evolution specified by the solution of the Lindblad master equation (Eq.~\eqref{eq:lind_threelevel}), we apply a $\pige$-pulse to swap the populations of the $\ket{\gstate}$ and $\ket{\estate}$ manifolds. We then perform a dispersive-shift readout. The resulting decay signal $\rdsgkq{\Tonekq{12}} (t)$ is given by
\begin{equation}
\begin{aligned}
\rdsgkq{\Tonekq{12}} (t) &= \rdckq{2} \, e^{-\kappaonekq{12} t} \\
&\quad + \rdckq{0} \, \frac{\kappaonekq{12}}{\kappaonekq{01} - \kappaonekq{12}} \left( e^{-\kappaonekq{12} t} - e^{-\kappaonekq{01} t} \right) \\
&\quad + \rdckq{1}  \left( 1 - e^{-\kappaonekq{12} t} - \frac{\kappaonekq{12}}{\kappaonekq{01} - \kappaonekq{12}} \left( e^{-\kappaonekq{12} t} - e^{-\kappaonekq{01} t} \right) \right),
\end{aligned}
\end{equation}
where
\begin{equation}
\rdckq{2}  \propto \fixedbra{\psikq{2}{+}}{} a^\dagger a \fixedket{\psikq{2}{+}}{} \approx \fixedbra{\psikq{2}{-}}{} a^\dagger a \fixedket{\psikq{2}{-}}{}.
\end{equation}

We now analyze the dynamics of a Ramsey-type interference experiment used to extract $\Tphikq{12}$. The system is prepared in the initial state
\begin{equation}
\hat{\rho}(0) = \frac{1}{2} \left( \fixedket{\phi^{\mathrm{A}}_{\mathrm{12}}}{} \fixedbra{\phi^{\mathrm{A}}_{\mathrm{12}}}{} + \fixedket{\phi^{\mathrm{B}}_{\mathrm{12}}}{} \fixedbra{\phi^{\mathrm{B}}_{\mathrm{12}}}{} \right),
\end{equation}
where
\begin{equation}
\fixedket{\phi^{\mathrm{A}}_{\mathrm{12}}}{} = \frac{1}{\sqrt{2}} \left( \fixedket{\psikq{1}{+}}{} + \fixedket{\psikq{2}{-}}{} \right)
\end{equation}
\begin{equation}
\fixedket{\phi^{\mathrm{B}}_{\mathrm{12}}}{} = \frac{1}{\sqrt{2}} \left( \fixedket{\psikq{1}{-}}{} + \fixedket{\psikq{2}{+}}{} \right).
\end{equation}
The dynamics of the system can be again separated into two independent subspaces, \(S_1\) and \(S_2\). We first solve for the evolution in the \(S_1\) subspace, obtaining the following time-dependent density matrix elements
\begin{equation}
\begin{aligned}
 \rhokq{22}{++}(t) &= \frac{1}{4} e^{-\kappaonekq{12} t}, \quad & 
  \rhokq{12}{-+}(t) &= \frac{1}{4} e^{\left[ -(\BigGamma{01}  + \BigGamma{12} ) + i \delta \omega \right] t},\\
 \rhokq{11}{--}(t) &= \frac{\kappaonekq{12}/4}{\kappaonekq{01} - \kappaonekq{12}} \left( e^{-\kappaonekq{12} t} - e^{-\kappaonekq{01} t} \right), \quad & 
\rhokq{21}{+-}(t) &= \frac{1}{4} e^{\left[ -(\BigGamma{01}  + \BigGamma{12} ) - i \delta \omega \right] t},\\
\rhokq{00}{++}(t) &= \frac{1}{2} -  \rhokq{11}{--}(t) -  \rhokq{22}{++}(t).
\end{aligned}
\end{equation}
The dynamics in the \(S_2\) subspace are identical, and the corresponding density matrix elements satisfy the following relations
$$
 \rhokq{22}{--}(t) =  \rhokq{22}{++}(t), \quad
 \rhokq{11}{++}(t) =  \rhokq{11}{--}(t), \quad
 \rhokq{00}{--}(t) =  \rhokq{00}{++}(t), \quad
\rhokq{12}{+-}(t) = \rhokq{12}{-+}(t), \quad
\rhokq{21}{-+}(t) = \rhokq{21}{+-}(t).
$$
All other density matrix elements are given by $\hat{\rho}_{xy}^{zf}(t) = 0$, where \( x, y \in \{ \mathrm{0}, \mathrm{1}, \mathrm{2}, \dots \} \) and \( z, f \in \{ +, - \} \) denote all other combinations of manifold and parity indices not explicitly listed above.
These equations describe the free evolution of the system. In the experimental sequence described in the main text, after the free evolution we apply a $\pief/2$-pulse on the $\transef$ transition, followed by a $\pige$-pulse on the $\transge$ transition, and then by a dispersive-shift readout.
The $\pief/2$-pulse on the $\transef$ transition is modeled by the unitary transformation
\[
R^\mathrm{12}\left( \frac{\pi}{2} \right) = \frac{1}{\sqrt{2}}
\begin{pmatrix}
\sqrt{2} & 0 & 0 & 0 & 0 & 0 \\
0 & \sqrt{2} & 0 & 0 & 0 & 0 \\
0 & 0 & 1 & 0 & 0 & -1 \\
0 & 0 & 0 & 1 & -1 & 0 \\
0 & 0 & 0 & 1 & 1 & 0 \\
0 & 0 & 1 & 0 & 0 & 1
\end{pmatrix}.
\]
After applying these two rotations, the resulting readout signal is given by
\begin{equation}
    \rdsgkq{\Tphikq{12}} (t) = 2 \rdckq{1}   \rhokq{00}{++}(t) + (\rdckq{2}  + \rdckq{0} ) \left[  \rhokq{11}{--}(t) +  \rhokq{22}{++}(t) \right] + \frac{\rdckq{2}  - \rdckq{0} }{2} \cos(\delta \omega t) e^{-(\BigGamma{01}  + \BigGamma{12} ) t}.
\end{equation}

\subsection{Leakage population $\pe$ fitting protocol}
\label{sec:p1_fitting}

In this subsection, we present the method used to extract the leakage population $\pe$ from fits to the Rabi-contrast measurements shown in the main text (see Fig.~\figtwo c). 
We begin by describing the numerical simulation that reproduces the full experimental sequence (see Fig.~\figtwo a). In particular, we explicitly simulate initialization into the steady state of the driven oscillator, coherent control of the manifold populations, and dispersive readout (as described in the main text).
Then, we explain how the simulated traces are fitted to the measured data to extract $\pe$. 
Finally, we discuss how the independently-measured $\pf$, presented in Section~IV.B, is used to refine the fitting of $\pe$.

We first simulate the initialization stage of the measurement protocol. 
In the experimental sequence, we adiabatically ramp the squeezing-drive amplitude and frequency and then wait for a time $\tauwait \gg T_{1}^{ij}, T_{\phi}^{ij}$ to reach the steady state in the driven oscillator. As discussed in the introduction of this section, this long waiting time generates an incoherent mixture of the states $\psii$ within each manifold $i$. Accordingly, in the simulation we define the initial state as
\begin{equation}
    \hat{\rho}(0) = \sum_{\pm}\frac{1}{2} \left( \pf \ket{\fstate}\bra{\fstate} 
    + \pe \ket{\estate}\bra{\estate}
    + (1-\pe-\pf)\ket{\gstate}\bra{\gstate} \right),
    \label{eq:p1_fit_init}
\end{equation}
with populations $\pf$, $\pe$, and $(1-\pe-\pf)$ in the $\ket{\fstate}$, $\ket{\estate}$, and $\ket{\gstate}$ manifolds, respectively.  

We then simulate the coherent-control stage of the $\pe$ measurement. We define the Hamiltonian
\begin{equation}
\hat{H}/\hbar = \Hkcq/\hbar + \Rabikq{01}(t)\left(\hat{a}\,e^{+i\omegage t} + \hat{a}^\dagger e^{-i\omegage t}\right) + \Rabikq{12}(t)\left(\hat{a}\,e^{+i\omegaef t} + \hat{a}^\dagger e^{-i\omegaef t}\right),
\label{eq:p1_Ham}
\end{equation}
where $\Hkcq$ denotes the \KQ~Hamiltonian (Eq.~\eqref{eq:Ham_kcq}), and $\Rabikq{01}(t)$ and $\Rabikq{12}(t)$ are the time-dependent Rabi drive amplitudes for the $\transge$ and $\transef$ transitions, respectively. The drive frequencies $\omegage$ and $\omegaef$ correspond to the transition frequencies between these manifolds, defined in a frame rotating at $\omegasq/2$ (see Fig.~\figone~of main text). 
As discussed in the main text, decoherence processes also influence the system dynamics during the $\pe$ measurement. 
To account for these effects, we simulate the pulse sequence defined in Fig.~\figtwo c, using the Lindblad master equation

\begin{equation}
\frac{\mathrm{d}\hat{\rho}}{\mathrm{d}t} = -\frac{i}{\hbar} [\hat{H},\hat{\rho}] + \sum_{\substack{i=0,1 \\ j=i+1}} \left[2\kappaphikq{ij}D\left(\sum_{\pm} \fixedket{\psikq{j}{\pm}}{} \fixedbra{\psikq{j}{\pm}}{} \right) +  \kappaonekq{ij} D\left(\sum_{\pm} \fixedket{\psikq{i}{\pm}}{} \fixedbra{\psikq{j}{\mp}}{} \right) \right]. 
\label{eq:lind_epopn}
\end{equation}
Here, $\kappa_{1}^{ij} = 1/T_{1}^{ij}$ and $\kappa_{\phi}^{ij} = 1/T_{\phi}^{ij}$ are the energy-relaxation and pure-dephasing rates for transitions between manifolds $i$ and $j$, respectively. 
Starting from $\hat{\rho}(0)$, we evolve the system according to the Lindblad master equation, yielding a final state $\hat{\rho}_{\mathrm{f}}$.

Finally, we  simulate the dispersive readout sequence, by defining the measurement observable
\begin{equation}
    \hat{O} = \sum_{\pm} \left( \rdckq{0} \ket{\gstate}\bra{\gstate} 
    + \rdckq{1} \ket{\estate}\bra{\estate} 
    + \rdckq{2}\ket{\fstate}\bra{\fstate} \right),
\end{equation}
where $\rdckq{i}$ denotes the readout amplitude associated with $\psii$. 
We compute the expectation value $\langle \hat{O}\rangle = \mathrm{tr}[\hat{O}\rhof]$, which returns a simulated Rabi-contrast trace (see Fig.~\figtwo c) for a given set of parameters.

We now describe the fixed parameters of the fit, and the methods used to extract them.
All parameters of $\Hkcq/\hbar$ are independently measured (see Section~II). To calibrate the Rabi drive amplitudes we first model them as $\Omega_{ij}(t) = V_{ij} \, \xi(t)$, where \(\xi(t)\) is the experimental pulse shape with amplitude normalized to unity (see pulse sequence in Fig.~\figtwo c), and \(V_{ij}\) are scaling factors. We then fit amplitude Rabi experiments on the $\transge$ and $\transef$ transitions and obtain the amplitudes \(V_{\mathrm{01}}\) and \(V_{\mathrm{12}}\).
To extract the decoherence times $T_{1}^{ij}$ and $T_{\phi}^{ij}$, we first use the relaxation (Fig.~\figtwo e) and Ramsey interference (Fig.~\figtwo f) measurements on the $\transge$ transition to fit $T_{1}^{01}$ and $T_{\phi}^{01}$. 
These rates are then fixed, and the relaxation (Fig.~\figtwo g) and Ramsey interference (Fig.~\figtwo h) measurements on the $\transef$ transition are used to fit $T_{1}^{12}$ and $T_{\phi}^{12}$.

With these parameters fixed, we fit the simulated Rabi-contrast traces to the experimental data and extract $\pe$ as a function of $\gdiss$. The fit uses a single set of global parameters $\rdckq{0}$, $\rdckq{1}$, and $\rdckq{2}$ for all datasets, while $\pe$ is fitted independently for each value of $\gdiss$. 
We repeat this procedure for two values of the population of the $\ket{\fstate}$ manifold, $\pf$. The results of the fitting procedure for $\pf=0$ are shown by the black diamonds in Fig.~\figthree b of the main text. In Section~IV.B we independently determined a population $\pf = (1.33 \pm 0.52)\%$ from incoherent spectroscopy at $\gdiss = 0$.
To account for this, we repeat the fitting procedure for this value of $\pf$, shown as blue dots in Fig.~\figthree b of the main text.
Uncertainty in $\pf$ is propagated to $\pe$ using a Monte Carlo approach~\cite{Zhang2020,Possolo2017}. We first sample $\pf$ from a Gaussian distribution with a mean and standard deviation set by the experimental value. We then perform the fit procedure for $\pe$ using this sampled $\pf$ value, yielding a $\pe$ value and a statistical fit error $\sigma_{\pe|\pf}$. Repeating the fit procedure $N = 270 \times 10^3$ times yields a distribution of fitted $\pe$ values, each with an associated $\sigma_{\pe|\pf}$. 
The mean of this distribution defines the expected $\pe$ value. To evaluate the uncertainty in $\pe$ we use the law of total variance~\cite{Champ2024}
\begin{equation}
    \sigma^2_{\pe} = E\!\left[\sigma^2_{\pe|\pf}\right] + \mathrm{Var}\!\left(\pe|\pf\right),
\end{equation}
where $E[\sigma^2_{\pe|\pf}]$ is the average fit error and $\mathrm{Var}(\pe|\pf)$ is the variance of the distribution of fitted $\pe$ values. This gives an uncertainty $\sigma_{\pe}$ associated with the expected $\pe$ value.

\subsection{Effect of $Z$-state readout on the measurement of $\pe$} 
In this subsection, we discuss the role of the \CQR~in the protocol used to measure the leakage population $\pe$, as presented in Figs.~\figtwo~and~\figthree~of the main text. The \CQR~is applied during the initialization stage of the sequence where we include it to ensure consistency with the $\Tz$ measurements shown in Fig.~\figfive~of the main text.

The \CQR~projects the \KQ~into $\ket{\pm Z}$. 
Since we do not condition the subsequent pulse sequence on this measurement outcome, the ensemble-averaged density matrix of the system after the \CQR~is an incoherent mixture of the two projected states $\ket{\pm Z}$,
\begin{equation}
\rho = \frac{1}{2} \left( \ket{+Z}\bra{+Z} + \ket{-Z}\bra{-Z} \right) = \frac{1}{2} \left( \ket{\psikq{0}{+}}\bra{\psikq{0}{+}} + \ket{\psikq{0}{-}}\bra{\psikq{0}{-}} \right).
\end{equation}
Following this projection, the system evolves during the wait time $\tauwait \gg T_{1}^{ij}, T_{\phi}^{ij}$ and reaches a steady state, which we model as a statistical mixture of the lowest energy manifolds: $\ket{\gstate}$, $\ket{\estate}$, and $\ket{\fstate}$ (see Sections~IV.D,E,F for details). After waiting for $\tauwait$, the statistical mixture of populations between manifolds is independent of the initial state. From these considerations, we do not expect the \CQR~to affect the measurement of $\pe$.

\subsection{Effect of excitation processes on the decoherence model}
The effective model used in Sections~IV.D,E,F (Fig.~\figtwo d of the main text) describes decoherence in the $\{\ket{\gstate},\ket{\estate},\ket{\fstate}\}$ subspace using only energy relaxation ($T_{1}^{ij}$) and pure dephasing ($T_{\phi}^{ij}$).
This is a simplification, since a finite leakage population $\pe>0$ implies the presence of excitation processes. 
In this subsection, we investigate how such processes affect the decoherence model used to fit the decoherence rates and to extract $\pe$. 
Specifically, we test the accuracy of the effective model in fitting a simulated dataset generated from a more complete model that explicitly includes excitation processes. This illustrative example allows us to benchmark our method to extract $\pe$ and identify potential systematic offsets arising from the simplifications of the model.

We first evaluate the impact of excitation channels ($\kappa_{\uparrow}^{ij}$) on the estimated decoherence times $T_{1}^{ij}$ and $T_{\phi}^{ij}$.
To do this, we simulate $T_1$- and $T_2$-like experiments in a three-level system ($\ket{i}$, $i=\{0,1,2\}$) that includes relaxation ($\kappa_1^{ij}$), dephasing ($\kappaphikqv^{ij}$), and excitation ($\kappa_{\uparrow}^{ij}$) channels. These processes are characterized by the following dissipators
\begin{equation}
    \kappaonekq{01} D\!\left(\ket{0}\bra{1}\right), \quad
    2\kappaphikq{01} D\!\left(\ket{1}\bra{1}\right), \quad
    \kappaonekq{12} D\!\left(\ket{1}\bra{2}\right), \quad
    2\kappaphikq{12} D\!\left(\ket{2}\bra{2}\right), \quad
    \kappaheatingkq{01} D\!\left(\ket{1}\bra{0}\right), \quad
    \kappaheatingkq{12} D\!\left(\ket{2}\bra{1}\right). \quad
    \label{eq:estatediss3}
\end{equation}
We fix the decoherence rates $\kappaonekq{ij}$ and $\kappaphikq{ij}$ to values consistent with those extracted in Fig.~\figtwo~of the main text (see Table~\ref{tab:fitted_rates}), and choose excitation rates $\kappaheatingkq{ij} = \kappaonekq{ij}/10$. This choice yields steady-state populations $\pe \simeq 9\%$ and $\pf \simeq 1\%$, close to the values observed experimentally at $\gdiss=0$ for the \KQ~(Fig.~\figtwo~of main text). The result of this simulation defines a dataset that incorporates excitation between manifolds. To characterize the simplified model, which does not account for excitations, we fit this simulated dataset using analytical results derived following the methods outlined in Sections~IV.D,E. From these fits, we extract $\kappaonekq{ij,\mathrm{fit}}$ and $\kappaphikq{ij,\mathrm{fit}}$ (see Table~\ref{tab:fitted_rates}). We find that the fitted values deviate from their set values by approximately $10$-$15\%$.

We now perform a similar test on the method used to estimate $\pe$. We simulate the Rabi-contrast protocol using the full model with excitation, relaxation, and dephasing, to generate a synthetic dataset. As described above, the fixed steady-state population of this synthetic dataset is $\pe \simeq 9\%$ and $\pf \simeq 1\%$. We then fit this data, using two approaches to find $\pe$. In both approaches we fix $\pf=1\%$ and fit $\pe$ together with the readout contrasts $A_i$. 
In the first approach, we fit $\pe$ directly from the Rabi oscillation amplitudes without accounting for decoherence, obtaining $(10.1 \pm 0.8)\%$. 
In the second, we fit the simulated data following the approach in Section~IV.F, using $\kappaonekq{ij,\mathrm{fit}}$ and $\kappaphikq{ij,\mathrm{fit}}$ extracted assuming $\kappaheatingkq{ij}=0$ (second row in Table~\ref{tab:fitted_rates}).
This second approach returns an estimated population of $(9.5\pm0.05)\%$, significantly closer to the true value of the synthetic dataset.
The relative error of this population estimate is $5.5\%$, which is of the same order as the standard deviation of our measured $\pe$ value. This small relative error justifies the approximations made in our effective model.

\begin{table}[h!]
\centering
\renewcommand{\arraystretch}{1.3}
\begin{tabular}{|c|c|c|c|c|c|c|}
\hline
 & $\kappaonekq{01}/2\pi$& $\kappaonekq{12}/2\pi$ & $\kappaheatingkq{01}/2\pi$ & $\kappaheatingkq{12}/2\pi$ & $\kappaphikq{01}/2\pi$ & $\kappaphikq{12}/2\pi$ \\ \hline
Fixed (incl. excitation) & 3.18 kHz & 15.92 kHz & 1.591 kHz & 0.318 kHz  & 10.61 kHz & 31.83 kHz \\ \hline
Fitted (no excitation) & 
$(3 \pm 0.004)~\mathrm{kHz}$ & 
$(18.38 \pm 0.016)~\mathrm{kHz}$ & 
- & 
- & 
$(11.702 \pm 0.07)~\mathrm{kHz}$ & 
$(30.18 \pm 0.034)~\mathrm{kHz}$ \\ \hline
Relative error & 5\% & 15\% & - & - & 10\% & 5\% \\ \hline
\end{tabular}
\caption{Comparison between the fixed decoherence rates, including excitation, and the values fitted with a model neglecting excitation ($\kappaheatingkq{ij}=0$).}
\label{tab:fitted_rates}
\end{table}

\section{Engineered Dissipation in the Driven Oscillator}
\subsection{Lindblad master equation for the engineered dissipation}
\label{subsec:noisemodels}

We define here the Lindblad master equation used to simulate the effect of engineered dissipation on the driven oscillator. 
We previously introduced this equation in the Methods (Eqs.~3 and~4) and restate it here for clarity. 
We further describe the noise processes used to model the driven oscillator, and state in each case whether the corresponding rate derives from an independent measurement or an effective parameter adjusted to reproduce the data (see Table~\ref{tab:noise_models}).

The engineered dissipation arises from a parametrically activated interaction between the driven oscillator and the readout cavity, characterized by a rate~$\gdiss$. This process is modeled by the effective Hamiltonian
\begin{equation}
    \frac{\Hdiss}{\hbar} = \frac{\Hkcq}{\hbar} + \frac{\Hstark}{\hbar} + \gdiss (\hat{a} \hat{b}^\dagger + \hat{a}^\dagger \hat{b}) + \Deltab \hat{b}^\dagger \hat{b},\label{eq:diss_KCQ_hamiltonian}
\end{equation}
where $\Hkcq$ is the \KQ~Hamiltonian (Eq.~\eqref{eq:Ham_kcq}), $\Hstark$ accounts for the Stark shift induced by the squeezing-drive pump (Eq.~\eqref{eq:Ham_stark} with $\xicqr{eff}=0$).  
We use here an effective description in which the engineered interaction term is time-independent. In this way, the detuning of the engineered interaction process is absorbed into the detuning of the cavity mode $\Deltab = \omegage + \delta\omegadiss = \frac{1}{2}(\omega_{1}^+ + \omega_{1}^- - \omega_{0}^+ - \omega_{0}^-) + \delta\omegadiss$.

The time evolution of the system density matrix~$\hat{\rho}$ is then governed by the Lindblad master equation
\begin{equation}
    \frac{\mathrm{d}\hat{\rho}}{\mathrm{d}t} = -\frac{i}{\hbar} \left[ \Hdiss, \hat{\rho} \right] + \sum_{k} \gamma_{k} D[\hat{L}_{k}] \hat{\rho},
    \label{eq:lindblad_noise_general}
\end{equation}
where each noise process contributes one or more collapse operators $\hat{L}_{k}$ with an associated rate $\gamma_{k}$. Several mechanisms that limit \KQ~coherence have been identified~\cite{frattini_observation_2024,Venkatraman2024_nl_diss,Su2024,BlaisRescueKCQ}, but experimental results have so far been described using effective or fitted parameters~\cite{frattini_observation_2024,venkatraman_driven_2024,albornoz_oscillatory_2024,hajr_high-coherence_2024,qing_benchmarking_2024,Ding2024}. 
We therefore keep the two classes of parameter explicitly separated throughout, and summarize them in Table~\ref{tab:noise_models}.

\textbf{Single-photon loss (independently measured).} The oscillator and cavity modes both lose photons to their environment, which we describe by introducing the following dissipators in Eq.~\eqref{eq:lindblad_noise_general}:
\begin{equation}
    \kappaqubit D[\aop], \quad \kappacav D[\bo].
    \label{eq:noise_loss}
\end{equation}
We set $\kappaqubit = 1/\Tone$ and $\kappacav = \kappa_{\mathrm{b,out}} + \kappa_{\mathrm{b,l}}$, using the values measured on the undriven system ($\epstwo = 0$) reported in Table~\ref{tab:params}.

\textbf{Thermal photon number of the undriven oscillator (independently measured).} The oscillator couples to an environment at finite temperature, which drives both excitation and loss. We model this by adding the following dissipators to Eq.~\eqref{eq:lindblad_noise_general}:
\begin{equation}
    \kappaqubit \nth D[\aop], \quad \kappaqubit \nth D[\aod].
    \label{eq:noise_nth}
\end{equation}
We measure the population of the undriven oscillator using a Rabi-contrast protocol~\cite{Geerlings2013,Jin2015} and obtain $\pe(\epstwo = 0) = 0.7\%$. This defines the residual thermal photon number in the undriven oscillator, $\nth = \nthval$, corresponding to a mode temperature of $\Tkcq \approx\SI{62}{\milli\kelvin}$.

\textbf{Thermal photon number of the readout cavity (effective, supported by an independent measurement).} The cavity mode likewise couples to an environment at finite temperature, which we model by adding the following dissipators to Eq.~\eqref{eq:lindblad_noise_general}:
\begin{equation}
    \kappacav \nthb D[\bo], \quad \kappacav \nthb D[\bod].
    \label{eq:noise_nthb}
\end{equation}
For the data presented in the main text we do not have an independent measurement of $\nthb$ on this device. We instead obtain $\nthb = \nthbval$ by reproducing the measured $\pe$ as a function of $\gdiss$ (Fig.~\figthree b of the main text, see Section V.F), and $\nthb = \nthbvalfigthree$ for the threshold simulations of Figs.~\figfour d,e, which were taken at a lower measurement duty cycle~\citeMethods. As described in Section V.G, we have measured $\nthb$ directly on a second, comparable device under similar drive conditions to Fig.~\figthree b and obtain $\nthb = \nthbvaldevtwo$. This supports the effective value used for Fig.~\figthree b.

\textbf{Higher-order terms of static-effective Hamiltonian (independently measured or estimated from numerical simulations).} The \KQ~Hamiltonian of Eq.~\eqref{eq:Ham_kcq} is derived within the RWA to second order in perturbation theory (Section~III.A)~\cite{frattini_observation_2024}. Higher orders of the expansion give additional terms that modify the eigenstates of the driven oscillator, and therefore their response to noise. Following Ref.~\cite{frattini_observation_2024}, the third-order expansion yields
\begin{equation}
    \frac{\hat{H}^{(3)}}{\hbar} = \varepsilon_{31} \left( \hat{a}^{\dag 3} \aop + \aod \aop^{3} \right),
    \label{eq:ham_o3}
\end{equation}
and the fourth-order expansion yields
\begin{equation}
    \frac{\hat{H}^{(4)}}{\hbar} = \lambda_{4} \hat{a}^{\dag 3} \aop^{3} + \varepsilon_{2n} \left( \hat{a}^{\dag 3} \aop + \aod\aop^{3} \right) + \varepsilon_{4} \left( \hat{a}^{\dag 4} + \aop^{4} \right).
    \label{eq:ham_o4}
\end{equation}
The coefficients are defined as 
\begin{align}
    \varepsilon_{31} &= \left( 20 g_{5} - \frac{126 \gthre \gfour}{\omegaqubit} + \frac{6 \gthre^{3}}{\omegaqubit^{2}} \right) \xisq{eff},
    \label{eq:eps31} \\
    \lambda_{4} &= -20 g_{6} + \frac{280 \gthre g_{5}}{\omegaqubit} + \frac{68 \gfour^{2}}{\omegaqubit} - \frac{1800 \gthre^{2} \gfour}{\omegaqubit^{2}} + \frac{2820 \gthre^{4}}{\omegaqubit^{3}},
    \label{eq:lam4} \\
    \varepsilon_{2n} &= - \left( \frac{353 \gthre \gfour}{2 \omegaqubit^{2}} + \frac{521 \gthre^{3}}{2 \omegaqubit^{3}} \right) \left( \Delta^{0} + 12 \gfour - \frac{60 \gthre^{2}}{\omegaqubit} \right) \xisq{eff},
    \label{eq:eps2n} \\
    \varepsilon_{4} &= \left( 15 g_{6} + \frac{2 \gthre g_{5}}{\omegaqubit} - \frac{132 \gfour^{2}}{\omegaqubit} - \frac{303 \gthre^{2} \gfour}{2 \omegaqubit^{2}} + \frac{1989 \gthre^{4}}{2 \omegaqubit^{3}} \right) \left( \xisq{eff} \right)^{2},
    \label{eq:eps4}
\end{align}
where $\xisq{eff} = \epstwo/3\gthre$ (Section~III.A), $g_{m}$ is the $m$-th order nonlinearity of the undriven oscillator, defined by extending the expansion of the SNAIL potential given in Section~III.A to fifth and sixth order, $\hat{H}_{\mathrm{a}}/\hbar = \omegaqubit^{0} \aod_{0} \aop_{0} + \sum_{m \geq 3} g_{m} (\aod_{0} + \aop_{0})^{m}$, and $\Delta^{0} \equiv \omegaqubit^{0} - \omegasq/2$ is the detuning referenced to the bare oscillator frequency.
All of these coefficients follow from independent measurements or from numerical simulation. The oscillator frequency $\omegaqubit$ is measured directly (Table~\ref{tab:params}). The fourth-order nonlinearity $\gfour/2\pi = \SI{-257}{\kilo\hertz}$ follows from the measured Kerr nonlinearity through $K = -6\gfoureff$ with $\gfoureff = \gfour - 5\gthre^{2}/\omegaqubit$ (Section~III.A). The third-, fifth- and sixth-order nonlinearities, $g_{3}/2\pi = \gthreeval$, $g_{5}/2\pi = \SI{3.71}{\kilo\hertz}$ and $g_{6}/2\pi = \SI{39}{\hertz}$, are obtained from a numerical model of the SNAIL array.

\textbf{Two-photon heating beyond the RWA (estimated based on independently measured parameters).} The effective master equation of the driven oscillator also contains nonlinear dissipative terms beyond the RWA, which correspond to multi-photon processes~\cite{Venkatraman2024_nl_diss}. We consider the leading correction terms, which respectively describe two-photon excitation and two-photon loss:
\begin{equation}
    \kappatwophoton \, \nthtwophoton \left( \frac{8\gthre}{3\omegasq} \right)^{2} D[\hat{a}^{\dag 2}], \quad
    \kappatwophoton \left( 1 + \nthtwophoton \right) \left( \frac{8\gthre}{3\omegasq} \right)^{2} D[\aop^{2}].
    \label{eq:noise_twophoton}
\end{equation}
Besides the measured system parameters $\gthre$ and $\omegasq$, these dissipators depend on the coupling rate of the oscillator to the bath at $\omegasq$, $\kappatwophoton$, and on the average photon number in the bath at that frequency, $\nthtwophoton$. 

We determine both $\kappatwophoton$ and $\nthtwophoton$ from independent characterization of our setup. Since the drive is delivered through the cavity mode, $\kappatwophoton \approx \kappa_{\mathrm{b,out}} (g/\Deltaab)^{2}$, which gives $\kappatwophoton/2\pi = \SI{13.41}{\kilo\hertz}$ using the parameters of Table~\ref{tab:params}.
We estimate $\nthtwophoton$ from the measured noise of our room-temperature microwave instruments.
We directly measure the single-sideband phase noise of our OPX+ and Octave setup for a carrier tone at the squeezing-drive frequency, at a carrier power of $2.4~\mathrm{dBm}$ approximately corresponding that used in the experiment.
We evaluate the phase noise at offsets of $1-100~\mathrm{MHz}$ from the carrier, which span the frequency gap $\omegage$ between the $\ket{\gstate}$ and $\ket{\estate}$ manifolds used in the experiment.
In this range, the phase noise is $-144~\mathrm{dBc}$. This corresponds to a noise power of $-141.6~\mathrm{dBm/Hz}$, and therefore to a noise temperature of $5.01\times10^{5}~\mathrm{K}$.
The noise is attenuated by $35.2~\mathrm{dB}$ along the lines down to the $\SI{4}{\kelvin}$ stage and by a further $24.8~\mathrm{dB}$ at the $\SI{10}{\milli\kelvin}$ stage before reaching the sample.
This gives a noise temperature at the sample of $\SI{515}{\milli\kelvin}$, and hence $\nthtwophoton = 0.439$.

\textbf{Additional thermal heating of the oscillator (effective).} Strong parametric drives may raise the effective temperature of the oscillator mode above that of the undriven system, for example through heating of imperfectly thermalized dissipative components close to the sample. We describe this with an additional thermal photon number $\nthplus$, corresponding to the added dissipators
\begin{equation}
    \kappaqubit \nthplus D[\aop], \quad \kappaqubit \nthplus D[\aod].
    \label{eq:noise_nthplus}
\end{equation}
Similar effective models have been used to describe \KQ~bit-flip times~\cite{frattini_observation_2024,venkatraman_driven_2024,hajr_high-coherence_2024,albornoz_oscillatory_2024}. In particular, Ref.~\cite{albornoz_oscillatory_2024} found that inter-well tunneling in a \KQ~could be described by an effective mode temperature scaling quadratically with $\epstwo$, as expected for heating induced by the parametric drive. 

\textbf{Dephasing noise (effective).} Dephasing noise also generates leakage, because the action of $\aod\aop$ on the \KQ~$Z$-states generates a component outside the qubit manifold that is exponentially suppressed with \KQ\ photon number. Such noise can originate from the external magnetic flux bias or from local fluctuators on the oscillator chip. We describe it by an effective dephasing rate $\kappaphi$, corresponding to the added dissipator
\begin{equation}
    \kappaphi D[\aod \aop].
    \label{eq:noise_kappaphi}
\end{equation}
Effective dephasing of this form has likewise been used to model \KQ~bit-flip times~\cite{frattini_observation_2024,hajr_high-coherence_2024,albornoz_oscillatory_2024}. This rate is not equal to the dephasing rate of the undriven oscillator, because the \KQ~encoding filters the noise spectral density at a different frequency than the Fock-state encoding~\cite{Grimm2020}. 

\begin{table}[h!]
	\centering
    \squeezetable
    \renewcommand{\arraystretch}{1.2}
	\begin{tabular}{l|c|c|l}
		\textbf{Noise process}
		& \textbf{Collapse operators}
		& \textbf{Parameter}
		& \textbf{Status}\\ \hlineB{2}
        Single-photon loss, oscillator & $\kappaqubit D[\aop]$ & $\kappaqubit = 1/\Tone$, $\Tone = \bareTone$ & Measured \\ \hline
        Single-photon loss, cavity & $\kappacav D[\bo]$ & $\kappacav/2\pi = \kappacavval$ & Measured \\ \hline
        Oscillator thermal photon number & $\kappaqubit \nth D[\aop]$, $\kappaqubit \nth D[\aod]$ & $\nth = \nthval$ & Measured \\ \hline
        Cavity thermal photon number & $\kappacav \nthb D[\bo]$, $\kappacav \nthb D[\bod]$ & $\nthb = \nthbval$ (Fig.~\figthree b) & Effective / measured \\ \hline
        Higher-order Hamiltonian terms & Eqs.~\eqref{eq:ham_o3},~\eqref{eq:ham_o4} (Hamiltonian) & $g_{m}$, $\xisq{eff}$ & Measured / modeled \\ \hline
        Two-photon heating & Eq.~\eqref{eq:noise_twophoton} & $\kappatwophoton/2\pi = \SI{13.41}{\kilo\hertz}$, $\nthtwophoton = 0.439$ & Measured \\ \hline
        Additional oscillator heating & $\kappaqubit \nthplus D[\aop]$, $\kappaqubit \nthplus D[\aod]$ & $\nthplus = \nthplusval$ & Effective \\ \hline
        Dephasing & $\kappaphi D[\aod \aop]$ & $\kappaphi/2\pi = \kappaphival$ & Effective \\ \hline
	\end{tabular}
	\caption{\textbf{Noise models for the driven oscillator.} Summary of the noise processes used to model the \KQ, the corresponding collapse operators or Hamiltonian terms, the parameter values, and whether each parameter is independently measured or is an effective parameter adjusted to reproduce the data. The cavity thermal photon number is an effective parameter for the device of the main text, but is measured independently on a second device (Section~V.G). The two effective parameters $\nthplus$ and $\kappaphi$ are set to values that reproduce the measured leakage population in the \KQ\ (see Section~V.F).}
	\label{tab:noise_models}
\end{table}

\subsection{Calibration of $\gdiss$ and simulation of $\kappadiss$}
\label{sec:g_diss}

In this subsection, we describe the calibration experiment for the engineered interaction rate $\gdiss$ using the oscillator in absence of the squeezing drive ($\epstwo=0$), and relate $\gdiss$ to the dissipation rate $\kappadiss$ on the $\transeg$ transition for the driven oscillator at \workingpoint.

\begin{figure*}[b]
    \includegraphics[angle = 0, width = \figwidthWide]{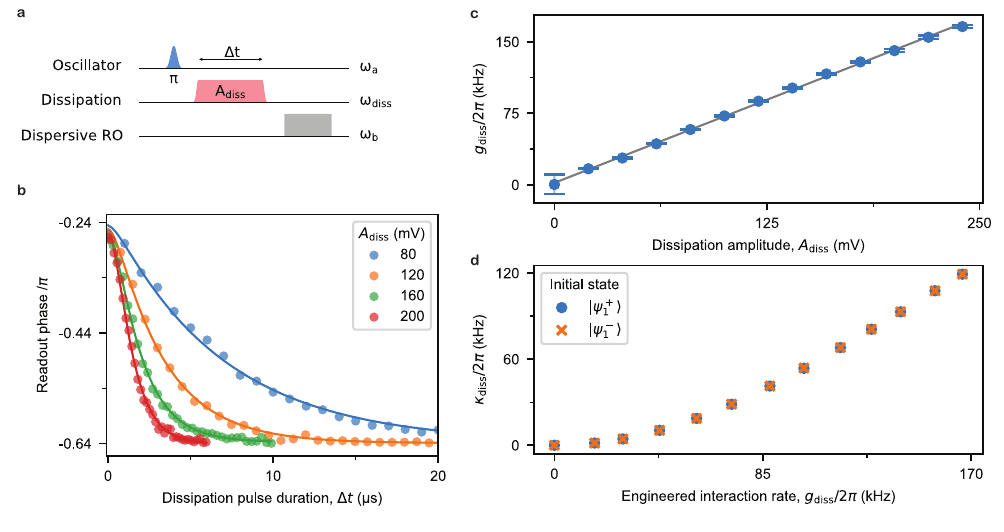}
    \caption{\label{fig_SI_gdiss} 
    \textbf{Calibration of the engineered dissipation rate. a}, Pulse sequence used to calibrate the engineered interaction rate $\gdiss$: a \(\pi\)-pulse on the oscillator in absence of the squeezing drive($\epstwo =0$) is followed by a dissipation pulse of amplitude \( \Adiss \) and duration \( \dt \), and then by a readout pulse at the cavity frequency \( \omegacav \).
    \textbf{b}, Measured decay curves as a function of the dissipation pulse duration \( \dt \) for different dissipation amplitudes \( \Adiss \). The y-axis shows the phase of the signal reflected from the readout cavity. Solid lines are fits of the oscillator population \( \bar{n}(t) \) using the coupled equations of motion for the oscillator-cavity dynamics (Eq.~\eqref{eq:diss_eq_of_motion_a}).
    \textbf{c}, Extracted engineered interaction rate \( \gdiss \) as a function of \( \Adiss \) (blue markers) with linear fit (gray line). Error bars are given by the fit uncertainties of the decay curves in b.
    \textbf{d}, Effective dissipation rate \( \kappadiss \) extracted from the $1/e$ decay time of $\bra{\estate}\hat{\rho}_{\mathrm{a}}(t)\ket{\estate}$, as a function of \( \gdiss \). Here, we set \workingpoint.
    }
\end{figure*}

In the frame rotating at the oscillator and cavity frequencies \( \omegaqubit \) and \( \omegacav \), respectively, the system Hamiltonian is
\begin{equation}
    \frac{\hat{H}}{\hbar} = \frac{\Hk}{\hbar} + \gdiss\left(\hat{a} \hat{b}^\dagger + \hat{a}^\dagger \hat{b}\right),
\end{equation}
where $\Hk$ is the Kerr Hamiltonian (Eq.~\eqref{eq:Ham_k}), and the interaction is taken to be resonant and therefore time independent. We assume \( \gdiss \) is real without loss of generality. Including single-photon loss rates \( \kappaqubit \) and \( \kappacav \) for the oscillator and cavity, respectively, the equations of motion for operators \( \hat{a}(t) \) and \( \hat{b}(t) \) are
\begin{equation}
\begin{aligned}
\frac{\mathrm{d}}{\mathrm{d}t} \hat{a}(t) &= -i \gdiss \hat{b}(t) - \frac{\kappaqubit}{2} \hat{a}(t), \\
\frac{\mathrm{d}}{\mathrm{d}t} \hat{b}(t) &= -i \gdiss \hat{a}(t) - \frac{\kappacav}{2} \hat{b}(t).
\end{aligned}
\end{equation}
The solutions to the coupled differential equations, with initial condition $\hat{a}(0) = \hat{a}_0$ and $\hat{b}(0) = 0$, are
\begin{align}
\label{eq:diss_eq_of_motion_a}
    \hat{a}(t) &= \hat{a}_0\, e^{-\kappagdiss t}\left[\cosh\left(\frac{\Lambda}{4} t \right)+\frac{\kappacav-\kappaqubit}{\Lambda}\,\sinh\left(\frac{\Lambda}{4} t \right)\right], \\
    \label{eq:diss_eq_of_motion_b}
    \hat{b}(t) &= -i\frac{4g}{\Lambda}\hat{a}_0 e^{-\kappagdiss t} \sinh\left(\frac{\Lambda}{4} t \right),
\end{align}
with the definitions
$$\kappagdiss = \frac{\kappaqubit+\kappacav}{4}, \quad
\Lambda = \sqrt{(\kappacav-\kappaqubit)^2-16\gdiss^2}.
$$
We characterize \( \gdiss \) with a \( \Tone \) measurement in presence of the engineered dissipation (see pulse sequence in Fig.~\ref{fig_SI_gdiss}a). First, we initialize the oscillator in Fock state $\ket{1}$, then we apply the dissipation pulse with variable amplitude \( \Adiss \) and duration \( \dt \), and finally we measure the oscillator population \( \bar{n}(t) = \langle \hat{a}^\dagger(t) \hat{a}(t) \rangle \). We fit the data using Eq.~\eqref{eq:diss_eq_of_motion_a}, extracting $\gdiss$ together with a global rescaling factor and a constant offset, the latter two accounting for the readout contrast of the oscillator states $\ket{0}$ and $\ket{1}$. The decay curves obtained for different dissipation pulse amplitudes \( \Adiss \), are shown in Fig.~\ref{fig_SI_gdiss}b. The extracted values of \( \gdiss \) with respect to \( \Adiss \) are shown in Fig.~\ref{fig_SI_gdiss}c. 
We find the expected linear dependence between the pulse amplitude and the interaction strength~\cite{Grimm2020} (see Section~III.A for an equivalent derivation of $\gcqr$).

We now use the Lindblad master equation described in the previous subsection (Eq.~\eqref{eq:lindblad_noise_general}) to extract the engineered dissipation rate $\kappadiss$ acting on the $\transeg$ transition of the driven oscillator at \workingpoint, corresponding to the data shown in Figs.~\figthree,~\figfour, and~\figfive~of the main text. To this end, we set $\delta\omegadiss = 0$, which corresponds to a rotating frame where the cavity mode rotates at frequency \( \Deltab = \omegage \) (Eq.~\eqref{eq:diss_KCQ_hamiltonian}). For this detuning value, the engineered interaction term becomes resonant with the \( \transeg \) transition, enabling selective dissipation from the $\ket{\estate}$ manifold. In this simulation, we set $\kappaqubit = \nth = \nthb = 0$, while $\kappacav$ is fixed to its experimentally measured value. We initialize the system in the state \( \ket{\psi(0)} = \ket{\psikq{1}{\pm}} \otimes \ket{0} \), where the second ket denotes to the cavity vacuum state.
We evolve the system for a variable delay time, and evaluate $\bra{\estate}\hat{\rho}_{\mathrm{a}}(t)\ket{\estate}$, with $\hat{\rho}_{\mathrm{a}}$ the density matrix of the oscillator. This results in a decaying value as a function of the variable delay time, from which we extract the $1/e$ time to estimate $\kappadiss$. This procedure is necessary because, for \( \gdiss \lesssim \kappacav \), the decay deviates from a simple exponential, rendering an exponential fit unreliable.
The extracted values (presented in Fig.~\ref{fig_SI_gdiss}d) then map $\gdiss$ onto the dissipation rate $\kappadiss$ in the driven oscillator.

\subsection{Modeling the leakage population $\pe$ of the driven oscillator}
\label{sec:pe_model}

In this subsection, we aim to reproduce in simulations the leakage population of the driven oscillator measured at \workingpoint~(Fig.~\figtwo~of the main text) and its suppression under engineered dissipation (Fig.~\figthree~of the main text). We define the driven oscillator 
population for states $\ket{\psi_i^\pm}$ with $i=\{0,1,2\}$ as
\begin{align}
    p^{\mathrm{sim}}_{i} = \sum_\pm \bra{\psi_{i}^{\pm}} \rhokq{\mathrm{a}}{} \ket{\psi_{i}^{\pm}},
    \label{eq:pe_sim_def}
\end{align}
where $\rhokq{\mathrm{a}}{}$ is the density matrix of the nonlinear oscillator.

\noindent
\paragraph{\textbf{Steady-state population without engineered dissipation.}}
We begin by modeling the steady state of the oscillator system in the absence of engineered dissipation. Specifically, we solve the Lindblad master equation Eq.~\eqref{eq:lindblad_noise_general} in the case of $\gdiss=0$ and after tracing out the cavity mode.
As discussed in the main text, we now introduce each noise mechanism independently to isolate its specific contribution to $\pe$ (see Subsec.~\ref{subsec:noisemodels} for a full description of the different noise processes).

\begin{figure}
    \centering
    \includegraphics[width=\linewidth]{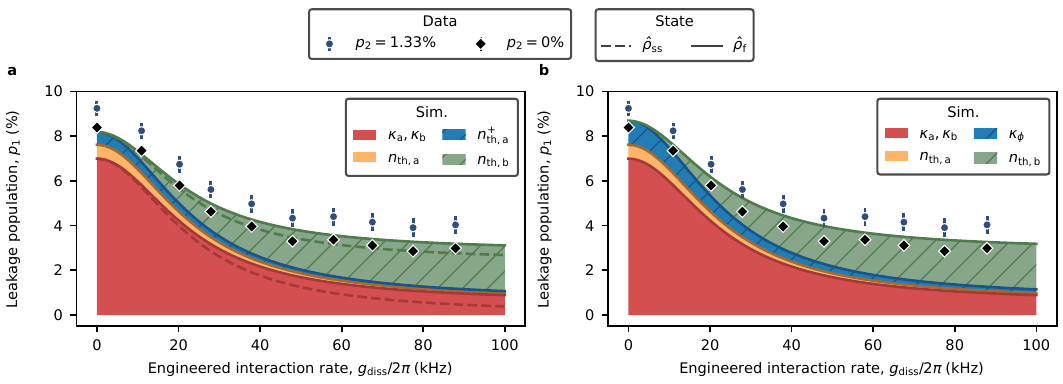}
    \caption{\textbf{Contributions to the leakage population.} Measured $\pe$ as a function of $\gdiss$ (markers, data from Fig.~3b of the main text) and simulated $\pe$ from different noise processes (shaded regions). Blue circles and black diamonds assume $\pf=\pfvalfit\%$ and $\pf=0\%$, respectively, with error bars denoting one standard deviation. 
    \textbf{a}, Single-photon loss (red), measured thermal heating in the undriven oscillator (yellow), and effective additional thermal heating of the oscillator (blue hatched) and of the cavity (green hatched). Figure~\figthree b in the main text reproduced here for convenience.
    Solid curves are obtained from the simulated state $\hat{\rho}_{\mathrm{f,a}}$ after the experimental wait sequence, while dashed curves correspond to the oscillator steady state $\hat{\rho}_{\mathrm{ss,a}}$.
    \textbf{b}, As \textbf{a}, using an effective dephasing noise instead of an effective additional thermal heating of the oscillator. Simulations are performed at $\epstwo=\epstwofigtwo K$ and $\Delta=\Deltafigtwo K$.}
    \label{fig:p1_noise_models}
\end{figure}

We first consider only noise parameters based on independent measurements, and evaluate their contributions to $\pe$ at $\gdiss=0$.  
As in the main text, we include single-photon loss of the oscillator, $\kappaqubit$, and cavity, $\kappacav$ (Eq.~\eqref{eq:noise_loss}). We find that it accounts for a leakage population of $\pesim = \peqhval$ (red shaded region in Fig.~\ref{fig:p1_noise_models}a). The presence of non-zero $\pesim$ despite the absence of thermal excitation arises because $\ket{\pm\alphaD}$ are not eigenstates of the annihilation operator $\hat{a}$. As a result, single-photon loss induces quantum heating in the system~\cite{Dykman2011, Ruiz2023}.
Including higher-order Hamiltonian terms as in Ref.~\cite{frattini_observation_2024} has only a small effect on the leakage population under single-photon loss: including third-order terms (Eq.~\eqref{eq:ham_o3}) yields $\pesim=6.8\%$, and additionally including fourth-order terms (Eq.~\eqref{eq:ham_o4}) gives $\pesim=6.7\%$. 
Since the correction to $\pesim$ is small when including these higher-order Hamiltonian terms, we use the second-order effective Hamiltonian for the rest of the analysis (Eq.~\eqref{eq:diss_KCQ_hamiltonian}). Next we add the measured thermal photon number in the undriven oscillator $\nth$ (Eq.~\eqref{eq:noise_nth}), giving $\pesim=\penthval$ (see orange shaded region in Fig.~\ref{fig:p1_noise_models}a). This still underestimates the experimental value by approximately $1.6\%$. We therefore investigate the contribution of two-photon heating to leakage (Eq.~\eqref{eq:noise_twophoton}), since such terms have been shown to generate additional leakage mechanisms that limit bit-flip times at large $\epstwo$~\cite{Venkatraman2024_nl_diss}. Using noise parameters derived from independent measurements, we find a negligible impact on the total leakage population. 
Based on this result, we conclude that the measured leakage population cannot be accounted for by considering two-photon heating and therefore neglect it from the following analysis.
More generally, these results show that we cannot reproduce the measured $\pe$ using independently measured parameters alone.

We therefore investigate two phenomenological noise models and the extent to which they account for the measured $\pe$ at $\gdiss=0$. Specifically, we consider additional thermal heating using an effective additional thermal photon number $\nthplus$  (Eq.~\eqref{eq:noise_nthplus}), and dephasing noise with an effective rate $\kappaphi$ (Eq.~\eqref{eq:noise_kappaphi}). 
As stated in the main text, the measured $\pe$ can be accounted for with $\nthplus=0.018$, corresponding to an overall mode temperature of $\approx\SI{82}{\milli\kelvin}$. This value is consistent with effective mode temperatures used elsewhere to describe \KQ~bit-flip times~\cite{frattini_observation_2024,venkatraman_driven_2024,hajr_high-coherence_2024,albornoz_oscillatory_2024}. This yields simulated populations of $\pesim = 9.1\%$ (blue hatched region in Fig.~\ref{fig:p1_noise_models}a). Equivalently, the measured $\pe$ can be accounted for with dephasing noise at rate $\kappaphi/2\pi=\kappaphival$ (blue hatched region in Fig.~\ref{fig:p1_noise_models}b). Since $\pe$ at $\gdiss=0$ can be described equally well by considering additional thermal heating or dephasing noise, we cannot distinguish between these two noise models from $\pe$ measurements alone. In the experiment, the system likely experiences a combination of additional thermal heating and dephasing, in addition to others not considered here such as spurious multi-photon transitions to higher-lying leakage states~\cite{BlaisRescueKCQ}.

\noindent
\paragraph{\textbf{\KQ\ population with engineered dissipation.}}
We now investigate the impact of engineered dissipation on $\pesim$. To this end, we model the full system Hamiltonian describing the nonlinear oscillator coupled to the readout cavity via the engineered single-photon interaction, as given by Eqs.~\eqref{eq:diss_KCQ_hamiltonian} and~\eqref{eq:lindblad_noise_general}.

As mentioned in the Methods, to accurately model the effect of dissipation on the measured $\pe$ we need to account for idle times in the experimental protocol where the dissipation pulse is not activated. 
Specifically, the experiment includes a delay $\taucav = \taucavval$ after the dissipation pulse to allow residual cavity photons to decay, a conditional $\pige$-pulse of duration $\tau_{\mathrm{R},01} = \taupulseval$, and a variable-amplitude Rabi drive on the $\transef$ transition of duration $\tau_{\mathrm{R},12} = \taupulseval$. We account for this in the simulation by incorporating a delay time $\taudelay=\taucav+\tau_{\mathrm{R},01}+\tau_{\mathrm{R},12}/2=\SI{4.2}{\micro\second}$, where the final term accounts for excitations induced during the $\transef$ Rabi drive by including half of its duration.  

To incorporate the effect of this finite delay time, $\taudelay$, into our model, we first perform a steady-state simulation at a fixed interaction rate $\gdiss$, obtaining the corresponding steady-state density matrix $\rhoss$. Starting from this state, $\hat{\rho}(t=0)=\rhoss$, we then evolve the system for a duration $\taudelay$ with $\gdiss=0$, yielding the final state $\rhof=\hat{\rho}(t=\taudelay)$. Finally, to compute the population in the nonlinear oscillator, we trace out the cavity mode and calculate $p_i^{\mathrm{sim}}$ using Eq.~\eqref{eq:pe_sim_def}. 

The results of the simulations are shown in Fig.~\ref{fig:p1_noise_models}a, which presents the results of Fig.~\figthree b here for convenience. For the case of quantum heating (red shaded region), the solid (dashed) line indicates the value of $\pesim$ extracted from the final state $\hat{\rho}_{\mathrm{f,a}}$ (steady-state $\hat{\rho}_{\mathrm{ss,a}}$). For the cases including the measured $\nth$ (orange shaded region) and the effective $\nthplus$ (blue hatched region), we show only $\pesim$ extracted from $\hat{\rho}_{\mathrm{f,a}}$ for clarity (solid lines). In Fig.~\ref{fig:p1_noise_models}b, we similarly show $\pesim$ as a function of $\gdiss$ obtained by considering an effective dephasing rate $\kappaphi$ (blue hatched region) in addition to the contributions from quantum heating and the measured $\nth$.
In all cases, $\pesim$ decreases with increasing $\gdiss$, qualitatively reproducing the experimental trend. However, despite matching the experimental values at $\gdiss=0$, an offset remains between the simulated and measured values at large $\gdiss$. In the simulations, $\pe$ tends to saturate toward zero, whereas experimentally it saturates at $\pe\sim3\%$.

The discrepancy between simulated and measured $\pe$ at large $\gdiss$ can be accounted for by considering a finite temperature of the cavity mode, $\nthb$. Here, we reproduce the experimental results for $\nthb=\nthbval$ (green hatched regions in Fig.~\ref{fig:p1_noise_models}), corresponding to to an effective temperature of $\Tcav \approx \Tcavval$. 
This temperature is significantly higher than the mixing-chamber temperature of our dilution refrigerator, which is below $\SI{10}{\milli\kelvin}$. 
However, we cannot exclude that the elevated $\nthb$ originates from noise injected through the strongly coupled pump ports used for the parametric drives, possibly coming from the amplification stages located at room temperature. 
Furthermore, spurious parametric processes activated by the high-power drives could provide additional heating that increases $\nthb$~\cite{BlaisRescueKCQ,Dai_Hazra_2025}.
Note that the value of $\nthb$ adopted in our simulations is consistent with those reported in related works~\cite{Venkatraman2024_nl_diss,qing_benchmarking_2024,hajr_high-coherence_2024}. 
In the following section, we present independent measurements of the cavity population on a similar device which show remarkable agreement with the effective value used here, and self-consistently describe the \KQ\ population $\pe$.

\subsection{Experimental verification of cavity thermal photon number $\nthb$}
\label{sec:cav_pop}

We support the effective cavity population $\nthb$ used to describe the data in the main text with measurements performed on a second device. In this section, we present measurements of \KQ\ leakage population $\pe$ (as in Fig. 3b of the main text) and complimentary measurements of the cavity thermal photon number $\nthb$ in presence of strong parametric drives. We use these results to motivate the $\nthb$ used in the main text, and its relation to $\pe$ for strong engineered dissipation strengths $\gdiss$.

These measurements were performed on a different device, after the datasets presented in the main text and in the rest of the Supplementary Information were taken. The device used for the experiments presented in this section is characteristically similar to that of the main text, and measurements were performed in an experimental setup under similar conditions. 
The notable difference with respect to the device of the main text is $\Nsnail=2$ which, together with the charging energy $\Ec=\SI{135}{\mega\hertz}$, results in a larger Kerr-nonlinearity $K/2\pi=\SI{3.20}{\mega\hertz}$.
The complete system parameters are outlined in Table~\ref{tab:devicetwo_params}.
\begin{table}[h]
	\centering
    \renewcommand{\arraystretch}{1.2}
	\begin{tabular}{l|c}
		\textbf{Parameter} 
		& \textbf{Value}\\ \hlineB{2}   
        Nonlinear oscillator charging energy $\Ec/2\pi$   & $\SI{135}{\mega\hertz}$ \\ \hline
        Nonlinear oscillator number of SNAILs $\Nsnail$   & $2$ \\ \hline
        Asymmetry factor in the SNAIL $\jjasym$  & $\approx0.07$ \\ \hline
        
        Nonlinear oscillator frequency $\omegaqubit/2\pi$   & $\SI{6.27}{\giga\hertz}$ \\ \hline
        Kerr-nonlinearity $K/2\pi$ & $\SI{3.20}{\mega\hertz}$ \\ \hline
        Third-order nonlinearity $g_3/2\pi$ & $\approx\SI{7.5}{\mega\hertz}$ \\ \hline
        Undriven nonlinear oscillator thermal photon number $\nth$ & $0.017$ \\ \hline
        Nonlinear oscillator single-photon decay time $\Tone$ & $\SI{68.1}{\micro\second}$ \\ \hline
        Nonlinear oscillator Ramsey coherence time $\TtwoR$ & $\SI{3.54}{\micro\second}$ \\ \hline
        Nonlinear oscillator Ramsey coherence time (echo) $\TtwoE$ & $\SI{14.2}{\micro\second}$ \\ \hline
        Readout cavity frequency $\omegacav/2\pi$  & $\SI{9.01}{\giga\hertz}$ \\ \hline 
        Readout cavity linewidth (output coupling) $\kappa_{\mathrm{b,out}}/2\pi$ & $\SI{340}{\kilo\hertz}$ \\ \hline
        Readout cavity linewidth (other losses) $\kappa_{\mathrm{b,l}}/2\pi$ & $\SI{610}{\kilo\hertz}$ \\ \hline
        Cavity --- nonlinear oscillator dispersive shift $\chidisp/2\pi$ & $\approx\SI{160}{\kilo\hertz}$ \\ \hline
        
	\end{tabular}  
    \caption{\label{tab:devicetwo_params}\textbf{System parameters for second device.} Summary of main system parameters of the second device.}
\end{table}

For the results shown in this section, we stabilize a \KQ\ with $\epstwo=5.14K$ and $\Delta=8K$. The two-photon drive strength is chosen to enable the leakage population measurement, which was not possible at $\epstwo=2.4K$ due to large leakage-state decoherence. We expect that the slight difference in system parameters and operating point with respect to the main text do not change the main conclusions. 

We perform a direct measurement of the cavity thermal photon number $\nthb$ while strong drives are applied to the system. 
To measure $\nthb$, we follow the $\pe$ measurement protocol outlined in the main text (see Fig.~\figtwo a) after first swapping the \KQ\ and cavity populations. 
This \KQ-cavity swap is realized using a Gaussian pulse (total length $6\sigma=\SI{1}{\micro\second}$) at a frequency $\omega_{\mathrm{ab}}=\omegacav-(\omegasq/2-\omegage)$. We then use the Rabi-contrast protocol to obtain $\nthb=0.0242$ ($\Tcav\approx\SI{117}{\milli\kelvin}$). This is very close to the value used in Fig.~\figthree b of the main text ($\nthb=\nthbval$), strongly supporting such elevated cavity temperatures in the presence of strong parametric drives. 

\begin{figure}
    \centering
    \includegraphics[width=0.5\linewidth]{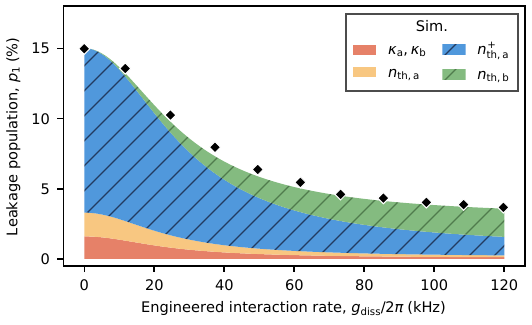}
    \caption{\textbf{Leakage population under engineered dissipation on the second device.} Measured $\pe$ as a function of $\gdiss$ (black diamonds), extracted as in Fig.~3b of the main text assuming $\pf=0$, with cumulative contributions from simulated processes (shaded regions, as in Fig.~\ref{fig:p1_noise_models}): measured single-photon loss on both modes ($\kappaqubit$, $\kappacav$: red) and the measured undriven oscillator thermal photon number ($\nth=\nthvaldevtwo$: orange), followed by the drive-induced terms (hatched) of an effective additional oscillator thermal photon number ($\nthplus=\nthplusvaldevtwo$: blue), set to match $\pe$ at $\gdiss=0$, and the independently measured cavity thermal photon number ($\nthb=\nthbvaldevtwo$: green). Measurements were performed at $\epstwo=\epstwodevtwo K$ and $\Delta=\Deltadevtwo K$, with a measurement duty cycle $d_{1}=0.48$.}
    \label{fig:p1_device2}
\end{figure}

On the same device, and interleaved with the $\nthb$ measurements, we directly measure the \KQ\ leakage population $\pe$ as a function of engineered dissipation interaction strength $\gdiss$, following the procedure outlined in the main text (Figs.~\figtwo a,c and Fig.~\figthree b). The results are shown in Fig.~\ref{fig:p1_device2}. As in Fig.~\figthree b of the main text, $\pe$ decreases with increasing $\gdiss$, from $\pe=\pezerodevtwo$ at $\gdiss=0$ to $\pe=\pesatdevtwo$ at the largest applied rate $\gdiss/2\pi=\gdissmaxdevtwo$. Note that a reliable estimate for $\pf$ (see Sec.~\ref{sec:incoherent_spec}) was not possible for this dataset because the corresponding spectroscopy peak could not be resolved above the experimental noise. We therefore report $\pe$ values considering $\pf=0$. This does not change the conclusions on decreasing $\pe$ with $\gdiss$, nor the saturation value of $\pe$ at large $\gdiss$, since $\pe$ decreases irrespective of the value chosen for $\pf$ and we expect $\pf\approx0$ for large $\gdiss$.

Following the discussion in Sec.~\ref{sec:pe_model}, we model the contributions to $\pe$ from different noise sources in the \KQ. We fix all experimentally measured values, namely the single photon loss on both modes ($\kappaqubit$, $\kappacav$), the undriven oscillator thermal photon number ($\nth$) and the driven cavity thermal photon number ($\nthb$). We introduce a single effective parameter: an additional \KQ\ thermal photon number due to the strong drive, $\nthplus$. In contrast to the results shown in the main text, single photon loss contributes only a small fraction to the leakage population ($\pe=\pelossdevtwo$ at $\gdiss=0$), since the effects of quantum heating are smaller at larger $\epstwo$ (see red shaded region in Fig.~\ref{fig:p1_device2}). Adding $\nth$ results in $\pe=\penthdevtwo$ at $\gdiss=0$ (orange shaded region in Fig.~\ref{fig:p1_device2}), which still does not account for the measured value. To match $\pe$ at $\gdiss=0$, we set an effective oscillator thermal photon number $\nthplus=\nthplusvaldevtwo$ (blue hatched region in Fig.~\ref{fig:p1_device2}), corresponding to an effective temperature of $\Tkcq\approx\Tkcqvaldevtwo$. The larger $\nthplus$ found for this sample compared to that of the main text could be due to by the larger $\epstwo$ compared to the working point of the main text~\cite{albornoz_oscillatory_2024}, or sample-dependent differences in the effective noise processes. Note that the measured $\pe$ could equivalently be accounted for using an effective dephasing parameter $\kappaphi/2\pi=\kappaphivaldevtwo$ (see Sec.~\ref{sec:pe_model}). Finally, $\pesim$ including the independently measured $\nthb=\nthbvaldevtwo$ self-consistently describes $\pe$ at large $\gdiss$ (green hatched region in Fig.~\ref{fig:p1_device2}). We highlight that the simulation reproduces the entire dependence of the measured $\pe$ at finite $\gdiss$, after fixing a single effective parameter $\nthplus$ at $\gdiss=0$. This demonstrates that an elevated cavity temperature significantly impacts the trend and saturation value of $\pe$ at large $\gdiss$, supporting the model used in Fig.~\figthree b of the main text.

To further understand the thermal noise processes affecting the \KQ\ and cavity populations, we repeated the experiments presented above using a different measurement duty cycle. We increase the wait time between measurement shots from $\SI{350}{\micro\second}$ to $\SI{2000}{\micro\second}$, such that the squeezing-drive pump is on for a smaller fraction of the experiment time. This change in duty cycle from $d_{1}=0.482$ to $d_{2}=0.138$ is comparable to the change in duty cycle between measurements presented in Figs.~\figthree b and~\figfour e of the main text. We obtain $\pe (\gdiss=0)=13.23\%$, $\pe(\gdiss/2\pi=\gdissmaxdevtwo)=2.54\%$ and $\nthb=0.019$ for the smaller duty cycle ($d_{2}$), indicative of thermal noise processes with long relaxation times compared to a single measurement shot. The reduction in $\nthb$ with duty cycle is consistent with the smaller effective $\nthb$ used to model the data in Fig.~\figfour e of the main text.

\subsection{Threshold between dissipation regimes as a function of $\epstwo$}
In this subsection we describe the procedure used to extract $\epstr$, whose results are presented in Fig.~\figfour~of the main text.

\begin{figure*}
    \includegraphics[angle = 0, width = \figwidthWide]{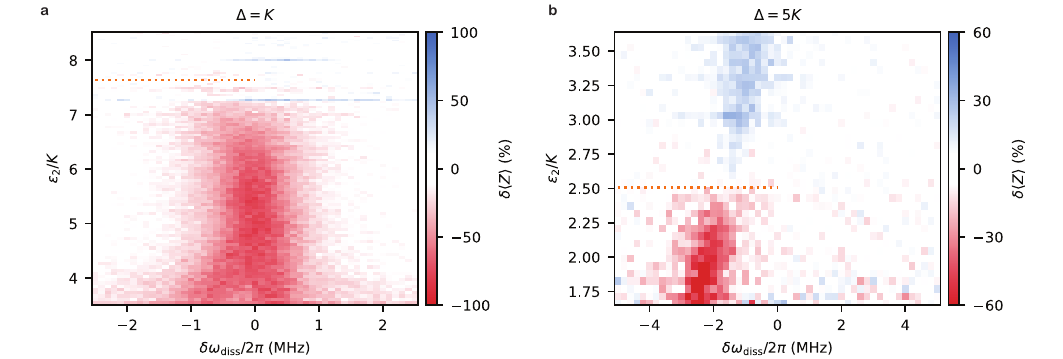}
    \caption{\label{fig_SI_e2tr} 
    \textbf{Change of $Z$-state readout (\CQR) contrast with engineered dissipation.}
    \textbf{a}(\textbf{b}) Relative change in the \CQR~contrast, $\delta \Zdiss$, as a function of dissipation detuning $\delta\omegadiss$ and squeezing-drive amplitude $\epstwo$, for $\Delta = K$ ($\Delta=5K$). 
    In both panels, the orange dotted lines indicate the extracted value of $\epstr$.
    }
\end{figure*}

We first perform an incoherent spectroscopy measurement for each set of parameters $\epstwo$ and $\Delta$ to extract the $\transge$ transition frequency $ \omegage $, using the same procedure as outlined in Refs.~\cite{frattini_observation_2024,venkatraman_driven_2024}~\citeMethods. We then carry out the measurements described by the pulse sequence shown in in Fig.~\figfour a of the main text. As mentioned there, we use a fixed dissipation pulse length of $\SI{50}{\micro\second}$ to give a change in $\Zdiss$ that is representative of a change in $\Tz$. This pulse length keeps the measurement overhead low while maintaining good readout contrast across the $\epstwo, \Delta$ parameter space. The readout signal changes as a function of $\epstwo$ because the cavity displacement is proportional to $\alphaD\propto\sqrt{\epstwo}$ and because $\Tz$, and therefore $\Zdiss$, also depends on $\epstwo$~\cite{Grimm2020, frattini_observation_2024}. Therefore, for each $\epstwo$, we fit the trace of $\Zdiss$ as a function of $\delta\omegadiss$ with a Lorentzian function to extract the signal background, $\Zdissbg$, and peak values. 
We then define the relative change with respect to the background as $\delta\Zdiss = (\Zdiss-\Zdissbg)/\Zdissbg$, shown in Fig.~\figfour b and  Figure~\ref{fig_SI_e2tr} for different values of $\Delta$.

In Fig.~\ref{fig_SI_e2tr}a ($\Delta=K$), $\delta\Zdiss$ gradually saturates to zero as $\epstwo$ increases. This dataset is representative of those where no positive $\delta\Zdiss$ (open markers in Fig.~\figfour e). In this case, the extracted $\epstr$ corresponds to the lowest $\epstwo$ where the fitted peak value is below the standard deviation in that measurement trace. In Fig.~\ref{fig_SI_e2tr}b ($\Delta=5K$), we observe a distinct change of $\delta\Zdiss$ from negative to positive values as $\epstwo$ increases. This is representative of the filled markers in Fig.~\figfour e. In this case, we define $\epstr$ as the $\epstwo$ for which this peak value changes sign. Note that the maximum value of $|\delta\Zdiss|$ does not occur at $\delta\omegadiss=0$, and in fact depends on $\epstwo$. This is due to a slight offset in the calibration of $\omegage(\epstwo)$, justifying the sweep in $\omegadiss$ for these measurements. Uncertainties in the extracted $\epstr$ originate from the fitted uncertainty in relative peak signal and uncertainty in the calibrated $\epstwo$ value obtained from incoherent spectroscopy measurements.

\subsection{Effective model for the engineered dissipation}
In this subsection, we use the effective dissipators describing the engineered dissipation (Eqs.~6 and~7 in the Methods) to study how the threshold $\epstr$ depends on the thermal photon numbers of the oscillator and cavity, $\nth$ and $\nthb$, respectively. We further compare these results with those obtained from the full model (Eq.~\eqref{eq:lindblad_noise_general}) and find good overall agreement between the two approaches in predicting $\epstr$. After validating the effective model, we then use it to develop an intuitive picture of how engineered dissipation affects the \KQ~bit-flip times and to explain the dependence of $\epstr$ on $\nthb$ and $\nth$.

As discussed in the Methods, starting from the Lindblad master equation describing the engineered interaction between the cavity and the oscillator (Eq.~\eqref{eq:lindblad_noise_general}), we derive effective dissipators acting on the driven oscillator (Eqs.~6 and~7). When the engineered dissipation is resonant with the $\transeg$ transition ($\delta \omegadiss = 0$), and $\dE_1 \neq 0$, these expressions yield the effective dissipators acting on $\transeg$:
\begin{align}
     \kappadiss(1+\nthb) D[ \hat{d} ] &= \frac{\kappacav \, \gdiss^{2}}{(\kappacav/2)^{2}+(\dE_1/2)^{2}}(1+\nthb)\, D[\Pi_0 \aop \Pi_1],
     \label{eq:eng_diss_dissipators1}\\
     \kappadiss\nthb D[ \hat{d}^\dagger ] &= \frac{\kappacav \, \gdiss^{2}}{(\kappacav/2)^{2}+(\dE_1/2)^{2}}\, \nthb \, D[\Pi_1 \aod \Pi_0],
    \label{eq:eng_diss_dissipators2}
\end{align}
where $\Pi_{i} = \sum_\pm \ket{\psi_{i}^\pm}\bra{\psi_{i}^\pm}$ and $\kappadiss$ is the dissipation rate. These terms describe relaxation and, for finite $\nthb$, excitation processes on the $\transge$ transition with linewidth $\kappacav$. The resulting effective dissipators lead to the effective master equation
\begin{align}
\frac{\mathrm{d}\hat{\rho}}{\mathrm{d}t} &= -\frac{i}{\hbar}\,[\Hkcq + \Hstark, \hat{\rho}] 
+ \kappaqubit (1+\nth) \,D[\aop] \hat{\rho} 
+ \kappaqubit \nth \,D[\aod] \hat{\rho} \notag \\
&\quad + \kappadiss (1+\nthb)\,D[\hat{d}] \hat{\rho} 
+ \kappadiss \nthb\,D[\hat{d}^\dagger] \hat{\rho},
\label{eq:KCQ_eng_diss_simple}
\end{align}
where \( \Hkcq \) is the \KQ~Hamiltonian (Eq.~\eqref{eq:Ham_kcq}) and  \( \Hstark \) accounts for the Stark shift arising from the squeezing-drive pump (Eq.~\eqref{eq:Ham_stark} with $\xicqr{eff}=0$).

\begin{figure*}[t]
\includegraphics[width=\figwidthWide]{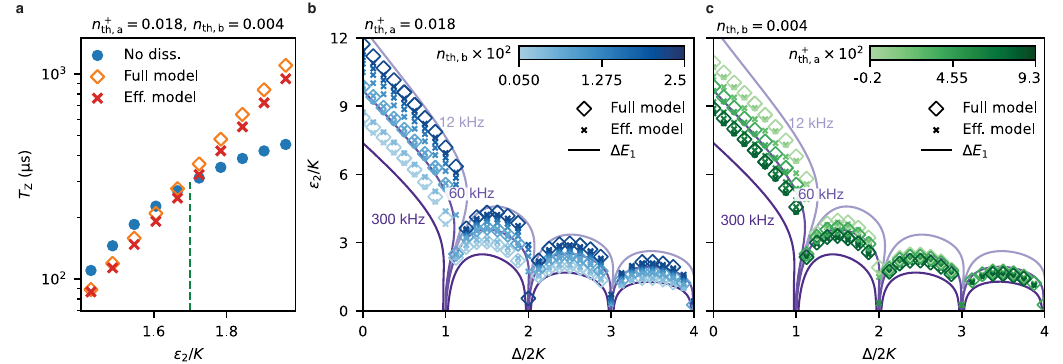}
\caption{\label{fig:SI_sim_diss_map} 
\textbf{Numerical simulation of $\epstr$.} 
\textbf{a}, Simulated $\Tz$ as a function of squeezing-drive amplitude $\epstwo$ without engineered dissipation (blue dots), with engineered dissipation obtained from the full master-equation model Eq.~\eqref{eq:lindblad_noise_general} (orange diamonds), and from the effective model (Eq.~\eqref{eq:KCQ_eng_diss_simple}) (red crosses). The green dashed line identifies $\epstr$ extracted from the full-model simulations. All simulations are performed at fixed squeezing-drive detuning $\Delta = 7K$.
\textbf{b,c}, Threshold $\epstr$ as a function of $\Delta$. In \textbf{b} (\textbf{c}), we fix $\nth = \nthvalsim$ ($\nthb = \nthbvalfigthree$) and vary $\nthb$ ($\nth$). Open diamonds (crosses) represent $\epstr$ obtained with the full (effective) model. Purple lines indicate isolines of the $\ket{\estate}$ splitting $\dE_1$. 
}
\end{figure*}

We benchmark the effective model (Eq.~\eqref{eq:KCQ_eng_diss_simple}) against the full model, in which the cavity mode is treated explicitly (Eq.~\eqref{eq:lindblad_noise_general}). We set $\Delta = 7K$ and simulate the bit-flip time as a function of $\epstwo$, both with and without engineered dissipation. The simulations use the measured parameters $\kappaqubit$, $\kappacav$, the effective thermal photon numbers $\nth = \nthvalsim$ and $\nthb = \nthbvalfigthree$, and the independently calibrated engineered coupling rate $\gdiss/2\pi = \gdissval$ ($\kappadiss/2\pi = \kappadissfigthree$), matching the conditions used in Fig.~\figfour~of the main text.
We initialize the oscillator in $\ket{+Z}$, and compute $\Zdiss$ as a function of the simulation time to extract $\Tz$. Figure~\ref{fig:SI_sim_diss_map}a shows the comparison between the simulated $\Tz$ obtained without engineered dissipation (setting $\gdiss = 0$) (blue dots), with engineered dissipation (orange diamonds), and with the effective model (red crosses). The effective model predicts an $\epstr$ within $3\%$ of the value extracted using the full model. The small deviation arises because $\gdiss$ is approaching the value of $\kappacav$, meaning that the adiabatic elimination approximation used in the derivation of Eqs.~\eqref{eq:eng_diss_dissipators1} and \eqref{eq:eng_diss_dissipators2} does not fully hold.

We now use the effective model to analyze how $\epstr$ depends on the thermal photon number of the oscillator and cavity, $\nth$ and $\nthb$.
In Figs.~\ref{fig:SI_sim_diss_map}b,c we compare simulations with the effective model (crosses, Eq.~\eqref{eq:KCQ_eng_diss_simple}) and the full model (open diamonds, Eq.~\eqref{eq:lindblad_noise_general}), showing good overall agreement between the two models. In both cases, $\epstr$ follows the trend of the $\dE_1$ isolines.  
We first perform simulations for a fixed $\nth$ and for different values of $\nthb$ (Fig.~\ref{fig:SI_sim_diss_map}b). We find that $\epstr$ increases with $\nthb$, following isolines of smaller $\dE_1$. The dependence is strongest at low $\nthb$ and gradually weakens as $\nthb$ increases.  
We also perform simulations for a fixed $\nthb$ and for different values of $\nth$ (Fig.~\ref{fig:SI_sim_diss_map}c). In this case, $\epstr$ decreases with $\nth$, following isolines of larger $\dE_1$. Here too, the effect is most pronounced at small $\nth$ and becomes weaker as $\nth$ increases.

Using the insights from the effective model described above, we now develop an intuitive description of how engineered dissipation affects the bit-flip time, and use this to explain the observed dependence of $\epstr$ on $\nthb$ and $\nth$.
The engineered dissipation introduces a cooling rate (Eq.~\eqref{eq:eng_diss_dissipators1}) and, for $\nthb > 0$, a corresponding excitation rate (Eq.~\eqref{eq:eng_diss_dissipators2}) on the $\transge$ transition. In our experiment, $\nthb \ll 1$, so the engineered cooling rate dominates over the excitation rate. As a result, the engineered dissipation reduces the leakage population $\pe$ (see Fig.~\figthree~of the main text). However, a smaller leakage population does not necessarily translate into a longer bit-flip time.
This is the case for $\dE_1/\hbar\gg\kappadiss$ ($\epstwo\ll\epstr$), where the tunneling rate between states in the $\ket{\estate}$ manifold is much larger than the dissipation rate $\kappadiss$. Since any leakage population undergoes tunneling much faster than it can be brought back to the \KQ~manifold, and the engineered interaction introduces additional excitations, the bit-flip probability in the system increases. This gives a reduced bit-flip time with engineered dissipation, compared to without engineered dissipation. 
Conversely, when $\dE_1/\hbar \ll \kappadiss$ ($\epstwo\gg\epstr$), leakage population is brought to the \KQ~manifold much faster than it can tunnel, meaning that engineered dissipation extends the bit-flip time.

We next use this intuition to interpret the dependence of $\epstr$ on $\nthb$ and $\nth$. 
The threshold, $\epstr$, corresponds to the $\epstwo$ for which the bit-flip time with engineered dissipation is equal to that without engineered dissipation. From the description above, we understand this as the $\epstwo$ for which the reduced tunneling probability due to the larger engineered cooling rate is exactly compensated for by the increased excitation rate when $\nthb>0$. This means that, while each excitation event has a smaller chance of leading to a bit-flip, the number of excitation events increases by enough to leave the total bit-flip rate unchanged.
Further, when $\nthb$ increases, the effective excitation rate induced by the engineered dissipation also increases. At $\epstr$, this larger excitation rate needs to be balanced by a smaller probability of tunneling, which requires a smaller energy splitting $\dE_1$ and therefore larger $\epstwo$. Consequently, $\epstr$ increases with $\nthb$.
Increasing $\nth$ has a different effect. The excitation and cooling rates, both with and without engineered dissipation, grow with $\nth$ by the same amount. While the excitation rate is larger, there is no change to the excitation introduced by the engineered interaction (which only depends on $\nthb$). However, at $\epstr$ the overall increase in the cooling rate needs to be balanced by a decrease in tunneling probability, which requires a larger energy splitting $\dE_1$ and therefore a smaller $\epstwo$. Hence, $\epstr$ decreases with $\nth$.

Note that we expect the threshold $\epstr$ to be insensitive to which noise model generates the leakage. Since $\epstr$ depends on the rates at which population is excited out of and cooled back into the \KQ~manifold, independent of the microscopic origin of the excitation, noise models that reproduce the same $\pe$ are expected to give the same $\epstr$. 

We confirm this intuition with numerical simulations of $\epstr$ for two different noise models: additional thermal heating ($\nthplus$) and dephasing noise ($\kappaphi$).
All parameters are independently measured, other than the effective parameters $\nthplus$ and $\kappaphi$ which are set to the fitted values from Section~\ref{sec:pe_model}. The results are shown in Table~\ref{tab:si_epstr}. 
\begin{table}
    \centering
    \setlength{\tabcolsep}{10pt}
    \setlength{\extrarowheight}{3pt}
    \begin{tabular}{|c||c|c|}
        \hline
         $\Delta$ & $\epstr$ for $\nthplus=\nthplusval$ & $\epstr$ for $\kappaphi/2\pi=\kappaphival$ \\
         \hline\hline
         $0$    & $9.71K$ & $9.19K$ \\
         \hline
         $1.5K$ & $6.76K$ & $6.42K$ \\
         \hline
         $3K$   & $3.59K$ & $3.55K$ \\
         \hline
         $7K$   & $1.67K$ & $1.68K$ \\
         \hline
    \end{tabular}
    \caption{\label{tab:si_epstr}\textbf{Dissipation threshold values $\epstr$ for different noise models.} Simulated $\epstr$ for different values of detuning $\Delta$, assuming additional thermal heating ($\nthplus$, Eq.~\eqref{eq:noise_nthplus}) or dephasing noise ($\kappaphi$, Eq.~\eqref{eq:noise_kappaphi}). Single-photon loss ($\kappaqubit$, $\kappacav$) and the thermal photon number of the undriven oscillator ($\nth$) are also included and set to their measured values. Effective parameters are set to the values that reproduce the leakage population $\pe$ (see Sec.~\ref{sec:pe_model}). An effective cavity thermal photon number of $\nthb=0.004$ is included in all cases (see Fig.~\figfour e of the main text).}
\end{table}
The threshold value does not change significantly between the different noise models. This is consistent with the results presented in Fig.~\ref{fig:SI_sim_diss_map}c, where $\epstr$ is relatively insensitive to the value of $\nthplus$. The noise model therefore does not change the conclusions of Fig.~\figfour e, and the insights into the \KQ-cavity interaction.

\subsection{Impact of dissipation frequency on \KQ~bit-flip time}

\begin{figure*}[t]
\includegraphics[width=\figwidthWide]{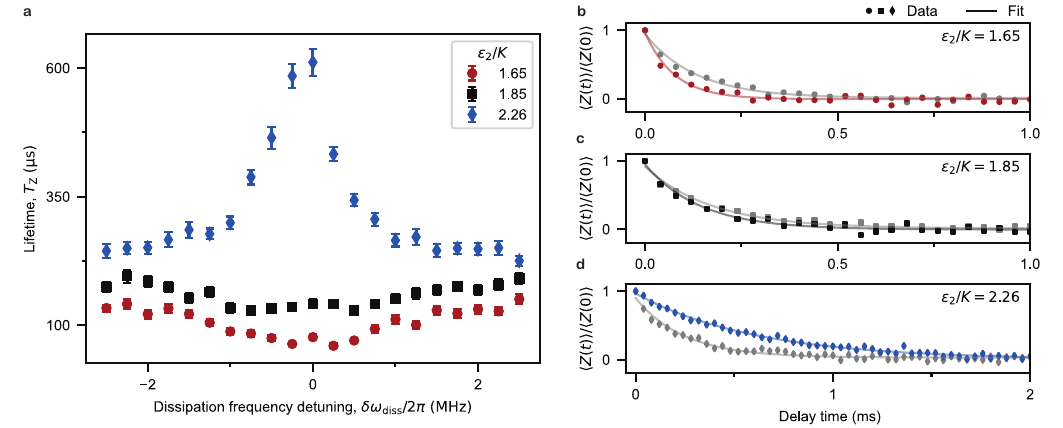}
\caption{\label{fig_SI_diss_Tz} 
\textbf{Effect of engineered dissipation on the bit-flip time $\Tz$.} 
\textbf{a}, Measured $\Tz$ at $\Delta=7K$ as a function of dissipation frequency detuning $\delta\omegadiss$ for three values of the squeezing strength $\epstwo = 1.65K$ (red circles), $1.85K$ (black squares), and $2.26K$ (blue diamonds). Error bars indicate the uncertainty obtained from fits to a single-exponential decay. 
\textbf{b–d}, Time-domain traces of $\langle Z(t)\rangle/\langle Z(0)\rangle$ for $\delta\omegadiss/2\pi = 0$ (colored markers) and $\SI{-2.5}{\mega\hertz}$ (gray markers), for the same values of $\epstwo/K$ as in \textbf{a}. Solid lines are single-exponential fits used to extract $\Tz$. Measurements are performed with an engineered dissipation rate $\kappadiss /2\pi = \kappadissfigthree$.}
\end{figure*}

In this subsection, we demonstrate that measurements of $\delta\Zdiss$ (see Fig.~\figfour b) are a reliable indicator of the bit-flip time $\Tz$.
To do this, we directly measure $\Tz$ at $\Delta=7K$ in the presence of engineered dissipation, as a function of dissipation frequency detuning $\delta\omegadiss$. The engineered interaction rate is set to $\gdiss/2\pi = \gdissval$ ($\kappadiss/2\pi = \kappadissfigthree$) such that the measurement parameters are identical to those in Fig.~\figfour b of the main text. We perform the $\Tz(\delta\omegadiss)$ measurement for three values of squeezing-drive amplitude $\epstwo$, corresponding to values below, at and above the transition $\epstr$. The results are shown in Fig.~\ref{fig_SI_diss_Tz}a for $\epstwo = 1.65K$ (red circles), $1.85K$ (black squares) and $2.26K$ (blue diamonds). For large dissipation frequency detuning $\delta\omegadiss/2\pi=\SI{-2.5}{\mega\hertz}$, the $\Tz$ values are similar -- small differences are attributed to the different $\epstwo$ values. However, when the engineered dissipation is on resonance with the $\transge$ transition ($\delta\omegadiss=0$), $\Tz$ decreases for $\epstwo<\epstr$ and increases for $\epstwo>\epstr$. This is illustrated in Figs.~\ref{fig_SI_diss_Tz}b-d, which show the time-domain traces of $\Tz$ measurements at each $\epstwo$ for $\delta\omegadiss=0$ (colored markers) and $\delta\omegadiss/2\pi=\SI{-2.5}{\mega\hertz}$ (gray markers). 
We conclude that the measured signal, after a fixed time, is therefore representative of a change in $\Tz$, verifying the approach used in Fig.~\figfour~of the main text. Note that the data presented in Fig.~\ref{fig_SI_diss_Tz}d is identical to that of Fig.~\figfour c of the main text.

\newpage
\subsection{Kerr-cat qubit coherence times at cardinal points in presence of engineered dissipation}

\begin{figure*}[t]
\includegraphics[width=\figwidthWide]{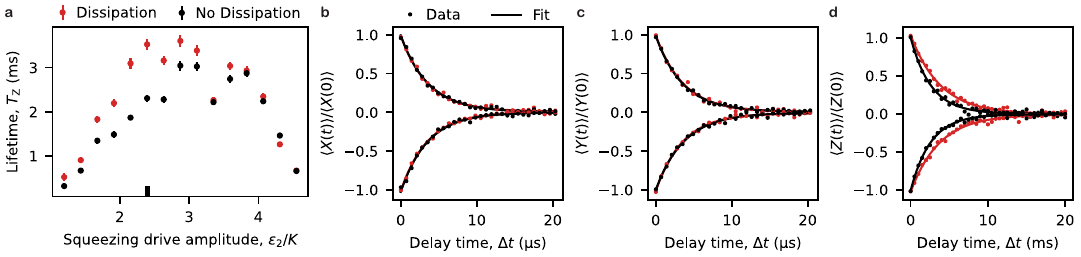}
\caption{\label{fig_SI_cardinalpts} 
\textbf{Kerr-cat qubit coherence times at \workingdelta~with and without engineered dissipation.}
\textbf{a}, Kerr-cat qubit bit-flip times $\Tz$ for increasing squeezing-drive amplitude $\epstwo$, with (red markers) and without (black markers) dissipation applied to the $\transge$ transition. Error bars indicate the uncertainty of an exponential fit to the experimental data. The marker on the x-axis identifies $\epstwo=2.4K$, corresponding to b-d.
\textbf{b-d}, Coherence times of the cardinal points of the Kerr-cat qubit Bloch sphere at $\epstwo=2.4K$, with (red) and without (black) dissipation applied to the $\transge$ transition. Data (markers) are normalized to the value at zero delay time, $\Delta t=0$, and are fitted with an exponential decay (solid lines). The engineered dissipation rate used for all results in this figure is $\kappadiss /2\pi = \SI{93}{\kilo \hertz}$.}
\end{figure*}

In this subsection, we show measurements of the coherence times at the \KQ~Bloch sphere cardinal points in the presence of engineered dissipation. Measurements were performed at a squeezing-drive detuning \workingdelta, as in Fig.~\figfive~of the main text. In particular, we show the bit-flip time enhancement due to engineered dissipation for different squeezing-drive amplitudes $\epstwo$ (Fig.~\ref{fig_SI_cardinalpts}a), and the effect of engineered dissipation on all cardinal points of the Bloch sphere at $\epstwo=2.4K$ (Figs.~\ref{fig_SI_cardinalpts}b-d).

We first investigate the effect of engineered dissipation on $\Tz$ as a function of $\epstwo$, for a fixed engineered interaction rate $\gdiss/2\pi = \SI{140}{\kilo\hertz}$ ($\kappadiss/2\pi = \SI{93}{\kilo\hertz}$). Figure~\ref{fig_SI_cardinalpts}a shows $\Tz$ for different $\epstwo$, measured using the pulse sequence presented in Fig.~\figfive a of the main text. The measured $\Tz$ in the absence of engineered dissipation is indicated by black markers. For increasing $\epstwo$, $\Tz$ increases up to a maximum of $\SI{3}{\milli\second}$ at $\epstwo=2.9K$. The $\Tz$ value in the presence of engineered dissipation is indicated by red markers. Dissipation is applied to the $\transge$ transition, with the transition frequency calibrated using incoherent spectroscopy measurements of the oscillator spectrum~\cite{frattini_observation_2024}. Similar to the case without dissipation, $\Tz$ increases to a peak value around $\epstwo=2.9K$. However, $\Tz$ values with engineered dissipation are markedly larger than those without dissipation, reaching a maximum value of $\SI{3.6}{\milli\second}$ at $\epstwo=2.9K$. For larger $\epstwo$, $\Tz$ decreases for the cases with and without dissipation. This might be explained by additional leakage processes that suppress $\Tz$ and are not corrected by the engineered dissipation \cite{Venkatraman2024_nl_diss, BlaisRescueKCQ,Dai_Hazra_2025}.

The working point \workingepstwo~corresponds to the maximum observed increase in $\Tz$ due to engineered dissipation, increasing from $\SI{2.3}{\milli\second}$ to $\SI{3.5}{\milli\second}$. This corresponds to the working point of results presented in Figs.~\figtwo, \figfour~and \figfive~of the main text. Figures~\ref{fig_SI_cardinalpts}b-d show measurements of the coherence times of the Bloch sphere cardinal points at this value of $\epstwo$, with and without dissipation. Measurements on the equator of the Bloch sphere, i.e. $\ket{\pm X}$ (Fig.~\ref{fig_SI_cardinalpts}b) and $\ket{\pm Y}$ (Fig.~\ref{fig_SI_cardinalpts}c), were performed using the pulse sequence presented in Fig.~\figfive b of the main text. We did not observe any change in coherence times for states on these axes due to the engineered dissipation. Data (markers) were fitted with an exponential decay (solid lines), with the results for the coherence times shown in Table~\ref{tab:Tz_cardinal}. Coherence-time values presented in Table~\ref{tab:Tz_cardinal}, Fig.~\ref{fig_SI_cardinalpts}a and Fig.~\figfive~of the main text differ slightly due to fluctuations over the three days during which these measurements were taken.

\begin{table}[h!]
	\centering
	\begin{tabular}{|l||c|c|c|c|c|c|}
		\hline
        & \textbf{$T_{\mathrm{+X}}~(\mathrm{\mu s})$} 
		& \textbf{$T_{\mathrm{-X}}~(\mathrm{\mu s})$}
		& \textbf{$T_{\mathrm{+Y}}~(\mathrm{\mu s})$}
		& \textbf{$T_{\mathrm{-Y}}~(\mathrm{\mu s})$} 
		& \textbf{$T_{\mathrm{+Z}}~(\mathrm{ms})$}
		& \textbf{$T_{\mathrm{-Z}}~(\mathrm{ms})$}\\ \hline   
        \textbf{No Dissipation} & $3.6\pm0.1$ & $3.5\pm0.1$ & $3.6\pm0.2$ & $3.5\pm0.1$ & $2.6\pm0.1$ & $2.6\pm0.1$\\ \hline 
        \textbf{Dissipation} & $3.6\pm0.1$ & $3.5\pm0.1$ & $3.6\pm0.1$ & $3.8\pm0.1$ & $3.8\pm0.1$ & $3.7\pm0.1$ \\ \hline
       
	\end{tabular}  
	\caption{\textbf{Coherence times of Kerr-cat qubit (\KQ) cardinal points.} Summary of the measured coherence times at the cardinal points of the \KQ~Bloch sphere with and without engineered dissipation, for a squeezing-drive detuning \workingdelta~and strength $\epstwo=2.4K$. The engineered interaction rate is $\gdiss/2\pi = \SI{140}{\kilo\hertz}$ ($\kappadiss/2\pi=\SI{93}{\kilo\hertz}$). }
	\label{tab:Tz_cardinal}
\end{table}

\subsection{Modeling the \KQ~bit-flip time}
\label{sec:Tz_model}

In this subsection, we simulate the bit-flip time $\Tz$ under the noise models of Section~\ref{subsec:noisemodels} and compare the results with the measurements of Fig.~\figfive c of the main text. We then use the simulations to identify the two processes that limit $\Tz$ in our model, namely the thermal photon number of the cavity and leakage beyond the first excited manifold. 

We simulate $\Tz$ for each of the noise models introduced in Section~\ref{subsec:noisemodels}, using the effective dissipators of Eqs.~\eqref{eq:eng_diss_dissipators1} and~\eqref{eq:eng_diss_dissipators2} to describe the engineered dissipation. The effective parameters $\nthplus$ and $\kappaphi$ are fixed to the values that reproduce the measured $\pe$ at $\gdiss = 0$ (Section~\ref{sec:pe_model}). We initialize the oscillator in $\ket{+Z}$ and extract $\Tz$ from the decay of $\Zdiss$ as a function of the delay time, for each engineered interaction rate $\gdiss$. Delay times shorter than $5/\kappaqubit \approx \SI{350}{\micro\second}$ are excluded from the fit, so that the transient intra-well relaxation of the initial state does not bias the extracted $\Tz$~\cite{BlaisRescueKCQ}.

\begin{figure*}
\includegraphics[width=\figwidthWide]{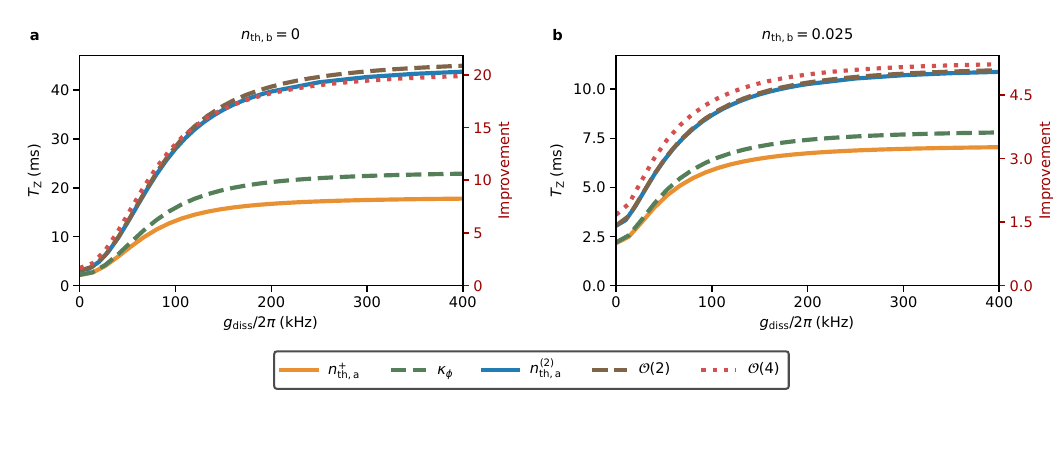}
\caption{\label{fig_SI_Tz_models}
\textbf{Simulated bit-flip time for the different noise models.}
Simulated bit-flip time $\Tz$ as a function of the engineered interaction rate $\gdiss$, for a cavity thermal photon number $\nthb = 0$ (\textbf{a}) and $\nthb = \nthbval$ (\textbf{b}). Every curve includes single-photon loss on both modes ($\kappaqubit$, $\kappacav$) and the measured thermal photon number in the undriven oscillator ($\nth = \nthval$), together with one additional noise model, which is either additional thermal heating of the oscillator ($\nthplus$, orange solid), effective dephasing ($\kappaphi$, green dashed) or two-photon heating beyond the RWA ($\nthtwophoton$, blue solid). The brown dashed and red dotted curves instead add Hamiltonian terms up to second ($\mathcal{O}(2)$) and fourth ($\mathcal{O}(4)$) order in the expansion of the Josephson potential. Effective parameters are fixed to the values that reproduce the measured $\pe$ at $\gdiss = 0$ (Section~\ref{sec:pe_model}). The right y-axes show the same curves as a relative improvement with respect to $\Tz$ at $\gdiss = 0$. Simulations are performed at the working point of Fig.~\figfive c of the main text, \workingpoint.}
\end{figure*}

The results are shown in Fig.~\ref{fig_SI_Tz_models}, both without ($\nthb = 0$) and with ($\nthb = \nthbval$) a finite cavity thermal photon number. For every noise model, $\Tz$ increases at small $\gdiss$ and then saturates, following the trend of the measurement. The saturation value is markedly lower for $\nthb = \nthbval$ than for $\nthb = 0$, bringing the simulation closer to the measured values. Beyond this qualitative agreement, however, the simulations do not quantitatively reproduce the data. The simulated $\Tz$ at $\gdiss = 0$ does not match the measured value. The simulated $\Tz$ at large $\gdiss$ is consistently higher than measured. The measured saturation also occurs at smaller $\gdiss$ than in the simulation. The noise models considered here are therefore incomplete as a description of bit flips in our device, when constrained to effective values that reproduce the measured leakage population.

Some noise mechanisms not considered here may have a significant effect on $\Tz$ while making only a small contribution to $\pe$. For example, high-order processes mediated by an additional mode can induce leakage to delocalized states above the double-well potential barrier~\cite{BlaisRescueKCQ}. Even when the excitation rate into these states is small, interwell transitions can occur rapidly while the system occupies them, resulting in a high probability of a bit flip per leakage event. At the same time, their low excitation rate and short steady-state occupation does not produce an appreciable change in the population of the first leakage manifold, $\pe$. Such processes can therefore substantially reduce $\Tz$ while having little effect on the measured $\pe$. 

\begin{figure*}[h]
\includegraphics[width=\figwidthWide]{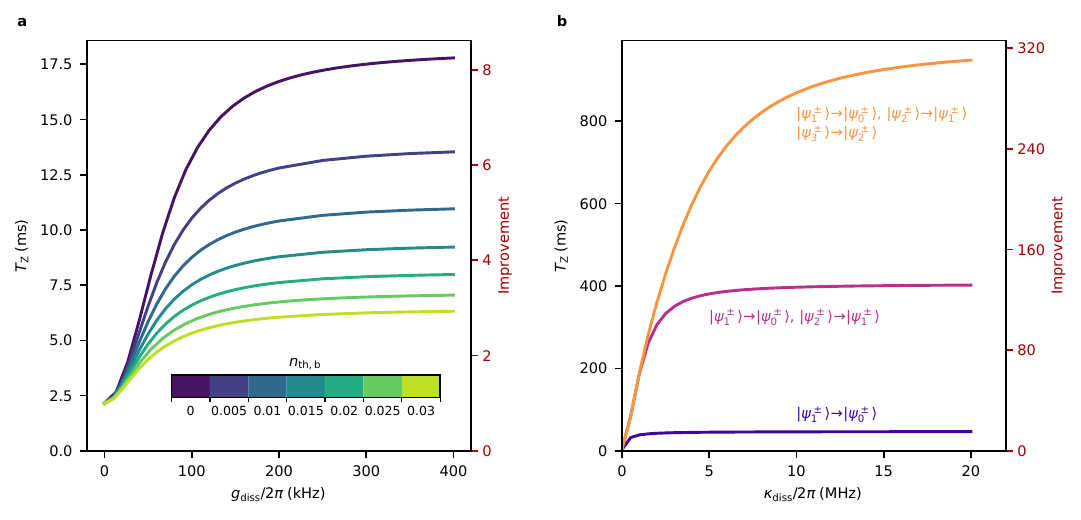}
\caption{\label{fig_SI_Tz_limits}
\textbf{Limits to the bit-flip time set by the cavity population and by higher leakage manifolds.}
\textbf{a}, Simulated bit-flip time $\Tz$ as a function of the engineered interaction rate $\gdiss$, for cavity thermal photon numbers between $\nthb = 0$ and $\nthb = 0.03$ (color scale).
\textbf{b}, Simulated $\Tz$ as a function of the engineered dissipation rate $\kappadiss$, with dissipation acting on the $\transeg$ transition only (purple), on $\transeg$ and $\transfe$ (magenta), and additionally on $\transhf$ (orange), for $\nthb = 0$.
Both panels use the independently measured parameters together with the additional thermal heating $\nthplus = \nthplusval$ that reproduces the measured $\pe$ at $\gdiss = 0$, at the working point \workingpoint. The right y-axes show the same curves as a relative improvement with respect to $\Tz$ in the absence of engineered dissipation. Simulations are performed for \workingpoint.}
\end{figure*}

Although our model does not reproduce the measured bit-flip times, it identifies two distinct limits to $\Tz$, shown in Fig.~\ref{fig_SI_Tz_limits}. The first is the cavity thermal photon number $\nthb$. Increasing $\nthb$ lowers the value at which $\Tz$ saturates (Fig.~\ref{fig_SI_Tz_limits}a), because the engineered interaction excites population to the leakage manifold at a rate proportional to $\nthb$ (Eq.~\eqref{eq:eng_diss_dissipators2}). Reducing the cavity population is therefore a prerequisite for benefiting from stronger engineered dissipation.
The second limit persists at $\nthb = 0$, where $\Tz$ still saturates. This is set by population reaching leakage manifolds above $\ket{\estate}$, which our dissipation does not address. Figure~\ref{fig_SI_Tz_limits}b supports this interpretation. Adding dissipation on the $\transfe$ and then the $\transhf$ transitions raises the saturation value of $\Tz$ by more than an order of magnitude compared with dissipating on the $\transeg$ transition alone.
Within this model, reducing the cavity thermal photon number and extending the engineered dissipation to higher leakage manifolds are thus both routes to longer bit-flip times. 

\end{document}